\documentclass[11pt,fleqn,reqno]{amsart}

\usepackage[margin=1in,headsep=18pt]{geometry}
\usepackage[T1]{fontenc}
\usepackage{amssymb}
\usepackage{mathtools}
\usepackage{nicematrix}
\usepackage{mathrsfs}
\usepackage{enumitem}
\usepackage{microtype}
\usepackage{needspace}
\usepackage{setspace}  
\usepackage{xcolor}
\usepackage{hyperref}
\hypersetup{
  plainpages=false,
  bookmarksopen=true,
  colorlinks=false,
  pdfborder={0 0 0},
  pdftitle={Optimal XOR Lemmas for Information Complexity},
}

\newtheoremstyle{myplain}{10pt}{10pt}{\itshape}{}{\scshape}{.}{.5em}{}
\newtheoremstyle{mydefinition}{10pt}{10pt}{}{}{\scshape}{.}{.5em}{}

\theoremstyle{myplain}
\newtheorem{theorem}{Theorem}[section]
\newtheorem{lemma}[theorem]{Lemma}
\newtheorem{corollary}[theorem]{Corollary}
\newtheorem{proposition}[theorem]{Proposition}
\newtheorem{fact}[theorem]{Fact}
\newtheorem{claim}[theorem]{Claim}
\newtheorem*{restatedthm}{\restatedheading}
\newcommand{\restatedheading}{}
\newenvironment{restated}[2]
  {\renewcommand{\restatedheading}{#1}\begin{restatedthm}[#2]}
  {\end{restatedthm}}

\theoremstyle{mydefinition}
\newtheorem{definition}[theorem]{Definition}
\newtheorem{remark}[theorem]{Remark}

\numberwithin{equation}{section}

\setlist[enumerate]{label=\normalfont(\roman*), labelsep=5mm}

\makeatletter
\def\@seccntformat#1{%
  \protect\textup{%
    \protect\@secnumfont
    \expandafter\protect\csname format#1\endcsname
    \csname the#1\endcsname
    \protect\@secnumpunct
  }%
}

\newcommand{\SparseDotfill}{\leavevmode\leaders\hb@xt@.7em{\hss.\hss}\hfill\kern\z@}
\def\@tocline#1#2#3#4#5#6#7{\relax
  \ifnum #1>\c@tocdepth 
  \else
    \par \addpenalty\@secpenalty\addvspace{\ifnum #1=1 2mm \else #2\fi}%
    \begingroup \hyphenpenalty\@M
    \@ifempty{#4}{%
      \@tempdima\csname r@tocindent\number#1\endcsname\relax
    }{%
      \@tempdima#4\relax
    }%
    \parindent\z@ \leftskip#3\relax \advance\leftskip\@tempdima\relax
    \rightskip\@pnumwidth plus4em \parfillskip-\@pnumwidth
    \ifnum #1=1 \bfseries #5\else #5\fi
    \leavevmode\hskip-\@tempdima
    \ifcase #1
      \or\or \hskip 1em \or \hskip 2em \else \hskip 3em \fi%
    #6 \nobreak\relax
    {\ifnum #1=1\hfill \else \SparseDotfill\fi}
    \hbox to\@pnumwidth{\@tocpagenum{
      \ifnum #1=1 \bfseries \fi #7}}\par
    \nobreak
    \endgroup
  \fi}
\makeatother

\renewcommand{\qedsymbol}{\leavevmode\hbox to.77778em{%
  \hfil\vrule\vbox to.875em{\hrule width.35em\vfil\hrule}\vrule\hfil}}

\renewcommand{\mathcal}[1]{\mathscr{#1}}
\renewcommand{\epsilon}{\varepsilon}
\renewcommand{\phi}{\varphi}
\renewcommand{\le}{\leqslant}
\renewcommand{\leq}{\leqslant}
\renewcommand{\ge}{\geqslant}
\renewcommand{\geq}{\geqslant}

\newcommand{\1}{\mathbf{1}}
\DeclareMathOperator*{\Exp}{\mathbf{E}}
\DeclareMathOperator*{\Prob}{\mathbf{P}}
\DeclareMathOperator{\IC}{IC}
\DeclareMathOperator{\err}{err}
\DeclareMathOperator{\supp}{supp}
\DeclareMathOperator{\IP}{IP}
\newcommand{\AND}{\mathrm{AND}}
\DeclareMathOperator{\dom}{dom}
\newcommand{\Dcal}{\mathcal{D}}
\newcommand{\Xcal}{\mathcal{X}}
\newcommand{\Ycal}{\mathcal{Y}}
\newcommand{\Zcal}{\mathcal{Z}}

\DeclareMathOperator{\SIC}{SIC}
\DeclareMathOperator{\VIC}{VIC}
\DeclareMathOperator{\LIC}{LIC}
\DeclareMathOperator{\TV}{TV}
\DeclareMathOperator{\disc}{disc}
\DeclareMathOperator{\margin}{m}
\DeclareMathOperator{\sgn}{sgn}
\DeclareMathOperator{\Var}{Var}

\newcommand{\Ccal}{\mathcal{C}}
\DeclarePairedDelimiterX{\kldelim}[2]{(}{)}{#1\,\delimsize\Vert\,#2}
\newcommand{\kl}[2]{\operatorname{KL}\kldelim{#1}{#2}}

\newcommand{\klnat}[2]{\operatorname{KL}_{\rm nat}\kldelim{#1}{#2}}

\DeclarePairedDelimiter{\abs}{\lvert}{\rvert}
\begin{document}

\title{Optimal XOR Lemmas for Information Complexity}
\author{Alexander A. Sherstov}
\thanks{Computer Science Department, UCLA, Los Angeles, CA~90095.
  Work supported by NSF grant CCF-2220232.\\
  Email: \texttt{sherstov@cs.ucla.edu}}
\date{}

\begin{abstract}
Information complexity studies how much information the parties to a communication
protocol must reveal to each other about their inputs to accomplish a given
task. For a Boolean-valued communication problem $F,$ let $\IC_\epsilon(F)$
denote its information complexity with error $\epsilon,$ and let $F^{\oplus
n}$ denote the XOR of $n$ independent instances of $F.$ A fundamental question
is whether information complexity obeys an XOR lemma, $\IC_{1/3}(F^{\oplus
n})=\Omega(n\IC_{1/(3n)}(F)),$ which would match the trivial upper bound.
Previous work either established the $n$-fold increase in information cost
without the $n$-fold decrease in error (Bar-Yossef et al., FOCS 2002) or
proved a lower bound with a subtracted penalty proportional to the input
length of $F$ (Sawettamalya and Yu, STOC 2025). In particular, the question
has remained open even for the XOR of $n$ or fewer copies of a function
$F\colon\{0,1\}^n\times\{0,1\}^n\to\{0,1\}.$

We completely resolve the question of XOR lemmas for information complexity,
for every error $0\leq\epsilon\leq1/3.$ For \emph{total} functions $F,$ we
prove that $\IC_\epsilon(F^{\oplus n})=\Theta(n\IC_{\epsilon/n}(F))$ unless
$F$ has the degenerate form $F(x,y)=a(x)\oplus b(y),$ in which case
$\IC_\epsilon(F^{\oplus n})=O(1).$ For the richer class of \emph{partial}
functions, we obtain the characterization $\IC_\epsilon(F^{\oplus
n})=\Theta(\IC_\epsilon(F)+n\SIC_{\epsilon/n}(F)),$ where $\SIC$ is the
information complexity of protocols in which Alice and Bob privately output
bits whose XOR is the answer.  Finally, we prove an optimal \emph{distributional}
XOR lemma, for every function $F$ and every distribution. We obtain an
optimal direct sum theorem to accompany each of our XOR lemmas. All implicit
constants are absolute. Our work contributes a novel combination of 
information-theoretic and geometric techniques.
\end{abstract}

\maketitle
\thispagestyle{empty}

\newpage
\thispagestyle{empty}
\begingroup
\tableofcontents
\endgroup
\newpage
\thispagestyle{empty}

\setcounter{page}{1}
\section{Introduction}\label{sec:intro}

Information complexity quantifies how much information the parties to a communication
protocol must reveal to each other about their inputs to accomplish a
given task. In Yao's two-party model~\cite{Yao79}, Alice and Bob receive
inputs $x\in\Xcal$ and $y\in\Ycal,$ respectively, and communicate by means
of a randomized protocol $\pi$ whose purpose is to compute the value
$F(x,y)$ of a given communication problem $F.$ For a probability
distribution $\mu$ on the domain of $F,$ the \emph{internal information cost} of
$\pi$ under $\mu$ is
\[
  \IC(\pi,\mu)=I(M;X\mid Y,R)+I(M;Y\mid X,R),
\]
where $(X,Y)\sim\mu,$ and where $R$ and $M$ denote the public randomness
and transcript of $\pi$ on input $(X,Y).$ The first summand is the
information that Bob learns about Alice's input from the transcript, and
the second is the information that Alice learns about Bob's input. The
\emph{information complexity} $\IC_\epsilon(F)$ is the infimum of
$\sup_\mu\IC(\pi,\mu)$ over all protocols $\pi$ that compute $F$ with
error at most $\epsilon$ on every input~\cite{Braverman15}. Its
distributional counterpart $\IC_\epsilon(F,\mu)$ is the infimum of
$\IC(\pi,\mu)$ over the protocols whose average error is at most
$\epsilon$ when the input is distributed according to $\mu.$ These
worst-case and distributional notions of information complexity are
related by a minimax theorem of Braverman~\cite{Braverman15}
(Fact~\ref{fact:minimax}), analogous to Yao's minimax
principle~\cite{Yao83} for randomized and distributional
communication complexity. We give the formal definitions in
Section~\ref{sec:preliminaries}. Since a transcript cannot reveal more
information than its length, information complexity is a lower bound on
randomized communication complexity. This subject grew out of the work of
Chakrabarti, Shi, Wirth, and Yao~\cite{CSWY01}, who measured how much an
external observer learns about the inputs from the transcript. The
internal measure above, which is now the standard notion of information
complexity, was used implicitly by Bar-Yossef, Jayram, Kumar,
and Sivakumar~\cite{BYJKS04} and was defined explicitly by Barak,
Braverman, Chen, and Rao~\cite{BBCR10}.

Information complexity has since become a central area of communication
complexity, with a substantial theory of its own; see the
surveys~\cite{Wei15,Braverman19,Braverman23}. Much of this theory concerns
the relationship between information and communication.  Information
complexity equals the amortized communication complexity of many
independent copies~\cite{BR14,Braverman15}. A protocol with information
cost $I$ can be compressed to communication cost
$2^{O(I)}$~\cite{Braverman15}, and this exponential dependence on $I$ is
the best possible in general~\cite{GKR16}, whereas much better compression
is possible when Alice and Bob's inputs are independent random
variables~\cite{BBCR10,Kol16,She18}. Information-theoretic arguments have
determined the communication complexity of set disjointness with vanishing
error up to lower-order terms~\cite{BGPW13}, and nearly all known lower
bound techniques for randomized communication complexity apply to
information complexity as well~\cite{BW,KLLRX15}. Applications of
information complexity elsewhere in complexity theory include data
streams~\cite{BYJKS04}, multiparty message passing~\cite{BEOPV13},
extended formulations of polytopes~\cite{BM13}, formula lower
bounds~\cite{GMWW17}, and direct product theorems~\cite{BRWY13}.

\looseness=-1 Among the most basic questions about a computational resource is the cost
of solving several independent instances of a problem at once, compared
with the cost of a single instance. For a communication problem $F,$ this
question takes two standard forms: the \emph{direct sum problem} $F^n$
asks for the answers to $n$ independent instances of $F,$ and the
\emph{XOR problem} $F^{\oplus n}$ asks for the XOR of these answers.
Solving the instances one at a time costs roughly $n$ times as much as
solving one, and a \emph{direct sum theorem} or \emph{XOR lemma} asserts
that one cannot do much better. Statements of this kind go back to Yao's
XOR lemma for circuits~\cite{Yao82,Lev87,GNW11}. In communication
complexity, the study of direct sum theorems was initiated by Karchmer,
Raz, and Wigderson~\cite{KRW95}, who proved that a direct sum theorem for
certain relations would imply super-logarithmic lower bounds on circuit
depth. Despite their intuitive appeal, direct sum theorems and XOR lemmas
can be very difficult to establish, and in some settings they fail. For
example, the amortized communication complexity of a function can be much
smaller than its communication complexity~\cite{FKNN95,GKR16}, and
Shaltiel~\cite{Sha03} showed that strong direct product assertions, in
their basic form, are false in every reasonable model of computation.

Direct sum phenomena lie at the heart of information complexity and are
the main reason for its success in applications. Unlike communication,
information is additive over independent instances: a protocol that solves
$n$ independent instances of a problem must reveal, in total, $n$ times
the information needed to solve one. This principle originates in the work
of Chakrabarti et al.~\cite{CSWY01} on simultaneous message protocols and
the work of Bar-Yossef et al.~\cite{BYJKS04}, who used it to derive lower
bounds for set disjointness and data streams from the information
complexity of the two-bit AND function; see also~\cite{JRS03,BBCR10,BR14}.
In these direct sum theorems, however, the single instance is solved with
the same error as the $n$ instances. This does not match the obvious
protocol, which solves each of the $n$ instances with error $1/(3n),$ so
that by the union bound all $n$ answers are correct with probability at
least $2/3.$ This raises the natural question of whether computing $n$
instances of $F$ with overall error $1/3$ requires $\Omega(n)$ times the
information complexity of a single instance with error $1/(3n),$ and
likewise for the XOR of $n$ instances. Recent composition theorems for
other combining functions, such as gap majority~\cite{SY26,Wan26,Saw26a}
and arbitrary functions of the $n$ answers~\cite{Saw26b}, also measure the
single instance with constant error, and so they do not address this
question.

The XOR version of this question
should be contrasted with the more traditional form of an
XOR lemma~\cite{Sha03,She12,Yu22,IR24}, which states that even a slight
advantage over random guessing (i.e., a success probability of
$1/2+2^{-\Omega(n)}$ for $F^{\oplus n}$) costs $\Omega(n)$ times the
resources needed to compute $F$ with error $1/3.$ As Sawettamalya and
Yu~\cite{SY} observe, this more familiar form of the XOR lemma is false
for information complexity. Consider the following protocol for
$F^{\oplus n}.$ Using public randomness, the parties decide with
probability $2^{1-n}$ that Alice reveals all of her $n$ inputs to Bob, who
then announces the answer. In the complementary case, they output a
uniformly random answer chosen with public randomness. On every input, this
protocol succeeds with probability $1/2+2^{-n},$ and its information cost
under every input distribution is at most $2^{1-n}(n\log|\Xcal|+1).$ Thus,
for large $n,$ a success probability of $1/2+2^{-n}$ costs almost no
information, even though the information complexity of computing a
function with error $1/3$ can be as large as $\Theta(\log|\Xcal|)$
(Corollary~\ref{cor:ip-lower}).

Sawettamalya and Yu~\cite{SY} recently made substantial progress on XOR
lemmas for information complexity. In our notation, their main
result~\cite[Thm.~2]{SY} is the lower bound
\begin{equation}\label{eq:sy-bound}
  \IC_{1/10}(F^{\oplus n})
  \geq cn\left(\IC_{1/n}(F)-O(1)
       -O\Bigl(\frac{\log(|\Xcal||\Ycal|)}{n^{0.41}}\Bigr)\right)
\end{equation}
for every total function $F\colon\Xcal\times\Ycal\to\{-1,1\},$ where $c>0$
is an absolute constant. The same lower bound applies to
$\IC_{1/10}(F^n),$ since the parties can compute $F^{\oplus n}$ from the
output of any protocol for $F^n$ without further communication.
Sawettamalya and Yu derive~\eqref{eq:sy-bound} from their distributional
XOR lemma~\cite[Thm.~4]{SY}, which states that
\begin{equation}\label{eq:sy-distributional}
  \IC_{1/10}(F^{\oplus n},\mu^n)
  \geq cn\left(\IC_{n^{-0.41}}(F,\mu)-O(1)
       -O\Bigl(\frac{\log(|\Xcal||\Ycal|)}{n^{0.41}}\Bigr)\right)
\end{equation}
for every probability distribution $\mu$ on $\Xcal\times\Ycal,$ where
$\mu^n$ denotes the joint distribution of $n$ independent pairs, each
distributed according to $\mu.$ The passage
from~\eqref{eq:sy-distributional} to~\eqref{eq:sy-bound} uses Braverman's
minimax theorem (Fact~\ref{fact:minimax}) and a
majority vote over constantly many independent runs.

These lower bounds fall short of the well-known upper bound of
$n\IC_{1/(3n)}(F)$ in several respects. Most importantly, they contain a
penalty term proportional to the input length $\log(|\Xcal||\Ycal|),$
which makes them vacuous unless the number of copies $n$ is sufficiently
large compared to the input length. In particular, \eqref{eq:sy-bound} and
\eqref{eq:sy-distributional} do not establish an XOR lemma for $n$ or
fewer copies of an arbitrary function
$F\colon\{0,1\}^n\times\{0,1\}^n\to\{-1,1\},$ which is arguably the most
natural setting. There are also more subtle limitations to these bounds.
An unspecified constant is subtracted from the information complexity of a
single copy, which makes the bounds vacuous for functions of small
information complexity; and the distributional
bound~\eqref{eq:sy-distributional} involves the error $n^{-0.41}$ rather
than the error of order $1/n$ that the obvious protocol uses.

In this paper, we settle the question of XOR lemmas for information
complexity completely and definitively, for total functions as well as for
the mathematically richer class of partial functions. We obtain optimal
XOR lemmas and direct sum theorems for every
function, with no additive loss and no dependence on the input length. In
particular, for every communication problem $F,$ we determine the
information complexity of $F^n$ and $F^{\oplus n}$ up to a constant factor
in terms of a single copy of $F.$ In addition, we obtain an optimal
distributional XOR lemma and direct sum theorem, which hold under every
input distribution.

\subsection{Our results}\label{subsec:results}

We first consider communication problems defined on every input pair,
which we call total for emphasis, then turn to partial communication
problems, and conclude with the distributional setting.
Our results for total and partial functions allow any 
error $\epsilon\in[0,1/3]$ for the $n$
instances, with error $\epsilon/n$ for a single instance.
This includes the zero-error case $\epsilon=0,$ and the
constants in our results do not depend on $\epsilon.$
Analogously, our results in the distributional setting hold
uniformly for any error $\epsilon\in[0,1/400]$ for the $n$
instances, with error of order $\epsilon/n$ for a single instance.

\subsubsection*{Total functions.}

We start with the direct sum problem. Recall that, for a communication problem $F$ and
an integer $n\geq1,$ the problem $F^n$ asks for the values of $F$ on $n$
input pairs $(x_1,y_1),\ldots,(x_n,y_n).$ Running a protocol for $F$ with
error $\epsilon/n$ independently on each pair and applying the union bound
shows that $\IC_\epsilon(F^n)\leq n\IC_{\epsilon/n}(F).$ Our first result is that
this upper bound is tight up to a constant factor.

\begin{theorem}[Direct sum theorem]\label{main:ds}
There is a universal constant $c>0$ such that for every
communication problem $F\colon\Xcal\times\Ycal\to\{-1,1\},$
every integer $n\geq1,$ and every $\epsilon\in[0,1/3],$
\begin{equation}\label{eq:total-direct-sum}
  cn\IC_{\epsilon/n}(F)
    \leq\IC_\epsilon(F^n)
    \leq n\IC_{\epsilon/n}(F).
\end{equation}
\end{theorem}

It is natural to expect a similar result for the information cost of
$F^{\oplus n},$ the problem of computing the XOR of $n$ copies of $F.$ In
other words, one might na\"ively expect that $\IC_\epsilon(F^{\oplus n})$ is
within a constant multiple of $n\IC_{\epsilon/n}(F).$ A moment's thought,
however, shows that this is false for the two-bit XOR function. More
generally, we call a function $F\colon\Xcal\times\Ycal\to\{-1,1\}$
\emph{XOR-reducible} if $F(x,y)=a(x)b(y)$ for some functions
$a\colon\Xcal\to\{-1,1\}$ and $b\colon\Ycal\to\{-1,1\}$ and all inputs
$(x,y).$ Besides the XOR function, this class contains the constant
functions and the functions that depend on only one party's input.
This degenerate class of functions requires separate treatment, as given
by the following proposition.

\begin{proposition}[XOR-reducible functions]
\label{main:xor-reducible}
Let $F\colon\Xcal\times\Ycal\to\{-1,1\}$ be XOR-reducible, and let
$n\geq1$ be an integer.
\begin{enumerate}
\item If $F$ is constant, then $\IC_0(F^{\oplus n})=\IC_0(F)=0.$
\item If $F$ is not constant, then
$\IC_\epsilon(F^{\oplus n})$ and $\IC_{\epsilon/n}(F)$ lie in
$[1-h_2(1/3),2]$ for every $\epsilon\in[0,1/3].$
\end{enumerate}
\end{proposition}

Here $h_2$ denotes the binary entropy function, and
$1-h_2(1/3)=0.0817\ldots$ is a positive constant. Thus, the XOR analogue
of~\eqref{eq:total-direct-sum} holds trivially for constant functions but
fails for every nonconstant XOR-reducible function. Our main result for
total functions is that, excepting this degenerate class of XOR-reducible
functions, a perfect XOR lemma holds for all other functions.

\begin{theorem}[XOR lemma]\label{main:xor}
There is a universal constant $c>0$ such that for every
communication problem $F\colon\Xcal\times\Ycal\to\{-1,1\}$
that is not XOR-reducible, and for every integer $n\geq1$ and every
$\epsilon\in[0,1/3],$
\begin{equation}\label{eq:total-xor}
  cn\IC_{\epsilon/n}(F)
    \leq\IC_\epsilon(F^{\oplus n})
    \leq n\IC_{\epsilon/n}(F).
\end{equation}
\end{theorem}

\smallskip
\subsubsection*{Partial functions.}

Apart from total functions, a question of significant interest is the
information complexity of partial functions, also known as promise
problems, in which the parties need to answer correctly only on a
prescribed set of legal inputs. Formally, a partial function is a mapping
$F\colon\Xcal\times\Ycal\to\{-1,1,*\},$ where the symbol $*$ marks the
inputs on which no answer is required. The \emph{domain} of
$F,$ written $\dom F = \{(x,y):F(x,y)\ne*\},$ is the set
of legal input pairs. Theorem~\ref{main:ds} on the direct
sum generalizes unchanged to partial functions:

\begin{theorem}[Direct sum theorem]\label{main:ds-partial}
There is a universal constant $c>0$ such that for every
communication problem $F\colon\Xcal\times\Ycal\to\{-1,1,*\},$
every integer $n\geq1,$ and every $\epsilon\in[0,1/3],$
\begin{equation}\label{eq:partial-direct-sum}
  cn\IC_{\epsilon/n}(F)
    \leq\IC_\epsilon(F^n)
    \leq n\IC_{\epsilon/n}(F).
\end{equation}
\end{theorem}

One might expect that generalizing our XOR lemma (Theorem~\ref{main:xor}) 
to partial functions would be a matter of excluding the analogous
degenerate class, namely, the partial functions that agree with an
XOR-reducible function on all legal inputs. This is not the case. Indeed,
let $m\geq2$ be an integer, and consider the communication problem
$F_m\colon\mathbb Z_m\times\mathbb Z_m\to\{-1,1,*\}$ whose legal inputs
are the pairs $(x,y)$ with $y\in\{x,x+1\},$ and whose value is $-1$ on
$(m-1,0)$ and $1$ on all other legal inputs. With rows indexed by $x$ and
columns indexed by $y,$ both in the order $0,1,\ldots,m-1,$ the
characteristic matrix of $F_m$ is
\begin{align*}
  \begin{bNiceMatrix}
    1  & 1 &   &        &        &   \\
       & 1 & 1 &        &        &   \\
       &   & 1 & 1      &        &   \\
       &   &   & \Ddots & \Ddots &   \\
       &   &   &        & 1      & 1 \\
    -1 &   &   &        &        & 1
  \end{bNiceMatrix},
\end{align*}
where the unmarked entries are $*.$ Every row and every column of this
matrix contains exactly two legal inputs. Hence, if $F_m$ agreed on its
domain with an XOR-reducible function $a(x)b(y),$ the product of the
values of $F_m$ over all legal inputs would be $1,$ because each factor
$a(x)$ and $b(y)$ would occur exactly twice. Since exactly one legal input
has value $-1,$ this product is in fact $-1,$ and therefore $F_m$
disagrees on its domain with every XOR-reducible function. However, once
any one of the $m$ legal inputs of the form $(x,x)$ is deleted from the
domain, $F_m$ does factor as $a(x)b(y)$ on the remaining legal inputs.
For $m>n,$ some $t\in\mathbb Z_m$ differs from all of Alice's inputs
$x_1,\ldots,x_n,$ and then the legal input $(t,t)$ occurs in none of
the $n$ coordinates. Alice can name the first such $t$ in a publicly chosen random
order, after which the parties compute $F_m^{\oplus n}$ without error by
announcing $\prod_ia(x_i)$ and $\prod_ib(y_i),$ respectively, for the
factorization of $F_m$ with $(t,t)$ deleted. For $m=8n,$ Alice's first
message reveals less than one bit of information, which leads to the
following theorem.

\begin{theorem}[Counterexample]\label{main:counterexample}
For every integer $n\geq1,$ there is a communication problem
$F\colon\Xcal\times\Ycal\to\{-1,1,*\}$ that disagrees on
$\dom F$ with every XOR-reducible function and yet satisfies
\begin{align}
  &\IC_0(F^{\oplus n})\leq3,\label{eq:counterexample-xor}\\
  &\IC_{1/3}(F)\geq1-h_2\Bigl(\frac13\Bigr).
     \label{eq:counterexample-single}
\end{align}
\end{theorem}

\noindent In particular, for every $\epsilon\in[0,1/3],$ there does 
not exist a constant $c>0$ such that $\IC_\epsilon(F^{\oplus n})\geq cn\IC_{\epsilon/n}(F)$
for every $n\geq1$ and every communication problem $F$ that disagrees on
$\dom F$ with every XOR-reducible function. Indeed, for each $n,$ the
problem $F$ in Theorem~\ref{main:counterexample} satisfies
$\IC_\epsilon(F^{\oplus n})\leq\IC_0(F^{\oplus n})\leq3,$ whereas
$n\IC_{\epsilon/n}(F)\geq n\IC_{1/3}(F)\geq(1-h_2(1/3))n.$ 

Importantly, however, the protocol for this counterexample suggests the right measure
for partial functions. In that protocol, for each copy $i,$ the answer
$F_m(x_i,y_i)$ is the product of a sign $a(x_i)$ computed by Alice and a
sign $b(y_i)$ computed by Bob; these $2n$ signs are never announced
individually. Accordingly, a \emph{split-output protocol} for $F$ ends
with Alice holding a sign $A$ and Bob holding a sign $B,$ each computed
from that party's view of the protocol, and it is correct on a legal input
$(x,y)$ if $AB=F(x,y).$ Neither sign is announced, or
appended to the transcript. This output convention
is also known as the XOR model~\cite{FLLN23}. The
\emph{split-output information complexity} $\SIC_\epsilon(F)$ is the
least information cost of a split-output protocol for $F$ whose error on
every legal input is at most $\epsilon$
(Definition~\ref{def:split-output}). Since an announced answer can serve
as Alice's sign, with Bob's sign equal to $1,$ we have
$\SIC_\epsilon(F)\leq\IC_\epsilon(F).$ Conversely, appending both signs to
the transcript of a split-output protocol makes the answer public and
increases the information cost by at most $2,$ so that
$\IC_\epsilon(F)\leq\SIC_\epsilon(F)+2.$ Our main result for partial
functions shows that the information complexity of the XOR of $n$ copies
is determined, up to constant factors, by the split-output information
complexity of a single copy with error $\epsilon/n,$ together with the
information complexity of a single copy with error $\epsilon.$

\begin{theorem}[XOR characterization for partial functions]
\label{main:characterization}
For every communication problem $F\colon\Xcal\times\Ycal\to\{-1,1,*\},$
every integer $n\geq1,$ and every $\epsilon\in[0,1/3],$
\begin{equation}\label{eq:shared-characterization}
  \IC_\epsilon(F^{\oplus n})
    =\Theta\!\left(\IC_\epsilon(F)+n\SIC_{\epsilon/n}(F)\right),
\end{equation}
with absolute implicit constants.
\end{theorem}

Both terms in~\eqref{eq:shared-characterization} are needed. A nonconstant
XOR-reducible function $F(x,y)=a(x)b(y)$ satisfies $\SIC_0(F)=0,$ since
the parties can output the signs $a(x)$ and $b(y)$ without communication;
yet $\IC_\epsilon(F)>0.$ Conversely, the term $\IC_\epsilon(F)$ alone does not
suffice: the two-bit AND function satisfies $\IC_\epsilon(\AND_2)\leq2,$
whereas $\IC_\epsilon(\AND_2^{\oplus n})$ grows linearly in $n$
(Lemma~\ref{lem:xor-linear}).

Theorem~\ref{main:characterization} subsumes our XOR lemma for total
functions. Its proof builds on the total case, using crucial
additional ideas from geometry.

\subsubsection*{Distributional direct sum theorem.}

In the distributional setting, the input is drawn from a probability
distribution $\mu$ on the domain of $F,$ and a protocol for $F$ 
needs to be correct only on average over $\mu.$ For $n$ copies, the input
pairs are independent and distributed according to $\mu,$ so that the
input has distribution $\mu^n.$ Solving each copy with error $\epsilon/n$
shows, as before, that
\[
  \IC_\epsilon(F^{\oplus n},\mu^n)\leq\IC_\epsilon(F^n,\mu^n)
  \leq n\IC_{\epsilon/n}(F,\mu).
\]

In the standard formalism of information complexity, the output of a
protocol is a function of its public randomness and transcript, so that
the protocol's answer is ``announced'' to both parties and to any observer of the
transcript. Analogous to our contributions above for total and
partial functions, our goal in the distributional setting is a
direct sum theorem and an XOR lemma that hold for every communication
problem and every input distribution. True multiplicativity
in these results cannot be achieved within the formalism of announced 
outputs. Indeed, consider the two-bit function $F(x,y)=(-1)^x$ for
$x,y\in\{0,1\},$ and let $\mu$ be the uniform distribution on
$\{0,1\}\times\{0,1\}.$ Then $\IC_0(F^{\oplus n},\mu^n)\leq1$ for every
$n,$ because Alice can announce the answer. On the other hand, a protocol
for $F$ with error $1/(3n)$ under $\mu$ reveals Alice's uniformly random
input bit to Bob with error at most $1/(3n),$ and it can be shown that
$\IC_{1/(3n)}(F,\mu)\geq1-h_2(1/(3n))$
(cf.~Proposition~\ref{prop:local-variation}), so that
$n\IC_{1/(3n)}(F,\mu)$ grows linearly in $n.$ In particular, there does
not exist a constant $c>0$ such that
$\IC_{1/3}(F^{\oplus n},\mu^n)\geq cn\IC_{1/(3n)}(F,\mu)$ for every
$n\geq1.$ This happens because announcing an answer costs information, and
$F^{\oplus n}$ needs only one announcement, whereas separately solving 
$n$ copies of $F$ would need $n$ announcements. We similarly prove that
for every $\epsilon\in(0,1/3],$ there does not exist a constant $c>0$ such that
$\IC_\epsilon(F^n,\mu^n)\geq cn\IC_{\epsilon/n}(F,\mu)$ for every $n\geq1$
and every input distribution~$\mu$ (Theorem~\ref{thm:announced-counterexample}).

Our strategy is therefore to identify, for the direct sum problem and for
the XOR problem, an output convention whose information complexity obeys a
truly multiplicative direct sum theorem or XOR lemma, respectively, and
differs from information complexity with announced outputs by at most an
additive constant. Only the output convention changes; protocols and
their information cost are defined as before. As a result, our
distributional theorems also determine the information complexity of
$F^n$ and $F^{\oplus n}$ with announced outputs in terms of a single copy,
up to a constant factor and an additive constant.

For the direct sum problem, we use the output convention of local outputs.  A 
\emph{local-output protocol} ends with Alice and Bob each holding an answer of his 
or her own, computed from that party's view of the protocol, and it is correct if both
answers are correct. Neither answer is announced. This output
convention has appeared in earlier work~\cite{BBCR10,BMY15,FLLN23}. The
\emph{local-output information complexity} $\LIC_\epsilon(F,\mu)$ is the
least information cost under $\mu$ of a local-output protocol for $F$ with
error at most $\epsilon$ under $\mu.$ Local-output
information complexity is always within an additive constant
of announced-output information complexity. Indeed,
an announced answer can serve as the answer of both parties,
so that $\LIC_\epsilon(F,\mu)\leq\IC_\epsilon(F,\mu).$ Conversely,
their common answer can be announced whenever they agree, 
at small additional information cost, regardless of the number of 
possible answers. Quantitatively, we prove the following
relationship (Theorem~\ref{thm:local-comparison}): for every
communication problem $F$ with finitely many possible answers, every
probability distribution $\mu$ on the domain of $F,$ and every
$\epsilon\in[0,1/2],$
\begin{equation}\label{eq:local-comparison}
  \LIC_\epsilon(F,\mu)\leq\IC_\epsilon(F,\mu)
  \leq\LIC_\epsilon(F,\mu)+2h_2(\epsilon)+4\epsilon.
\end{equation}
In particular, $\IC_0(F,\mu)=\LIC_0(F,\mu).$ Our distributional direct sum
theorem states that local-output information complexity is truly
multiplicative.

\begin{theorem}[Distributional direct sum theorem]\label{main:dist-direct-sum}
For every communication problem $F\colon\Xcal\times\Ycal\to\{-1,1,*\},$
every probability distribution $\mu$ on the domain of $F,$ every integer $n\geq1,$ and every
$\epsilon\in[0,1/100],$
\begin{equation}\label{eq:dist-direct-sum}
  \frac n{32}\LIC_{32\epsilon/n}(F,\mu)
    \leq\LIC_\epsilon(F^n,\mu^n)
    \leq n\LIC_{\epsilon/n}(F,\mu).
\end{equation}
Moreover,
\begin{equation}\label{eq:dist-direct-sum-zero}
  \LIC_0(F^n,\mu^n)=n\LIC_0(F,\mu).
\end{equation}
\end{theorem}

\noindent
The two bounds in~\eqref{eq:dist-direct-sum} involve 
error rates $32\epsilon/n$ and $\epsilon/n$ that differ by a constant factor.
This is because in the distributional setting, a protocol may err with certainty on a
small set of inputs, so errors cannot be reduced by repetition. 

Along with~\eqref{eq:local-comparison}, Theorem~\ref{main:dist-direct-sum}
determines the information complexity of $F^n$ with announced outputs, in
terms of the local-output information complexity of a single copy.

\begin{corollary}[Distributional direct sum for announced outputs]
\label{main:ds-announced}
For every communication problem $F\colon\Xcal\times\Ycal\to\{-1,1,*\},$
every probability distribution $\mu$ on the domain of $F,$ every integer
$n\geq1,$ and every $\epsilon\in[0,1/100],$
\begin{equation}\label{eq:ds-announced}
  \frac n{32}\LIC_{32\epsilon/n}(F,\mu)
    \leq\IC_\epsilon(F^n,\mu^n)
    \leq n\LIC_{\epsilon/n}(F,\mu)+2h_2(\epsilon)+4\epsilon.
\end{equation}
\end{corollary}

We also prove a direct sum theorem with constant additive loss that
involves announced outputs alone.

\begin{theorem}[Distributional direct sum with constant additive loss]
\label{main:ds-additive}
For every communication problem $F\colon\Xcal\times\Ycal\to\{-1,1,*\},$
every probability distribution $\mu$ on the domain of $F,$ every integer
$n\geq1,$ and every $\epsilon\in[0,1/3],$
\begin{equation}\label{eq:ds-additive}
  \frac n4\bigl(\IC_{4\epsilon/n}(F,\mu)-8\bigr)
  \leq\IC_\epsilon(F^n,\mu^n)
  \leq n\IC_{\epsilon/n}(F,\mu).
\end{equation}
\end{theorem}

\noindent
As mentioned earlier, the constant additive loss in this theorem cannot be
eliminated (Theorem~\ref{thm:announced-counterexample}).

\subsubsection*{Distributional XOR lemma.}
Recall that a split-output protocol ends with Alice
and Bob holding signs whose product is the answer, and write
$\SIC_\epsilon(F,\mu)$ for the least information cost under $\mu$ of a
split-output protocol for $F$ with error at most $\epsilon$ under $\mu.$
To achieve a truly multiplicative XOR lemma in the
distributional setting, we use a natural geometric generalization of split 
outputs.  A protocol with \emph{vector outputs} ends with Alice and
Bob holding unit vectors $u$ and $v,$ respectively, in a common Euclidean
space, each computed from that party's view of the protocol. Neither vector
is communicated. The \emph{deficit} of such a protocol under $\mu$ is the
expectation of $1-F(X,Y)\langle u,v\rangle,$ where $(X,Y)\sim\mu,$ and half
of the deficit is the \emph{vector error}. The \emph{vector-output information
complexity} $\VIC_\epsilon(F,\mu)$ is the least information cost under
$\mu$ of a protocol with vector outputs and vector error at most $\epsilon$
under $\mu.$ Vector-output information complexity is within
an additive constant of split-output information complexity, as
follows. For Alice and Bob's signs in a split-output
protocol, which are unit
vectors in dimension $1,$ the vector error is the probability that their
product differs from $F(X,Y).$ As a result,
$\VIC_\epsilon(F,\mu)\leq\SIC_\epsilon(F,\mu).$ For the
opposite direction, a rounding procedure
turns vector outputs into signs with no increase in error and at a small
cost in information. Specifically, we prove the following
equivalence of $\VIC_\epsilon,\SIC_\epsilon,$ and $\IC_\epsilon,$
up to an additive constant (Theorem~\ref{thm:vector-comparison}): for
every communication problem $F\colon\Xcal\times\Ycal\to\{-1,1,*\},$ every
probability distribution $\mu$ on the domain of $F,$ and every
$\epsilon\in[0,1/2],$
\begin{equation}\label{eq:vector-comparison}
  \VIC_\epsilon(F,\mu)\leq\SIC_\epsilon(F,\mu)\leq\IC_\epsilon(F,\mu)
  \leq\VIC_\epsilon(F,\mu)+5.
\end{equation}
We prove that vector-output information complexity obeys a
truly multiplicative XOR lemma. 

\begin{theorem}[Distributional XOR lemma]\label{main:dist-xor}
For every communication problem $F\colon\Xcal\times\Ycal\to\{-1,1,*\},$
every probability distribution $\mu$ on the domain of $F,$ every 
integer $n\geq1,$ and every $\epsilon\in[0,1/400],$
\begin{equation}\label{eq:dist-xor}
  \frac n{34}\VIC_{100\epsilon/n}(F,\mu)
    \leq\VIC_\epsilon(F^{\oplus n},\mu^n)
    \leq n\VIC_{\epsilon/n}(F,\mu).
\end{equation}
Moreover,
\begin{equation}\label{eq:dist-xor-zero}
  \VIC_0(F^{\oplus n},\mu^n)=n\VIC_0(F,\mu).
\end{equation}
\end{theorem}

For announced outputs, we obtain the following XOR lemma,
with the single copy measured using vector outputs or,
alternatively, using split outputs.

\begin{theorem}[Distributional XOR lemma for announced outputs]
\label{main:xor-announced}
For every communication problem $F\colon\Xcal\times\Ycal\to\{-1,1,*\},$
every probability distribution $\mu$ on the domain of $F,$ every integer
$n\geq1,$ and every $\epsilon\in[0,1/400],$
\begin{align}
  \frac n{34}\VIC_{100\epsilon/n}(F,\mu)
    &\leq\IC_\epsilon(F^{\oplus n},\mu^n)
    \leq n\VIC_{\epsilon/n}(F,\mu)+5,\label{eq:announced-xor-vic}\\
  \frac n{34}\SIC_{100\epsilon/n}(F,\mu)
    &\leq\IC_\epsilon(F^{\oplus n},\mu^n)
    \leq n\SIC_{\epsilon/n}(F,\mu)+2.\label{eq:announced-xor-sic}
\end{align}
\end{theorem}

We also prove an XOR lemma with constant additive loss in which both sides
use announced outputs.

\begin{theorem}[Distributional XOR lemma with constant additive loss]
\label{main:xor-additive}
For every communication problem $F\colon\Xcal\times\Ycal\to\{-1,1,*\},$
every probability distribution $\mu$ on the domain of $F,$ every integer
$n\geq1,$ and every $\epsilon\in[0,1/6],$
\begin{equation}\label{eq:xor-additive}
  \frac n4\bigl(\IC_{4\epsilon/n}(F,\mu)-16\bigr)
  \leq\IC_\epsilon(F^{\oplus n},\mu^n)
  \leq n\IC_{\epsilon/n}(F,\mu).
\end{equation}
\end{theorem}

By the example of the two-bit function $F(x,y)=(-1)^x,$ the
constant additive loss in this theorem cannot be removed. This lower bound
subsumes and substantially strengthens the distributional XOR
lemma~\eqref{eq:sy-distributional} of Sawettamalya and Yu, which concerns
the error $\epsilon=1/10.$ Unlike~\eqref{eq:sy-distributional}, it has no
penalty term that depends on the input length, and the error of its single copy is
of order $\epsilon/n$ instead of $n^{-0.41}.$

\subsection{Previous work}\label{subsec:previous-work}
Proofs of XOR lemmas for information complexity start from a classic
embedding argument that originated with Bar-Yossef, Jayram, Kumar, and
Sivakumar~\cite{BYJKS04} and that Barak, Braverman, Chen, and
Rao~\cite{BBCR10} stated explicitly for internal information cost. Let
$\pi$ be a protocol for $F^{\oplus n},$ and let $\mu$ be an input
distribution for $F.$ To solve a single instance of $F,$ the parties place
it in a uniformly random coordinate $j$ of an input for $\pi$ and fill in
the other coordinates with independent samples from $\mu.$ For each
coordinate before $j,$ Alice's input is sampled publicly and Bob's input
privately by Bob; for each coordinate after $j,$ the roles of the parties
are reversed. The parties then run $\pi$ on the completed input
(Definition~\ref{def:standard-embedding}). On average over $j,$ the
resulting protocol reveals a $1/n$ fraction of the information that $\pi$
reveals under $\mu^n$ (Fact~\ref{fact:embedding-cost}). With two
additional bits, which reveal the XOR of the answers on the coordinates 
other than $j,$ the parties recover the answer for their own instance from
the output of $\pi.$ However, this answer is correct exactly when the
output of $\pi$ is correct (Fact~\ref{fact:ordinary-embedding}). Thus, the
embedding preserves the error of $\pi$ instead of reducing it, whereas an XOR
lemma requires a protocol for $F$ with error smaller than that of $\pi$ by a 
factor of $n.$ The central difficulty in proving an XOR lemma is to reduce the 
error from a constant to $1/(3n)$ while keeping the information cost within a
constant factor of $I/n,$ where $I$ is the information cost of $\pi.$

Sawettamalya and Yu~\cite{SY} achieved this error reduction, up to the
penalty term in~\eqref{eq:sy-bound} proportional to the input length, by a 
sophisticated argument that involves a recursive splitting of $\pi$ and is inspired 
in part by the earlier work of Yu~\cite{Yu22}. We describe Sawettamalya and Yu's approach
in the notation of this paper. Let $\pi$ be a protocol for $F^{\oplus2k}$
with error $\epsilon$ and information cost $I$ under $\mu^{2k}.$ Write the
input of $\pi$ as $((X_0,X_1),(Y_0,Y_1))\sim\mu^{2k},$ where block
$j\in\{0,1\}$ consists of $k$ coordinates with inputs $X_j$ and $Y_j.$
Applying the embedding to blocks rather than to coordinates gives
protocols $\pi^{(0)}$ and $\pi^{(1)}$ whose information 
costs under $\mu^k$ sum to $I.$ In $\pi^{(0)},$ the
actual input occupies block $0,$ the input $Y_1$ is sampled publicly, and
Alice samples $X_1$ privately; in $\pi^{(1)},$ the actual input occupies
block $1,$ the input $X_0$ is sampled publicly, and Bob samples $Y_0$
privately. Let $R$ and $M$ be the public randomness and transcript of
$\pi,$ and put $V=(X_0,Y_1,R,M).$ Then Alice knows $V$ in $\pi^{(0)},$ and
Bob knows $V$ in $\pi^{(1)}.$ The key observation, due to Yu~\cite{Yu22},
is that the rectangle property of protocols makes the answers
$B_0=F^{\oplus k}(X_0,Y_0)$ and $B_1=F^{\oplus k}(X_1,Y_1)$ on the two
blocks conditionally independent given $V$
(Lemma~\ref{lem:cross-independence}).  As a result, their posterior biases
$\Gamma_j=\abs{\Exp[B_j\mid V]}$ satisfy
\begin{equation}\label{eq:intro-bias-product}
  \abs{\Exp[B_0B_1\mid V]}=\Gamma_0\Gamma_1.
\end{equation}
Let $\pi_j$ be the protocol for $F^{\oplus k}$ in which, at the end of
$\pi^{(j)},$ the party that knows $V$ announces the more likely value of
$B_j$ as the answer. Given $V,$ this answer is wrong with probability
$(1-\Gamma_j)/2.$ On the other hand, the output of $\pi$ is a function of
$V,$ so by~\eqref{eq:intro-bias-product} it errs with probability at least
$(1-\Gamma_0\Gamma_1)/2$ given $V.$

\looseness=1 When $\Gamma_0$ and $\Gamma_1$ are close to $1,$ the combined error
$(1-\Gamma_0)/2+(1-\Gamma_1)/2$ of $\pi_0$ and $\pi_1$ given $V$ exceeds
the lower bound $(1-\Gamma_0\Gamma_1)/2$ on the error of $\pi$ only by
$(1-\Gamma_0)(1-\Gamma_1)/2,$ a second-order quantity. If this were the
case on every transcript, the errors of $\pi_0$ and $\pi_1$ would sum to
little more than the error of $\pi.$ Their information costs sum to at most
$I+2,$ because those of $\pi^{(0)}$ and $\pi^{(1)}$ sum to $I$ and each
announcement costs at most one bit. After $\log n$ levels of splitting, a
uniformly random leaf would be a protocol for $F$ with error about
$\epsilon/n$ and information cost about $I/n$ plus a constant, as an
optimal XOR lemma requires. As Yu points out, however,
identity~\eqref{eq:intro-bias-product} holds only pointwise. On
transcripts with a small posterior bias $\Gamma_0\Gamma_1,$ it gives no
useful control: if $\Gamma_0=\Gamma_1=0,$ for instance, then $\pi_0$ and
$\pi_1$ err with probability $1/2$ each given $V,$ for a combined error of
$1,$ which is twice the error of $\pi.$ If such transcripts carry most of
the error at every level of the recursion, the total error doubles from
one level to the next, and the average error over the leaves does not
decrease at all.

\looseness=1 Sawettamalya and Yu overcome this difficulty by
\emph{conditioning} on the event that the 
posterior bias $\abs{\Exp[B_0B_1\mid V]}$ is at least $3/5$ and
splitting the conditioned protocol. This ensures that the total error
grows by a factor of at most $2/(1+\sqrt{3/5})=1.127\ldots$ from one level
to the next, rather than $2,$ but it comes at a considerable cost.
Conditioning on an event that depends on both parties' inputs destroys the
defining property of a protocol: after conditioning, a message may depend
on the receiver's input as well as on the sender's. Moreover, after
conditioning, the input no longer follows a power of $\mu,$ and such
deviations compound over the levels of the recursion. To control these
effects, Sawettamalya and Yu work with what they call \emph{generalized
protocols}. They measure by a relative entropy how far a generalized
protocol is from an ordinary protocol on the intended input distribution,
select a leaf of the recursion from a nonuniform distribution weighted by
the probabilities of the conditioning events, and finally convert the
selected leaf into an ordinary protocol for $F$ with input distribution
$\mu.$ At the selected leaf, both the error and the relative entropy are
about $n^{-0.83}.$ The conversion increases the error by an amount of
order the square root of the relative entropy, and the information cost by
this amount times the input length $\log(|\Xcal||\Ycal|);$ the announced
bits add a constant per copy. These costs are the subtracted terms
in~\eqref{eq:sy-bound} and~\eqref{eq:sy-distributional}. The square root
halves the exponent $0.83,$ which explains the exponent $0.41$ in these
bounds; the analysis of Sawettamalya and Yu allows any constant less than
$0.4137\ldots$ in its place.

\subsection{Our approach}\label{subsec:approach}

Despite our best efforts, we were unable to eliminate the dependence on
the input length within the framework of Sawettamalya and Yu, and this led
us to a different approach. We adopt their starting point, namely, the
recursive splitting of the copies into two blocks, but depart from
their approach in several fundamental ways. For total functions, the
result is a particularly short proof. Partial functions and the
distributional setting, by contrast, require much deeper ideas, both
information-theoretic and geometric.

\subsubsection*{Total functions.}

In outline, our proof for total functions splits the given protocol
recursively, as in Section~\ref{subsec:previous-work}, but keeps every
transcript. On the transcripts with a small posterior bias $\Gamma_0\Gamma_1,$ 
which Sawettamalya and Yu discard by conditioning, we run an additional
protocol that we call a \emph{fallback protocol}. Since no transcript is
discarded, every protocol in our recursion is an ordinary protocol,
analyzed under a power $\mu^k$ of the original input distribution, and no
conversion step is needed. The price is that the fallback protocols must
be paid for, under input distributions over which we have no control. In
return, our analysis has no loss that depends on the input length,
unlike~\eqref{eq:sy-bound}.

Let $\pi$ be a protocol for $F^{\oplus2k}$ with error $\epsilon$ and
information cost $I$ under $\mu^{2k},$ and let $\pi^{(j)},$ $V,$ $B_j,$
$\Gamma_j$ be as in Section~\ref{subsec:previous-work}. We call the event
that $\Gamma_0\Gamma_1\geq1/4$ the \emph{good event}. In our error
reduction step (Lemma~\ref{lem:one-step}), we modify $\pi_j$ as follows.
After $\pi^{(j)},$ the party that knows $V$ announces whether the good
event has occurred, together with the more likely value of $B_j.$ If the
good event has occurred, this value is the output of $\pi_j.$ In the
complementary case, the parties run a fallback protocol for $F^{\oplus k}$
on the actual input, with fresh randomness, and output its answer. On the
good event, an argument of Sawettamalya and Yu based on the concavity of
the square root function shows that the combined error of $\pi_0$ and
$\pi_1$ given $V$ is at most $4/3$ times that of $\pi.$ On the
complementary event, $\pi$ errs with probability more than $3/8$ given
$V,$ whereas a fallback protocol with error at most $1/5$ on every legal
input caps the combined error at $2/5.$ Altogether, the errors of $\pi_0$
and $\pi_1$ under $\mu^k$ sum to at most $4\epsilon/3,$ so that the total
error grows by a factor of at most $4/3,$ rather than $2,$ from one level
of the recursion to the next.

The fallback protocol is run with probability at most $8\epsilon/3.$
Conditionally on what has been communicated, however, its input can have
an arbitrary distribution rather than a power of $\mu.$ We therefore need
a bound $K$ on the information cost of the fallback protocol under every
input distribution. With such a bound, the information costs of $\pi_0$
and $\pi_1$ under $\mu^k$ sum to at most $I+4+16\epsilon K/3,$ where the
term $4$ accounts for the announced bits. This is how worst-case
information complexity enters our otherwise distributional argument.

The fallback protocols come from the very protocol that we decompose. Let
$\pi$ be a protocol for $F^{\oplus N}$ with worst-case error $1/10$ and
worst-case information cost $I,$ where $N$ is a power of two. For every
$k$ that divides $N,$ applying the embedding of
Section~\ref{subsec:previous-work} to blocks of $k$ coordinates, under an
arbitrary input distribution on a block, and then the minimax theorem of
Braverman~\cite{Braverman15}, we obtain
$\IC_{1/5}(F^{\oplus k})\leq2kI/N+4$ (Corollary~\ref{cor:fallback}). Thus,
the embedding, which falls short of an XOR lemma because it preserves the
error, supplies exactly what a fallback protocol needs: constant error and
an information cost proportional to the number of copies, up to an
additive constant.

We apply the error reduction step recursively for $\log N$ levels,
starting with $\pi$ under $\mu^N$ and using at depth $i$ a fallback
protocol for $F^{\oplus N/2^{i+1}}$ with worst-case error $1/5$ and
worst-case information cost $I/2^i+O(1).$ The errors at depth $i$ sum to
at most $(4/3)^i/10,$ so the fallback protocols at depth $i$ are run with
total probability $O((4/3)^i).$ Near the root, the fallback protocols are
expensive but rarely run; deeper in the recursion, they are run more often
but cost proportionally less, and the charges proportional to $I$ decrease
geometrically, as $(2/3)^iI.$ Summing over the levels, the fallback
protocols contribute $O(I)+o(N)$ to the total information cost, and the
announced bits contribute at most $4$ per leaf. A uniformly random leaf is
therefore a protocol for $F$ with error at most $N^{-\alpha}/10,$ where
$\alpha=\log(3/2)=0.5849\ldots,$ and information cost at most
$13I/(5N)+4+o(1)$ under $\mu$ (Lemma~\ref{lem:recursive-bound}). Since
this holds for every $\mu,$ the minimax theorem gives a protocol for $F$
with worst-case error $N^{-\alpha}/5,$ at the price of a factor of $2$ in
information cost. Because $2\alpha>1,$ a majority vote over three
independent runs of this protocol has worst-case error below $1/(6n)$ when
$N$ is the largest power of two not exceeding $n.$ After a routine change
of the initial error from $1/10$ to $1/3,$ we obtain
$\IC_{1/(6n)}(F)\leq480\IC_{1/3}(F^{\oplus n})/n+64$
(Lemma~\ref{lem:xor-conversion}). For a smaller error $\epsilon>0,$ we run
a protocol for $F^{\oplus n}$ with error $\epsilon$ independently on
$r=\lfloor1/(3\epsilon)\rfloor$ disjoint blocks of $n$ copies. The result
is a protocol for $F^{\oplus rn}$ with error at most $1/3$ and at most $r$
times the information cost. The previous bound, applied with $rn$ in place of
$n,$ gives a single copy with error $1/(6rn)\leq\epsilon/n,$ and the
factor $r$ cancels in the information cost. Finally, a direct embedding argument
handles $\epsilon=0.$ We thus arrive at an XOR lemma with constant additive
loss (Theorem~\ref{thm:additive-conversion}): for every
$\epsilon\in[0,1/3],$
\[
  \IC_{\epsilon/n}(F)\leq\frac{480}n\IC_\epsilon(F^{\oplus n})+64.
\]
Two features of this bound are essential. First, the additive loss is $64$
for every $F,$ every $n,$ and every $\epsilon,$ no matter how long the
inputs are. Second,
the proof never evaluates $F$ outside $\dom F,$ and therefore the bound
holds for partial functions as well.

For total functions, the first feature allows us to remove the additive
loss, because a constant is absorbed by any lower bound on the
information complexity of the $n$ copies that grows linearly in $n.$ For
the direct sum, Proposition~\ref{prop:local-variation} gives such a lower
bound, $\IC_\epsilon(F^n)\geq(1-h_2(1/3))n,$ for every nonconstant $F$
and every $\epsilon\in[0,1/3].$ Since
$\IC_\epsilon(F^{\oplus n})\leq\IC_\epsilon(F^n)$
(Proposition~\ref{prop:general-upper}),
Theorem~\ref{thm:additive-conversion} also gives
$\IC_{\epsilon/n}(F)\leq480\IC_\epsilon(F^n)/n+64.$ A weighted average of
these two bounds proves Theorem~\ref{main:ds}. For the XOR lemma,
$\IC_\epsilon(F^{\oplus n})$ grows linearly in $n$ precisely when $F$ is
not XOR-reducible. If $F$ is XOR-reducible, then
$\IC_\epsilon(F^{\oplus n})\leq2$ (Proposition~\ref{main:xor-reducible}).
Every other total function $F$ has a restriction to two rows and two
columns that is the two-bit AND function or its negation, so that
$F^{\oplus n}$ has, up to sign, the inner product function $\IP_n$ on
$n$-bit strings as a restriction. Since the information complexity of
$\IP_n$ is linear in $n$ (Corollary~\ref{cor:ip-lower}), so is that of
$F^{\oplus n}$ (Lemma~\ref{lem:xor-linear}), and a weighted average again
proves Theorem~\ref{main:xor}.

\subsubsection*{Partial functions.}

Our Theorem~\ref{thm:additive-conversion} applies verbatim to partial
functions, and so does the proof of the direct sum theorem, once the
constant functions are replaced by the communication problems whose rows
and columns are all constant on their legal inputs
(Theorem~\ref{main:ds-partial}). The XOR lemma is a very different matter.
Theorem~\ref{main:counterexample} shows that, for partial functions, the
additive loss in Theorem~\ref{thm:additive-conversion} reflects a genuine
phenomenon, even outside the analogous degenerate class:
$\IC_\epsilon(F^{\oplus n})$ can be at most $3$ while $\IC_{\epsilon/n}(F)$ is
bounded away from zero. Thus, as long as the single copy is measured by
$\IC_{\epsilon/n}(F),$ an additive loss is unavoidable.  As the protocol 
for the counterexample suggests, we measure the single copy with split outputs instead. The
lower bound in Theorem~\ref{main:characterization} then rests on an
extraction theorem with no additive loss at all
(Theorem~\ref{thm:extraction}): for every $\epsilon\in[0,1/3],$
\begin{equation}\label{eq:intro-extraction}
  \SIC_{\epsilon/n}(F)\leq\frac Cn\IC_\epsilon(F^{\oplus n}),
\end{equation}
where $C$ is an absolute constant. Combined with the bound
$\IC_\epsilon(F^{\oplus n})\geq\IC_\epsilon(F)$
(Proposition~\ref{prop:restriction-floor}), the extraction theorem gives
the lower bound in~\eqref{eq:shared-characterization}. The upper bound
in~\eqref{eq:shared-characterization} follows by 
running split-output protocols with error $\epsilon/n$ on the $n$
copies and letting each party announce the XOR of his or her \emph{shares},
that is, of the signs that he or she holds at the end of these protocols. The
two announced bits cost at most $2,$ which is $O(\IC_\epsilon(F))$ unless
both sides of~\eqref{eq:shared-characterization} vanish
(Proposition~\ref{prop:local-variation}).

We now focus on the proof of the extraction
theorem~\eqref{eq:intro-extraction}. Split outputs eliminate one source of
additive loss at no cost. Indeed, in the embedding of
Section~\ref{subsec:previous-work}, the two additional bits that reveal
the XOR of the answers on the other coordinates are no longer needed: each
party absorbs into his or her share the XOR of the answers on the
coordinates where he or she knows both inputs. Together with a minimax
theorem over the class of protocols with worst-case error at most
$\epsilon$ (Proposition~\ref{prop:class-minimax}), this gives the lossless
extraction $\SIC_\epsilon(F)\leq\SIC_\epsilon(F^{\oplus k})/k$
(Lemma~\ref{lem:tensor}). This extraction leaves the error unchanged,
however, and so it settles only the case $\epsilon=0$
of~\eqref{eq:intro-extraction}. In our proof overview so far, the only
mechanism that reduces the
error in passing from an XOR to a single copy is
Theorem~\ref{thm:additive-conversion}, and it incurs an additive loss
of~$64.$ To obtain the best of both worlds, namely, error reduction with
no additive loss, we develop two complementary mechanisms, grouping and
rounding. We describe them for $\epsilon=1/3,$ with the target error
$1/(6n)$ for the single copy; as for total functions, independent
executions on disjoint blocks then give every $\epsilon\in(0,1/3].$

Grouping applies our XOR lemma with constant additive loss to blocks of
copies (Lemma~\ref{lem:block}). We partition the $n$ copies into about
$n^{1/3}$ blocks of about $n^{2/3}$ copies each and apply
Lemma~\ref{lem:xor-conversion} to the XOR of the block problems.
This gives a protocol for a single block with error $O(n^{-1/3})$ and an
additive loss of $64,$ and a majority vote over five independent runs
reduces its error to below $1/(6n).$ Lemma~\ref{lem:tensor} then extracts
a single copy of $F$ from the block, dividing the additive loss by the
block size. The order of these steps matters: the majority vote needs
announced answers, so it comes before the extraction. The result is a
split-output protocol for $F$ with error at most $1/(6n)$ and information
cost $O(I/n+n^{-2/3}),$ where $I=\IC_{1/3}(F^{\oplus n}).$ The additive
term is $O(I/n)$ whenever $I=\Omega(n^{1/3}).$

Rounding handles the complementary case and is geometric in nature
(Lemma~\ref{lem:rounding}). The \emph{margin} $\margin(F)$ of $F$ is the
maximum, over all families of unit vectors $u_x$ and $v_y$ in a Euclidean
space, of the minimum of $F(x,y)\langle u_x,v_y\rangle$ over the legal
inputs $(x,y).$ The characteristic matrix of $F^{\oplus n}$ is the $n$th
tensor power of that of $F,$ and margin is multiplicative under tensor
products, as shown by Lee, Shraibman, and \v{S}palek~\cite{LSS08} for
total sign matrices.
Hence $\margin(F^{\oplus n})=\margin(F)^n$
(Proposition~\ref{prop:margin-tensor}). Since discrepancy is bounded by
margin (Proposition~\ref{prop:disc-margin}), the discrepancy bound of
Braverman and Weinstein~\cite{BW} now gives
\[
  \IC_{1/3}(F^{\oplus n})\geq cn\log\frac1{\margin(F)}
\]
for an absolute constant $c>0$ (Theorem~\ref{thm:margin-ic}). Thus, when
$I/n$ is small, the margin of $F$ is close to $1;$ more precisely, the
quantity $s=1-\margin(F)^2$ satisfies $s=O(I/n).$

Now fix unit vectors $u_x$ and $v_y$ that achieve the margin. For every
legal input $(x,y),$ the vectors $u_x$ and $F(x,y)v_y$ form an angle
$\theta$ with $\sin^2\theta\leq s.$ This suggests random hyperplane
rounding, which Goemans and Williamson~\cite{GW95} introduced for their
semidefinite approximation algorithms: the parties pick a uniformly random
unit vector $w$ using public randomness and take
$\sgn\langle u_x,w\rangle$ and $\sgn\langle v_y,w\rangle$ as their shares.
This protocol requires no communication, but its error $\theta/\pi$ can be
of order $\sqrt s,$ far above the error $s^{3/2}$ that our analysis
requires. More generally, every split-output protocol without
communication errs with probability at least $s/4$ on some legal input,
which is still much larger than $s^{3/2}.$ Our protocol sends a single
message. Using public randomness, the parties choose a uniformly random
orthonormal basis. Alice selects one of its vectors with probability equal
to the square of the corresponding coordinate of $u_x$ and sends its index
to Bob. The shares are the signs of the coordinates of $u_x$ and $v_y$
along the selected vector. Errors arise only from basis vectors whose
projections onto the plane spanned by $u_x$ and $F(x,y)v_y$ are nearly
orthogonal to $u_x,$ and the selection rule makes such vectors unlikely.
An exact calculation shows that the error becomes
$(\theta-\sin\theta\cos\theta)/\pi\leq s^{3/2},$ which is cubic in the
angle rather than linear, while the information cost of the message is
only $O(s).$

The two mechanisms combine as follows (Theorem~\ref{thm:extraction}). If
$s^{3/2}\leq1/(6n),$ rounding gives a split-output protocol for $F$ with
error at most $1/(6n)$ and information cost $O(s)=O(I/n).$ In the
complementary case, $s>(6n)^{-2/3},$ so that
$I/n=\Omega(s)=\Omega(n^{-2/3}),$ and the additive term in the grouping
bound is $O(I/n).$ Either way, $\SIC_{1/(6n)}(F)=O(I/n).$ The exponents
match precisely: rounding succeeds as long as $s\leq(6n)^{-2/3},$ and when
it fails, $I/n$ is large enough to pay for the additive loss of grouping.
In particular, rounding without communication, whose error is at least
$s/4$ on some legal input, would not suffice. This yields the extraction
theorem~\eqref{eq:intro-extraction} and, with it, the characterization in
Theorem~\ref{main:characterization}.

\subsubsection*{Distributional direct sum theorem.}

In the distributional setting, the given protocol is correct only on
average over the input distribution, and the main tools of our proofs for
total and partial functions break down completely. The fallback protocols 
of the error reduction step run under a distribution over which we have
no control, namely the conditional distribution of the input given
what has been communicated. Previously, we paid for
the fallback protocols with a bound on their information cost under every input
distribution, which came from a protocol with an error guarantee on every
legal input, through the minimax theorem (Corollary~\ref{cor:fallback}).
No such protocol is available in the distributional setting. Moreover, the
error at the leaves of our recursion was reduced by majority votes, but
errors under a given input distribution cannot be reduced by repetition.
Finally, the margin bound of Theorem~\ref{thm:margin-ic}, which drives the
rounding mechanism, concerns the worst legal input. The distributional
setting therefore requires new techniques, both for direct sums and for
XORs. Let $\pi$ be a local-output protocol for $F^n$ with error $\epsilon$
and information cost $I$ under $\mu^n.$ Our proof of the distributional
direct sum theorem distinguishes three regimes of $I.$

In the high-information regime $I\geq n,$ a constant additive loss per
copy is affordable, and we use our direct sum theorem with constant
additive loss for announced outputs (Theorem~\ref{main:ds-additive}),
after $\pi$ announces its answers at a small cost
(Lemma~\ref{lem:announcement}). The proof of
Theorem~\ref{main:ds-additive} is a far-reaching generalization of the
error reduction step. Instead of bounding the information cost and the
error of a protocol separately, we charge errors in units of information.
Formally, for a price $\beta>0,$ the \emph{priced information complexity}
$W_\beta(F,\mu)$ is the infimum, over all protocols for $F,$ of the
information cost under $\mu$ plus $\beta$ times the error under $\mu.$
This weighted combination of information and error is inspired by
Braverman's proof of the minimax theorem~\cite{Braverman15}.

To split a protocol for the direct sum $F_0\times F_1$ of two
communication problems, under the product distribution $\mu_0\otimes\mu_1,$
we use the standard embeddings into the two blocks, as
in Section~\ref{subsec:previous-work}. The fallback protocol $\rho_j$ for
block $j$ is now a protocol for $F_j$ whose priced cost under $\mu_j$ is
close to $W_\beta(F_j,\mu_j).$ It runs under the conditional distribution
of the input given the view $V,$ and we bound its priced cost there by its
\emph{cost profile} (Proposition~\ref{prop:cost-profile}). This is an
affine function of the input distribution that lies above the priced cost
of $\rho_j$ and coincides with it at $\mu_j,$ and its existence rests on
the concavity of information cost. The party that knows $V$ keeps the
answer of the given protocol for block $j$ if the price of its conditional
error given $V$ is at most the conditional expectation of the cost
profile, and runs $\rho_j$ otherwise. Unlike the error reduction step,
which lets the total error grow by a factor of $4/3,$ this split loses
almost nothing. By the conditional independence of the two blocks given
$V,$ the normalized slack $1-W_\beta/\beta$ is submultiplicative, up to an
additive term $2/\beta$ for the two announced decisions
(Lemma~\ref{lem:priced-product}):
\[
  1-\frac{W_\beta(F_0\times F_1,\mu_0\otimes\mu_1)}\beta
  \leq\Bigl(1-\frac{W_\beta(F_0,\mu_0)}\beta\Bigr)
      \Bigl(1-\frac{W_\beta(F_1,\mu_1)}\beta\Bigr)+\frac2\beta.
\]
Iterating this bound over $n$ blocks of one copy each
(Lemma~\ref{lem:block-choice}), with $\beta=(I+2n)/\epsilon,$ gives a
protocol for $F$ with error less than $4\epsilon/n$ and information cost
less than $4I/n+8.$ The term $8$ is affordable when $I\geq n.$

In the low-information regime $I\leq1,$ an additive loss is unaffordable,
and we use a completely different idea. Let $p_A$ and $p_B$ be the least
errors of predicting $F(X,Y)$ from $X$ alone and from $Y$ alone, where
$(X,Y)\sim\mu.$ We prove that these two parameters govern the low-information 
regime: a local-output protocol for $F^n$ with error $1/3$ under $\mu^n$ reveals at least
$2n(p_A+p_B)/(3\ln2)-2$ bits of information (Lemma~\ref{lem:input-only}).
Here, therefore, $p_A+p_B=O(1/n),$ and each party can predict the answer
on his or her own with error $O(1/n).$ Our extraction starts from the
standard embedding into a uniformly random coordinate $j,$ which divides
the information cost by $n.$ The error, however, does not decrease, since
$\pi$ may answer incorrectly on many coordinates at once. We exploit the
fact that, in the embedding, each party knows both inputs on some
coordinates: Alice on those after $j,$ and Bob on those before $j.$ Each
party checks the answers of $\pi$ on these coordinates and keeps his or her
answer for coordinate $j$ only if all of them are correct. A wrong answer
survives Alice's check only if $j$ is the last coordinate on which her
answers are wrong, and it survives Bob's check only if $j$ is the first
such coordinate for him; as $j$ is uniformly random, each of these events
has probability at most $\epsilon/n.$ The parties run this procedure twice,
independently, and each party falls back on his or her own prediction only
if both of his or her checks fail. A variance bound (Lemma~\ref{lem:variance}) shows that the probability that
a check fails varies little with the input, and the resulting protocol for
$F$ has information cost at most $2I/n$ and error at most $2\epsilon^2(p_A+p_B)+8\epsilon/n$
under $\mu$ (Lemma~\ref{lem:two-checks}), which is $O(\epsilon/n)$ in this
regime. The two parties may keep answers from different runs, and this is
why our multiplicative theorem uses local outputs.

In the intermediate regime $1<I<n,$ we combine the two techniques. We
partition the $n$ copies into $m=\lceil I\rceil$ blocks of about $n/m$
copies each and apply the pricing argument to these blocks rather than to
single copies. We obtain a block of some size $k$ and a protocol for $F^k$
with error $O(\epsilon k/n)$ and information cost $O(Ik/n)=O(1),$ which
places the block in the low-information regime: Lemma~\ref{lem:input-only},
applied to $\pi,$ gives $p_A+p_B=O(I/n)=O(1/k).$ Extracting a single copy
from the block by checks gives error $O(\epsilon/n)$ and information cost
$O(I/n),$ as in the other two regimes, and completes the proof of
Theorem~\ref{main:dist-direct-sum}.

\subsubsection*{Distributional XOR lemma.}

Our proof of the distributional XOR lemma follows the same three regimes,
but it is considerably deeper, because a protocol for $F^{\oplus n}$
commits only to the XOR of the $n$ answers. Pricing carries over to XORs
with announced outputs (Lemma~\ref{lem:priced-product-xor}). The default
answer for a block is now the more likely value of the answer on that block
given $V,$ which must be announced, and the errors of the two default
answers combine through the product of biases~\eqref{eq:intro-bias-product}.
This costs a factor of $2$ in the price and two more announced bits per
split, and gives our XOR lemma with constant additive loss
(Theorem~\ref{main:xor-additive}). By the example in
Section~\ref{subsec:results}, this additive loss cannot be removed with
announced outputs. For the optimal theorem, we work with vector outputs
throughout, and each step requires new ideas.

With vector outputs, we price deficits instead of errors
(Lemma~\ref{lem:priced-product-vector}). The default outputs for a block
are now vectors that need not be announced. For block $0,$ Alice uses the
conditional expectation, given $V,$ of the product of the answer on block
$0$ and Bob's output vector, normalized to unit length, and Bob keeps his
output vector; block $1$ is symmetric. Given $V,$ the deficit of the given
protocol is at least $1-(1-\Delta_0)(1-\Delta_1),$ where $\Delta_j$ is the
deficit of the default outputs for block $j.$ Thus, these deficits combine
exactly as errors do for direct sums, and the normalized slack of priced
vector-output information complexity is submultiplicative up to an
additive term $2/\beta,$ with no announcements other than the two
decisions.

The low-information regime requires the most new ideas. The checks of the
direct sum proof are unavailable, since a protocol for $F^{\oplus n}$
produces no answers on individual coordinates, and we use geometry
instead. The role of $p_A$ and $p_B$ is played by the least deficit
$\Delta_\mu(F)$ of a protocol with vector outputs and no communication,
which is the distributional analogue of $1-\margin(F);$ let $w_x$ and
$w'_y$ be unit vectors that attain it. Our extraction
(Lemma~\ref{lem:vector-lift}) again starts from the standard embedding
into a uniformly random coordinate $j.$ At its end, Alice can compute the
conditional expectation $a$ of Bob's output vector multiplied by the XOR of
the answers on coordinates $1,\ldots,j,$ and Bob can compute the analogous
vector $b$ for coordinates $1,\ldots,j-1.$ A telescoping argument shows
that the expected squared distance between $b$ and $F(x,y)a,$ for the
actual input $(x,y),$ is at most $4\epsilon/n,$ where $\epsilon$ is the
vector error of the given protocol. To produce unit vectors, the parties
apply an inverse stereographic projection, which combines a vector of norm
at most $1$ with a unit vector. The result stays close to $w_x$ or $w'_y$
when $a$ or $b$ is short, and close to the direction of $a$ or $b$ when it
is nearly a unit vector. As in the extraction by checks, two independent
runs square the weight of the communication-free vectors, and the variance
bound controls the fluctuations of this weight. The result is a protocol for $F$ with vector
outputs, information cost at most $2I/n,$ and deficit at most
$32\epsilon^2\Delta_\mu(F)+80\epsilon/n.$

It remains to bound $\Delta_\mu(F).$ For this, we show that every protocol
with vector outputs yields communication-free vectors: each party combines
his or her output vectors for all possible transcripts into a single
vector, scaling the vector for each transcript by the square root of its
conditional probability given that party's input. The deficit of the merged vectors is at most twice the
deficit of the protocol plus twice its information cost
(Lemma~\ref{lem:transcript-removal}). Applied to the protocol produced by
our extraction, this gives
$\Delta_\mu(F)\leq64\epsilon^2\Delta_\mu(F)+(160\epsilon+4I)/n,$ and hence
$\Delta_\mu(F)\leq(5I+2)/n$ for $\epsilon\leq1/100$
(Lemma~\ref{lem:deficit}). This bound is a distributional counterpart of
the margin bound of Theorem~\ref{thm:margin-ic}. In particular, for
$I\leq1/2,$ our extraction produces vector error below $1/(2n).$

For $I>1/2,$ we combine pricing with this extraction. We partition the $n$
copies into $m=\min\{n,\lceil I\rceil\}$ blocks, so that in the
high-information regime $I\geq n$ the blocks are single copies, and we
price the deficits of the blocks (Lemma~\ref{lem:block-choice-vector}).
This yields a block of some size $k$ and a protocol for $F^{\oplus k}$
with vector outputs, vector error $O(1/m),$ and information cost
$O(Ik/n).$ Since $\Delta_\mu(F)=O(m/n),$ our extraction applied to this
block gives a protocol for $F$ with vector error $O(1/n)$ and information
cost $O(I/n).$ Altogether, a protocol for $F^{\oplus n}$ with vector error
$1/100$ yields a protocol for $F$ with vector error less than $3/(4n)$ and
at most $34/n$ times the information cost
(Theorem~\ref{thm:dist-extraction}). To reach a smaller error $\epsilon,$
we run a given protocol independently on $r=\lfloor1/(100\epsilon)\rfloor$
disjoint blocks of $n$ copies, and each party outputs the tensor product of
his or her $r$ vectors. The signed inner products multiply, so that the
result is a protocol for $F^{\oplus rn}$ with vector error at most
$r\epsilon\leq1/100,$ and Theorem~\ref{main:dist-xor} follows. Finally, the
rounding protocol of Section~\ref{subsec:rounding}, run with optimal
vectors for each conditional distribution of the input given the
transcript, converts vector outputs into shares with no increase in error
and at a small cost in information (Lemma~\ref{lem:vector-rounding}). This
gives~\eqref{eq:vector-comparison}.

\section{Preliminaries}\label{sec:preliminaries}

This section establishes notation and provides the necessary background
on communication protocols, information complexity, and margins. The
information-theoretic facts are used throughout, whereas the
facts about margins and discrepancy are needed only for partial
functions; the less standard of these are proved in
Appendix~\ref{app:norms}.

\subsection{General notation}

All sets in this paper are finite and nonempty, unless explicitly stated
otherwise. We adopt the shorthand $[n]=\{1,2,\ldots,n\}$ and write $\log$ for
the binary logarithm and $\ln$ for the natural logarithm. For a tuple $x=(x_1,\ldots,x_n),$ we use the notation
$x_{<j}=(x_1,\ldots,x_{j-1})$ and $x_{>j}=(x_{j+1},\ldots,x_n),$ with
$x_{\leq j}$ and $x_{\geq j}$ defined analogously. An empty product has value
$1.$

We represent Boolean output values by $-1$ and $1,$ corresponding to
``true'' and ``false,'' respectively. In this representation, the XOR
operation is multiplication. Input bits, where they occur, take the
values $0$ and $1.$ A \emph{communication problem} with output set $\Zcal$ is a
function $F\colon\Xcal\times\Ycal\to\Zcal\cup\{*\},$ where the symbol
$*\notin\Zcal$ marks the inputs on which no answer is required. The
\emph{domain}, or \emph{promise domain}, of $F$ is the set
\[
  \dom F=\{(x,y)\in\Xcal\times\Ycal:F(x,y)\ne*\},
\]
which we require to be nonempty. A \emph{legal input} is any pair in
$\dom F.$ The problem $F$ is \emph{total} if $\dom F=\Xcal\times\Ycal,$ and
\emph{partial} otherwise; accordingly, we also speak of total and partial
functions. In particular, a Boolean function
$F\colon\Xcal\times\Ycal\to\{-1,1\}$ is a total communication problem.

For a set $\Dcal\subseteq\Xcal\times\Ycal$ and an integer $n\geq1,$ put
\[
  \Dcal^n=\{(x,y)\in\Xcal^n\times\Ycal^n:
                    (x_i,y_i)\in\Dcal\text{ for every }i\in[n]\}.
\]
For a communication problem $F\colon\Xcal\times\Ycal\to\{-1,1,*\}$ and an
integer $n\geq1,$ define communication problems
$F^{\oplus n}\colon\Xcal^n\times\Ycal^n\to\{-1,1,*\}$ and
$F^n\colon\Xcal^n\times\Ycal^n\to\{-1,1\}^n\cup\{*\}$ by
\begin{align*}
  F^{\oplus n}(x,y)&=\prod_{i=1}^n F(x_i,y_i),\\
  F^n(x,y)&=(F(x_1,y_1),\ldots,F(x_n,y_n))
\end{align*}
for $(x,y)\in(\dom F)^n,$ and by $F^{\oplus n}(x,y)=F^n(x,y)=*$ otherwise.
Thus $\dom F^{\oplus n}=\dom F^n=(\dom F)^n,$ and the promise holds
separately in each coordinate. The problem $F^{\oplus n}$ asks for the XOR
of $n$ answers, whereas $F^n$ asks for the entire tuple of answers. For
$n=1,$ we identify both problems with $F.$ 

A \emph{partial real matrix} with rows indexed by $\Xcal$ and
columns indexed by $\Ycal$ is a matrix
$A\in(\mathbb R\cup\{*\})^{\Xcal\times\Ycal}$ whose \emph{domain}
$\dom A=\{(x,y)\in\Xcal\times\Ycal:A_{xy}\ne*\}$ is
nonempty; the matrix entries
outside $\dom A$ are unspecified and are indicated with $*.$ 
We call $A$ a \emph{partial sign matrix} if its
entries lie in $\{-1,1,*\},$ and a real matrix if
$\dom A=\Xcal\times\Ycal.$ We distinguish communication
problems, which are functions, from matrices.  The
\emph{characteristic matrix} of a communication problem
$F\colon\Xcal\times\Ycal\to\{-1,1,*\}$ is the partial sign
matrix $A_F\in\{-1,1,*\}^{\Xcal\times\Ycal}$ with entries
$(A_F)_{xy}=F(x,y).$

We write $\Prob$ and $\Exp$ for probability and expectation, respectively,
$\1_E$ for the indicator of an event $E,$ and $\overline E$ for its
complement. For a set $S,$ we write
$\mathcal P(S)$ for the set of probability distributions on $S.$ When
$S\subseteq S',$ we regard $\mathcal P(S)$ as the set of distributions in
$\mathcal P(S')$ supported on $S.$ For $\mu\in\mathcal P(S)$ and a function
$f$ on $S,$ we write $\Exp_\mu f$ and $\Var_\mu(f)$ for the mean and variance
of $f$ under $\mu;$ a random variable in a subscript, as in $\Exp_X,$
indicates averaging over that variable. The notation $X\perp Y\mid Z$
means that $X$ and $Y$ are conditionally independent given $Z.$ For a set
$\Dcal\subseteq\Xcal\times\Ycal$ and $\mu\in\mathcal P(\Dcal),$ the notation
$\mu^n\in\mathcal P(\Dcal^n)$ refers to the distribution of $n$ independent
input pairs with common law $\mu.$ Independence between Alice's and Bob's
inputs within a pair is not assumed. Conditional distributions on events of
probability zero may be defined arbitrarily. Every input distribution for a
communication problem $F$ is understood to lie in $\mathcal P(\dom F).$

\subsection{Information theory}\label{subsec:information-theory}

For a finite random variable $X,$ its entropy is
\[
  H(X)=-\sum_x\Prob[X=x]\log\Prob[X=x],
\]
where $0\log0=0.$ We write
\[
  h_2(p)=-p\log p-(1-p)\log(1-p),\qquad 0\leq p\leq1,
\]
for binary entropy. Conditional entropy is defined by averaging over the conditioning variable. The conditional mutual information of $X$ and $Y$
given $Z$ is
\[
  I(X;Y\mid Z)=H(X\mid Z)-H(X\mid Y,Z).
\]
It is nonnegative and symmetric in $X$ and $Y.$ We use the chain rule
\begin{equation}\label{eq:chain-rule}
  I(T;X,Y\mid Z)=I(T;X\mid Z)+I(T;Y\mid X,Z).
\end{equation}

For probability distributions $p,q$ on a common finite set, their
\emph{relative entropy}, or Kullback--Leibler divergence, is
\begin{equation}\label{eq:relative-entropy}
  \kl{p}{q}=\sum_z p(z)\log\frac{p(z)}{q(z)}.
\end{equation}
A summand with $p(z)=0$ is zero, and the value is $+\infty$ if
$p(z)>0=q(z)$ for some $z.$ All relative entropies are in bits; a subscript
$\mathrm{nat}$ indicates natural logarithms. The \emph{total variation
distance} between $p$ and $q$ is
$\TV(p,q)=\sum_z(p(z)-q(z))_+=\frac12\sum_z\abs{p(z)-q(z)},$ where
$a_+=\max\{a,0\}.$

We use nonnegativity and the data-processing inequality for relative
entropy. We also use the following reference-distribution identity. If
$(p_x)_x$ are probability distributions, $X$ is random, and
$\bar p=\Exp_X p_X,$ then
\begin{equation}\label{eq:kl-mixture}
  \Exp_X\kl{p_X}{q}
    =\Exp_X\kl{p_X}{\bar p}+\kl{\bar p}{q}.
\end{equation}
The first term on the right is $I(X;Z)$ when the conditional law of $Z$
given $X=x$ is $p_x.$ The same identity can be applied conditionally on
additional observations. It follows by splitting
$\log(p_x/q)=\log(p_x/\bar p)+\log(\bar p/q),$ averaging over $z$ with
respect to $p_x,$ and then averaging over $X;$ in the second
term on the r.h.s., these two averages combine into
an average with respect to $\bar p.$

For a random variable $B\in\{-1,1\}$ and an observation $V,$ the least error
of any prediction of $B$ based on $V$ is
\begin{equation}\label{eq:prediction-error}
  \frac{1-\abs{\Exp[B\mid V]}}{2}.
\end{equation}
This error is attained by outputting the more likely
value of $B$ conditional on $V,$ with ties broken arbitrarily.

\subsection{Protocols and information complexity}\label{subsec:protocols}

We work in the two-party randomized model with both public and private
randomness. We first define protocols without public randomness. A
\emph{private-coin protocol} is given by a finite rooted tree in which every
internal node has two children and is owned by Alice or by Bob. Alice
receives an input $x\in\Xcal,$ and Bob receives an input $y\in\Ycal.$
Starting at the root, the owner of the current node sends a bit, and the
protocol moves to the corresponding child. At a node $v$ owned by Alice,
she sends $1$ with a probability $p_v(x)$ that depends only on $v$ and on
her input $x;$ likewise, at a node $v$ owned by Bob, he sends $1$ with a
probability $q_v(y)$ that depends only on $v$ and on his input $y.$ The
sequence of bits sent is the \emph{transcript}; it determines the leaf at
which the protocol ends. In particular, a private-coin protocol has finite
depth and finitely many transcripts. There is no restriction on local
computation, and the depth can be arbitrarily large. Unless another output
convention is explicitly specified, the output is a function of the
transcript, so that both parties know the answer. Under the other output
conventions, each party ends with an output of his or her own, drawn from
a distribution that depends only on the transcript and on his or her
input, independently of the other party's output. Each party's random choices
are made using his or her \emph{private randomness}, which is
independent of the input and of the other party's private randomness. For
example, at a node $v$ that she owns, Alice can draw a fresh random number
uniformly from $[0,1]$ and send $1$ if and only if it is less than
$p_v(x).$

A \emph{public-coin protocol} $\pi$ is a probability distribution over
private-coin protocols with a common input space $\Xcal\times\Ycal$ and a
common output convention. To run $\pi,$ the parties use public randomness
to draw a private-coin protocol from this distribution, and then they run
it. Notationally, the private-coin protocol drawn is indexed by a random
variable $R,$ the \emph{public randomness} of $\pi,$ which is independent of
the input, and we write $\pi_R$ for this private-coin protocol. The public
randomness may take infinitely many
values, as when the parties publicly choose a uniformly random real number
or unit vector. The transcript of $\pi$ given $R$ is that of $\pi_R;$ we do not regard
$R$ as part of it. Under the default output convention, the output of $\pi$
is thus a function of the public randomness and the transcript. Throughout
this paper, the word \emph{protocol} refers to a public-coin protocol; a
private-coin protocol is the special case of a constant $R.$ A protocol for
a communication problem $F$ is defined on the full input space
$\Xcal\times\Ycal,$ but no correctness requirement is imposed outside
$\dom F.$

Averaging over the private-coin protocols drawn by a protocol $\pi$ amounts
to averaging over $R.$ Thus, every probability associated with $\pi,$ such
as its probability of error on an input, is the average over $R$ of the
corresponding probability for $\pi_R.$ Likewise, in information-theoretic
expressions, conditioning on $R$ means averaging over $R$: a quantity such
as $I(M;X\mid Y,R),$ where $M$ is the transcript and $(X,Y)$ is the input, is
the average over $R$ of the corresponding quantity $I(M;X\mid Y)$ for
$\pi_R.$ When $R$ takes finitely many values, this agrees with the
definitions in Section~\ref{subsec:information-theory}.

For an input distribution $\mu,$ let $(X,Y)\sim\mu,$ let $R$ be the public
randomness, and let $M$ be the transcript. The
\emph{internal information cost} of $\pi$ under $\mu$ is
\begin{align}
  \IC(\pi,\mu)=I(M;X\mid Y,R)+I(M;Y\mid X,R).
\label{eq:information-cost}
\end{align}
In particular, $\IC(\pi,\mu)$ is the average over $R$ of $\IC(\pi_R,\mu).$
When $R$ takes finitely many values, the independence of $R$
and the input $(X,Y)$ yields
\begin{align}
  \IC(\pi,\mu)= I(R,M;X\mid Y) + I(R,M;Y\mid X).
\label{eq:information-cost-R}
\end{align}
We will also use
\[
  \IC(\pi)=\sup_\mu\IC(\pi,\mu),
\]
where the supremum ranges over all $\mu\in\mathcal P(\dom F)$ for the
communication problem $F$ under consideration. For a total problem, this is
all of $\mathcal P(\Xcal\times\Ycal).$

For a communication problem $F$ with output set $\Zcal,$ write
\begin{align*}
  \err_\mu(\pi,F)&=\Prob[\pi(X,Y)\ne F(X,Y)],\\
  \err(\pi,F)&=\max_{(x,y)\in\dom F}\Prob[\pi(x,y)\ne F(x,y)].
\end{align*}
The first probability includes both the input distribution
and the protocol's
public and private randomness; the second includes only the
protocol's public and private randomness. Define
\begin{align}
  \IC_\epsilon(F,\mu)
    &=\inf_{\pi:\,\err_\mu(\pi,F)\leq\epsilon}\IC(\pi,\mu),
      \label{eq:distributional-ic}\\
  \IC^{\mathsf D}_\epsilon(F)
    &=\sup_\mu\IC_\epsilon(F,\mu),
      \label{eq:max-distributional-ic}\\
  \IC_\epsilon(F)
    &=\inf_{\pi:\,\err(\pi,F)\leq\epsilon}\IC(\pi).
      \label{eq:worst-case-ic}
\end{align}
We refer to~\eqref{eq:worst-case-ic} as the \emph{information complexity} of
$F.$ In particular, the protocol in~\eqref{eq:worst-case-ic} must satisfy
its error bound on every legal input and its information bound under every
input distribution in $\mathcal P(\dom F).$ For $F^n,$ an error means that at
least one of the $n$ reported answers is incorrect.

We recall two elementary properties of protocols. First, after fixing the
public randomness $R=r,$ the probability of any transcript $m$ factors as
\begin{equation}\label{eq:rectangle}
  \Prob[M=m\mid X=x,Y=y,R=r]=a_{r,m}(x)b_{r,m}(y)
\end{equation}
for nonnegative functions $a_{r,m}$ and $b_{r,m}.$ This is the usual
\emph{rectangle property}, obtained by grouping the conditional message
probabilities according to their sender. Second, sending an additional bit
increases internal information cost by at most one. For example, if 
Alice appends a bit $B$ to the transcript $M,$ then
\[
  I(B;Y\mid X,R,M)=0,
  \qquad
  I(B;X\mid Y,R,M)\leq1,
\]
as one can verify from the chain rule. Announcing an answer that is a
function of $(R,M),$ on the other hand, adds no information
at all.

For a fixed protocol, information cost is a concave and continuous function
of the input distribution, with a modulus of continuity that does not depend
on the protocol.  Concavity is well known; see, e.g.,~\cite[Lemma~A.1]{BGPW13}.
A continuity bound of this kind was proved by Braverman, Garg, Pankratov, and
Weinstein~\cite[Lemma~B.1]{BGPW13}. We will use both properties in this paper.
For completeness, we include a short proof in the form that we need.

\begin{fact}[Concavity and continuity]\label{fact:concavity}
Let $\pi$ be a protocol with input space $\Xcal\times\Ycal.$ Then
$\IC(\pi,\mu)$ is a concave function of $\mu\in\mathcal P(\Xcal\times\Ycal).$
Moreover, for all $\mu,\nu\in\mathcal P(\Xcal\times\Ycal),$
\[
  \abs{\IC(\pi,\mu)-\IC(\pi,\nu)}\leq \Delta\log(|\Xcal||\Ycal|)+2h_2(\Delta),
\]
where $\Delta=\TV(\mu,\nu).$
\end{fact}

\begin{proof}
The information cost of $\pi$ under every input distribution is the
average of that of $\pi_R$ over the public randomness $R,$ and averaging
preserves both claims. We may therefore assume that $\pi$ is a
private-coin protocol. Let $Z$ be a random variable with finitely many
values, and let $(X,Y)$ be an input whose conditional distribution given
$Z=z$ is $\mu_z\in\mathcal P(\Xcal\times\Ycal),$ so that $(X,Y)$ has
distribution $\mu=\sum_z\Prob[Z=z]\,\mu_z.$ Let $M$ be the transcript of
$\pi$ on input $(X,Y).$ Since $M$ depends on $Z$ only through the input, we
have $I(M;Z\mid X,Y)=0,$ and the chain rule gives
\[
  I(M;X\mid Y)=I(M;X,Z\mid Y)=I(M;Z\mid Y)+I(M;X\mid Y,Z).
\]
Adding the analogous identity with $X$ and $Y$ interchanged, we obtain
\begin{align}\label{eq:ic-concave}
  \IC(\pi,\mu)=I(M;Z\mid Y)+I(M;Z\mid X)+\sum_z\Prob[Z=z]\IC(\pi,\mu_z),
\end{align}
where each of the first two terms lies in $[0,H(Z)].$ In particular,
$\IC(\pi,\cdot)$ is concave.

For the continuity bound, note that
$0\leq\IC(\pi,\cdot)\leq\log(|\Xcal||\Ycal|),$ because
$I(M;X\mid Y)\leq H(X)\leq\log|\Xcal|$ and likewise
$I(M;Y\mid X)\leq\log|\Ycal|.$ This settles the cases $\Delta=0$ and $\Delta=1.$ For
$0<\Delta<1,$ we have $\sum_{x,y}\min\{\mu(x,y),\nu(x,y)\}=1-\Delta,$ so that
$\mu=(1-\Delta)\omega+\Delta\mu'$ and $\nu=(1-\Delta)\omega+\Delta\nu'$ for some
$\omega,\mu',\nu'\in\mathcal P(\Xcal\times\Ycal).$ Equation~\eqref{eq:ic-concave},
applied with $Z\in\{0,1\}$ selecting the component, so that
$H(Z)=h_2(\Delta),$ shows that
$\IC(\pi,\mu)-(1-\Delta)\IC(\pi,\omega)-\Delta\IC(\pi,\mu')$ and
$\IC(\pi,\nu)-(1-\Delta)\IC(\pi,\omega)-\Delta\IC(\pi,\nu')$ both lie in
$[0,2h_2(\Delta)].$ Hence
\[
  \abs{\IC(\pi,\mu)-\IC(\pi,\nu)}
  \leq \Delta\abs{\IC(\pi,\mu')-\IC(\pi,\nu')}+2h_2(\Delta)
  \leq \Delta\log(|\Xcal||\Ycal|)+2h_2(\Delta).\qedhere
\]
\end{proof}

Finally, we say that a protocol has \emph{finite public randomness} if its public
randomness takes finitely many values. In that case, the public randomness
and the transcript jointly take finitely many values.
When designing protocols, we freely use public randomness with infinitely
many values. In many proofs, on the other hand, it is convenient to assume
finite public randomness, and Appendix~\ref{app:finite} shows that this
entails no loss of generality. Specifically, every protocol can be replaced
by a protocol with finite public randomness that errs on every legal input
with the same probability as the original. Moreover, the
replacement protocol can be chosen to have the same
information cost as the original under any finitely many
input distributions (Proposition~\ref{prop:finite-fixed}), or information cost
within any prescribed $\eta>0$ of the original under every
input distribution (Proposition~\ref{prop:finite-all}).

\subsection{The minimax theorem}\label{subsec:minimax}

The definitions immediately imply
$\IC^{\mathsf D}_\epsilon(F)\leq\IC_\epsilon(F).$ We will use the following
converse due to Braverman~\cite[Thm.~3.5]{Braverman15},
generalized here to partial functions. 

\begin{fact}[Minimax theorem]\label{fact:minimax}
Let $F\colon\Xcal\times\Ycal\to\Zcal\cup\{*\}$ be a communication problem.
Then for every $\epsilon>0$ and $\alpha\in(0,1),$
\[
  \IC_{\epsilon/\alpha}(F)
  \leq\frac{\IC^{\mathsf D}_\epsilon(F)}{1-\alpha}.
\]
In particular,
\begin{equation}\label{eq:minimax}
  \IC_{2\epsilon}(F)\leq2\IC^{\mathsf D}_\epsilon(F).
\end{equation}
\end{fact}

\begin{proof}
We adapt Braverman's proof to make explicit that the domain of $F$ need not
be a rectangle. Let $I>\IC^{\mathsf D}_\epsilon(F),$ and consider the cost
function
\[
  L(\pi,\mu)=\frac{1-\alpha}{I}\IC(\pi,\mu)
                   +\frac{\alpha}{\epsilon}\err_\mu(\pi,F),
\]
where $\pi$ ranges over protocols and $\mu$ over $\mathcal P(\dom F).$ For a
fixed protocol, $L$ is a concave function of $\mu$ by
Fact~\ref{fact:concavity}, since error is linear in $\mu.$

Let $0<\Delta\leq1/2.$ Since $\mathcal P(\dom F)$ is compact, there are finitely many
distributions $\mu_1,\ldots,\mu_N\in\mathcal P(\dom F),$ among them the point
masses on the legal inputs, such that every $\mu\in\mathcal P(\dom F)$
satisfies $\TV(\mu,\mu_j)\leq \Delta$ for some $j.$ For a protocol $\pi,$ put
$\ell(\pi)=(L(\pi,\mu_1),\ldots,L(\pi,\mu_N)),$ and consider the sets
\begin{align*}
  C&=\{\ell(\pi):\pi\text{ is a protocol}\},\\
  O&=(-\infty,1)^N.
\end{align*}
Both sets are convex, and $O$ is open. For $C,$ convexity holds because
public mixtures average both terms in $L.$ We claim that $C$ and $O$
intersect. Otherwise, by the separating hyperplane theorem, there is a
nonzero $\phi\in\mathbb R^N$ such that
\[
  \langle\phi,\ell(\pi)\rangle\geq\langle\phi,\omega\rangle
\]
for every protocol $\pi$ and every $\omega\in O.$ Letting a single coordinate
of $\omega$ tend to $-\infty$ shows that $\phi\geq0.$ Thus $\phi$ is a
positive multiple of a probability vector $p=(p_1,\ldots,p_N),$ and letting
$\omega$ tend to the constant vector with entries $1$ gives
$\langle p,\ell(\pi)\rangle\geq1$ for every protocol $\pi.$ On the other
hand, the distribution $\mu^*=\sum_jp_j\mu_j$ satisfies
$\IC_\epsilon(F,\mu^*)<I,$ so that some protocol $\pi^*$ satisfies
$L(\pi^*,\mu^*)<1.$ By concavity,
$\langle p,\ell(\pi^*)\rangle\leq L(\pi^*,\mu^*)<1,$ a contradiction.

Thus, some protocol $\pi$ satisfies $L(\pi,\mu_j)<1$ for every $j.$ The two
terms in $L$ are nonnegative, so $\IC(\pi,\mu_j)<I/(1-\alpha)$ and
$\err_{\mu_j}(\pi,F)<\epsilon/\alpha$ for every $j.$ The point masses among
the $\mu_j$ show that $\pi$ has worst-case error less than
$\epsilon/\alpha.$ For an arbitrary $\mu\in\mathcal P(\dom F),$ choose $j$
with $\TV(\mu,\mu_j)\leq\Delta.$ As $h_2$ is increasing on $[0,1/2],$
Fact~\ref{fact:concavity} gives
\[
  \IC(\pi,\mu)\leq\IC(\pi,\mu_j)+\Delta\log(|\Xcal||\Ycal|)+2h_2(\Delta)
  \leq\frac I{1-\alpha}+\Delta\log(|\Xcal||\Ycal|)+2h_2(\Delta).
\]
Hence
$\IC_{\epsilon/\alpha}(F)\leq I/(1-\alpha)+\Delta\log(|\Xcal||\Ycal|)+2h_2(\Delta).$
Letting $\Delta\to0$ and then $I$ decrease to $\IC^{\mathsf D}_\epsilon(F)$
completes the proof.
\end{proof}

We will also use a minimax theorem for information cost alone, over a
prescribed class of protocols that is closed under finite public
mixtures. A typical example is a class of protocols that compute a given
communication problem with worst-case error at most $\epsilon.$ Since
the error bound is then built into the class rather than traded against
information, the following proposition, unlike Fact~\ref{fact:minimax},
loses nothing in either error or information.

\begin{proposition}[The minimax theorem for a protocol class]
\label{prop:class-minimax}
Let $\Ccal$ be a nonempty class of protocols on a common finite promise
domain $\Dcal\subseteq\Xcal\times\Ycal,$ closed under finite public
mixtures. Then
\[
  \inf_{\pi\in\Ccal}\sup_{\mu\in\mathcal P(\Dcal)}\IC(\pi,\mu)
  =\sup_{\mu\in\mathcal P(\Dcal)}\inf_{\pi\in\Ccal}\IC(\pi,\mu).
\]
\end{proposition}

\begin{proof}
Let $I$ denote the right-hand side. It is immediate that the left-hand side
is at least $I.$ For the reverse inequality, let $\eta>0$ and
$0<\Delta\leq1/2,$ and
choose finitely many distributions $\mu_1,\ldots,\mu_N\in\mathcal P(\Dcal)$
such that every $\mu\in\mathcal P(\Dcal)$ satisfies $\TV(\mu,\mu_j)\leq \Delta$
for some $j.$ We argue as in the proof of Fact~\ref{fact:minimax}, with the
sets $C=\{(\IC(\pi,\mu_1),\ldots,\IC(\pi,\mu_N)):\pi\in\Ccal\}$ and
$O=(-\infty,I+\eta)^N.$ Here $C$ is convex because $\Ccal$ is closed under
finite public mixtures. If $C$ and $O$ were disjoint, the separating
hyperplane theorem would give a probability vector $p$ with
$\sum_jp_j\IC(\pi,\mu_j)\geq I+\eta$ for every $\pi\in\Ccal.$ However, by
the definition of $I,$ some $\pi\in\Ccal$ has cost less than $I+\eta$ under
$\sum_jp_j\mu_j,$ and by
concavity, $\sum_jp_j\IC(\pi,\mu_j)$ is at most this cost. Hence some
$\pi\in\Ccal$ satisfies $\IC(\pi,\mu_j)<I+\eta$ for every $j.$ As $h_2$ is
increasing on $[0,1/2],$ Fact~\ref{fact:concavity} shows that this protocol
has cost less than
$I+\eta+\Delta\log(|\Xcal||\Ycal|)+2h_2(\Delta)$ under every input distribution.
Letting $\eta\to0$ and $\Delta\to0$ shows that the left-hand side is at most
$I.$
\end{proof}

\subsection{Protocol composition}

Publicly choosing among protocols averages their information costs and
errors. More precisely, let $F\colon\Xcal\times\Ycal\to\Zcal\cup\{*\}$ be a
communication problem, and let $\pi_1,\ldots,\pi_t$ be protocols with input
space $\Xcal\times\Ycal$ and output space $\Zcal.$ Let $\pi$ publicly choose
an index $J$ independently of the input and run $\pi_J.$ Then for every
$\mu\in\mathcal P(\dom F),$
\begin{align}
  \IC(\pi,\mu)&=\sum_{j=1}^t\Prob[J=j]\IC(\pi_j,\mu),
    \label{eq:public-mixture-cost}\\
  \err_\mu(\pi,F)&=\sum_{j=1}^t\Prob[J=j]\err_\mu(\pi_j,F).
    \label{eq:public-mixture-error}
\end{align}
Equation~\eqref{eq:public-mixture-cost} follows by conditioning on the
public choice $J,$ and~\eqref{eq:public-mixture-error} is the law of total
probability.

The next proposition concerns running a protocol $\rho$ after a protocol $\pi,$
but only on some of the transcripts of $\pi.$ Conditioning on a transcript
changes the input distribution, so the proposition requires a bound on the
information cost of $\rho$ under every input distribution on the promise
domain.

\begin{proposition}[Conditional composition]\label{prop:continuation}
Let $F\colon\Xcal\times\Ycal\to\Zcal\cup\{*\}$ be a communication problem.
Let $\pi$ and $\rho$ be protocols with
input space $\Xcal\times\Ycal$ and output space $\Zcal,$ and let
$\mathcal M$ be a set of transcripts of $\pi.$ Let $\pi'$ be the protocol
that runs $\pi$ and then, if the transcript of $\pi$ lies in $\mathcal M,$
runs $\rho$ on the same input with fresh public and private randomness.
The output of $\pi'$ is that of $\rho$ if $\rho$ is run, and that of $\pi$
otherwise. Suppose that
\begin{align*}
  \IC(\rho,\nu)&\leq K\quad\text{for every $\nu\in\mathcal P(\dom F)$},\\
  \err(\rho,F)&\leq\epsilon.
\end{align*}
Then for every $\mu\in\mathcal P(\dom F),$
\begin{align*}
  \IC(\pi',\mu)&\leq\IC(\pi,\mu)+qK,\\
  \err_\mu(\pi',F)
    &\leq\Prob[M\notin\mathcal M,\ \pi(X,Y)\ne F(X,Y)]+q\epsilon,
\end{align*}
where $(X,Y)\sim\mu,$ $M$ is the transcript of $\pi$ on input $(X,Y),$
and $q=\Prob[M\in\mathcal M].$
\end{proposition}

\begin{proof}
Let $R$ be the public randomness of $\pi,$ and let $R_\rho$ be the fresh
public randomness for $\rho.$ Let $M_\rho$ be the transcript of $\rho$ if
$\rho$ is run, and a fixed null symbol otherwise. Thus $\pi'$ has public
randomness $(R,R_\rho)$ and transcript $(M,M_\rho).$ The public
randomness $R_\rho$ and the private randomness of $\rho$ are independent of
the input and of the randomness of $\pi,$ and hence of $(X,Y,R,M).$ As a
result, $I(M;X\mid Y,R,R_\rho)=I(M;X\mid Y,R)$ and
$I(M;Y\mid X,R,R_\rho)=I(M;Y\mid X,R).$ Together with the chain rule,
this gives
\[
  \IC(\pi',\mu)
  =\IC(\pi,\mu)+I(M_\rho;X\mid Y,R,R_\rho,M)+I(M_\rho;Y\mid X,R,R_\rho,M).
\]
For values $r$ of $R$ and $m$ of $M,$ let $\mu_{r,m}\in\mathcal P(\dom F)$
be the conditional distribution of $(X,Y)$ given $(R,M)=(r,m).$ If
$m\in\mathcal M,$ then conditional on $(R,M)=(r,m),$ the tuple
$(X,Y,R_\rho,M_\rho)$ has the joint distribution of the input, public
randomness, and transcript of $\rho$ when the input is drawn from
$\mu_{r,m},$ so that
\begin{align*}
  &I(M_\rho;X\mid Y,R_\rho,(R,M)=(r,m))+I(M_\rho;Y\mid X,R_\rho,(R,M)=(r,m))\\
  &\qquad=\IC(\rho,\mu_{r,m})\leq K.
\end{align*}
If $m\notin\mathcal M,$ then $\rho$ is not run, so $M_\rho$ is constant and
the left-hand side vanishes. Averaging over $(R,M),$ we obtain
\[
  \IC(\pi',\mu)
  =\IC(\pi,\mu)
   +\Exp\bigl[\1_{\{M\in\mathcal M\}}\IC(\rho,\mu_{R,M})\bigr]
  \leq\IC(\pi,\mu)+qK.
\]

The protocol $\pi'$ errs exactly when either (i) $M\notin\mathcal M$ and
$\pi$ errs, or (ii) $M\in\mathcal M$ and $\rho$ errs. Conditional on
$(X,Y)=(x,y)$ and $M=m\in\mathcal M,$ the protocol $\rho$ is run on the
legal input $(x,y)$ with fresh randomness, and therefore errs with
probability at most $\epsilon.$ Averaging over $(X,Y)$ and $M$ shows that
event (ii) has probability at most $q\epsilon,$ which proves the error
bound.
\end{proof}

We will also need information bounds for independent executions, both on a
common input and on different coordinates of a larger input.

\begin{lemma}[Independent executions]\label{lem:independent-executions}
Let $\pi_1,\ldots,\pi_t$ be protocols on a common input space
$\Xcal\times\Ycal,$ and let $\pi$ run them with independent randomness,
without using the private state or messages of one run in another. Then for
every $\mu\in\mathcal P(\Xcal\times\Ycal),$
\begin{equation}\label{eq:independent-common-input}
  \IC(\pi,\mu)\leq\sum_{i=1}^t\IC(\pi_i,\mu).
\end{equation}
Alternatively, suppose that $\pi_i$ runs only on the $i$th coordinate pair of
an input $(X,Y)\in\Xcal\times\Ycal,$ where $\Xcal=\prod_{i=1}^t\Xcal_i$ and
$\Ycal=\prod_{i=1}^t\Ycal_i.$ For an arbitrary
$\nu\in\mathcal P(\Xcal\times\Ycal),$ let
$\nu_i\in\mathcal P(\Xcal_i\times\Ycal_i)$ be the marginal law of
$(X_i,Y_i).$ In this case,
\begin{equation}\label{eq:independent-distinct-inputs}
  \IC(\pi,\nu)\leq\sum_{i=1}^t\IC(\pi_i,\nu_i).
\end{equation}
\end{lemma}

\begin{proof}
Write $R_i$ and $M_i$ for the public randomness and transcript of $\pi_i.$
Then $\pi$ has public randomness $R=(R_1,\ldots,R_t)$ and transcript
$M=(M_1,\ldots,M_t).$ Let $(X,Y)\sim\mu.$ For each $i\in[t],$
\begin{align*}
  I(M_i;X\mid Y,R,M_{<i})
    &=H(M_i\mid Y,R,M_{<i})-H(M_i\mid X,Y,R,M_{<i})\\
    &\leq H(M_i\mid Y,R_i)-H(M_i\mid X,Y,R_i)\\
    &=I(M_i;X\mid Y,R_i).
\end{align*}
The inequality holds because $H(M_i\mid Y,R,M_{<i})\leq H(M_i\mid Y,R_i),$
as conditioning does not increase entropy, and
$H(M_i\mid X,Y,R,M_{<i})=H(M_i\mid X,Y,R_i).$ This equality holds because,
given $(X,Y,R_i),$ the transcript $M_i$ is a function of the private
randomness of $\pi_i,$ which is independent of $(X,Y,R,M_{<i}).$
Summing over $i$ and using the chain rule, we obtain
$I(M;X\mid Y,R)=\sum_{i=1}^tI(M_i;X\mid Y,R,M_{<i})
\leq\sum_{i=1}^tI(M_i;X\mid Y,R_i).$
An analogous argument gives
$I(M;Y\mid X,R)\leq\sum_{i=1}^tI(M_i;Y\mid X,R_i).$
Summing these upper bounds yields
$\IC(\pi,\mu)\leq\sum_{i=1}^t\IC(\pi_i,\mu),$
which is~\eqref{eq:independent-common-input}.

For~\eqref{eq:independent-distinct-inputs}, let $(X,Y)\sim\nu.$ Since
$\pi_i$ runs only on $(X_i,Y_i),$ the transcript $M_i$ is a function of
$(X_i,Y_i,R_i)$ and the private randomness of $\pi_i,$ 
which is independent of $(X,Y,R,M_{<i}).$ The preceding
argument, with $(X_i,Y_i)$ in place of $(X,Y)$ in the last two lines of
the display, therefore gives
$I(M;X\mid Y,R)\leq\sum_{i=1}^tI(M_i;X_i\mid Y_i,R_i)$ and, analogously,
$I(M;Y\mid X,R)\leq\sum_{i=1}^tI(M_i;Y_i\mid X_i,R_i).$ Summing these
bounds yields~\eqref{eq:independent-distinct-inputs}, because
$(X_i,Y_i,R_i,M_i)$ has the joint distribution of the input, public
randomness, and transcript of $\pi_i$ when the input is drawn from
$\nu_i.$
\end{proof}

In particular, $t$ independent repetitions increase worst-case information
cost by at most a factor of $t.$ For a Boolean output, the majority of
the $t$ answers, with ties broken by a fixed rule, is determined by the
public randomness and transcripts of the repetitions, so outputting the
majority adds no information beyond that of the repetitions themselves. If each repetition has
worst-case error at most $p,$ the majority error is at most
\begin{equation}\label{eq:majority-error}
  \sum_{j\geq t/2}\binom tj p^j(1-p)^{t-j}.
\end{equation}
In particular, for $p\leq1/3,$ the Chernoff bound gives 
\begin{equation}\label{eq:majority-exponential}
  \Prob[\text{majority is incorrect}]
  \leq\exp\left(-\frac2t\left(\frac t6\right)^2\right)
  =\exp\left(-\frac t{18}\right).
\end{equation}

Independent executions on disjoint blocks of coordinates trade error for
the number of copies. Let $F\colon\Xcal\times\Ycal\to\{-1,1,*\}$ be a
communication problem, and let $\pi$ be a protocol for $F^{\oplus n}$ with
worst-case error at most $\epsilon.$ For an integer $r\geq1,$ run $\pi$ with
independent randomness on each of $r$ disjoint blocks of $n$ coordinate
pairs, and output the XOR of the $r$ answers. The result is a protocol for
$F^{\oplus rn}$ with worst-case error at most $r\epsilon,$ by the union bound.
Moreover,~\eqref{eq:independent-distinct-inputs} bounds its information cost
under every input distribution by $r\IC(\pi),$ even though the $r$ blocks of
the input need not be independent.

\subsection{Local variation and information}

The next elementary proposition distinguishes the communication problems
of zero information complexity from all others.

\begin{proposition}\label{prop:local-variation}
Let $F\colon\Xcal\times\Ycal\to\{-1,1,*\}$ be a communication problem. If
every row and every column has a constant value on its legal inputs, then
$\IC_0(F)=0.$ Otherwise, for every $0\leq\epsilon<1/2$ and every integer
$n\geq1,$
\begin{equation}\label{eq:local-variation-lower}
  \IC_\epsilon(F^n)\geq\bigl(1-h_2(\epsilon)\bigr)n.
\end{equation}
For a total function, the first case occurs if and only if $F$ is constant.
\end{proposition}

\begin{proof}
Suppose first that every row and every column is constant on its legal
inputs. Let $a(x)$ and $b(y)$ denote their respective values, with arbitrary
values assigned to empty rows and columns. Then
\begin{equation}\label{eq:locally-known}
  F(x,y)=a(x)=b(y),\qquad (x,y)\in\dom F.
\end{equation}
Alice announces $a(x),$ and both parties output the announced value. This
message is determined by each party's input on the promise domain.
Consequently, both mutual informations in the definition of
internal information cost are zero under every $\mu\in\mathcal P(\dom F).$
This proves $\IC_0(F)=0.$
If $F$ is total, \eqref{eq:locally-known} forces $F$ to be constant.

In the remaining case, a row or a column contains two legal inputs with
different function values. Interchanging the parties if necessary, there
are $x^*,y_0,y_1$ such that
\[
  (x^*,y_0),(x^*,y_1)\in\dom F,\qquad
  F(x^*,y_z)=(-1)^z\quad(z=0,1).
\]
In particular, $y_0\ne y_1.$ Let $\mu\in\mathcal P(\dom F)$ be the uniform
distribution on $\{(x^*,y_0),(x^*,y_1)\}.$ Fix $0\leq\epsilon<1/2,$ an
integer $n\geq1,$ and a protocol $\pi$ with $\err(\pi,F^n)\leq\epsilon.$
Since $\IC(\pi)\geq\IC(\pi,\mu^n),$ it suffices to show that
\begin{equation}\label{eq:local-variation-claim}
  \IC(\pi,\mu^n)\geq\bigl(1-h_2(\epsilon)\bigr)n;
\end{equation}
taking the infimum over $\pi$ then proves~\eqref{eq:local-variation-lower}.

Let $(X,Y)\sim\mu^n,$ and let $R$ and $M$ be the public randomness and
transcript of $\pi$ on input $(X,Y).$ Then $X=(x^*,\ldots,x^*)$ is
constant, and $Y=(y_{B_1},\ldots,y_{B_n})$ for independent uniform bits
$B_1,\ldots,B_n.$ Since $X$ is constant, $I(M;X\mid Y,R)=0$ and
$I(M;Y\mid X,R)=I(M;Y\mid R).$ Moreover, $Y$ and $B=(B_1,\ldots,B_n)$
determine each other, and $B$ is uniform on $\{0,1\}^n$ and independent
of $R.$ Hence
\begin{equation}\label{eq:local-variation-identity}
  \IC(\pi,\mu^n)=I(M;B\mid R)=H(B\mid R)-H(B\mid R,M)=n-H(B\mid R,M).
\end{equation}

To bound $H(B\mid R,M),$ fix $i\in[n].$ Write the $i$th answer output by
$\pi$ as $(-1)^{\widehat B_i},$ where $\widehat B_i\in\{0,1\};$ like the
output of $\pi,$ the bit $\widehat B_i$ is a function of $(R,M).$ The
correct $i$th answer is $F(X_i,Y_i)=(-1)^{B_i},$ so $\pi$ errs whenever
$B_i\oplus\widehat B_i=1.$ Hence $\Prob[B_i\oplus\widehat B_i=1]\leq\epsilon<1/2.$
Using that $\widehat B_i$ is a function of $(R,M),$ that conditioning
does not increase entropy, and that $h_2$ is increasing on $[0,1/2],$ we
obtain
\[
  H(B_i\mid R,M)=H(B_i\oplus\widehat B_i\mid R,M)
  \leq H(B_i\oplus\widehat B_i)
  =h_2\bigl(\Prob[B_i\oplus\widehat B_i=1]\bigr)\leq h_2(\epsilon).
\]
By the subadditivity of entropy, it follows that
$H(B\mid R,M)\leq\sum_{i=1}^nH(B_i\mid R,M)\leq h_2(\epsilon)n.$
Together with~\eqref{eq:local-variation-identity}, this
proves~\eqref{eq:local-variation-claim}.
\end{proof}

We now record a trivial lower bound for XOR powers. Its main use is to
ensure that no additive constants are lost. Specifically, suppose that a
row or a column of a communication problem
$F\colon\Xcal\times\Ycal\to\{-1,1,*\}$ contains two legal inputs with
different function values, and let $0\leq\epsilon<1/2.$ Then
Propositions~\ref{prop:local-variation} and~\ref{prop:restriction-floor}
give $\IC_\epsilon(F^{\oplus n})\geq1-h_2(\epsilon)$ for every $n\geq1.$
As a result, a lower bound $\IC_\epsilon(F^{\oplus n})\geq a-b$ with an
additive loss $b\geq0$ implies the purely multiplicative bound
$\IC_\epsilon(F^{\oplus n})\geq a/\bigl(1+b/(1-h_2(\epsilon))\bigr).$

\begin{proposition}\label{prop:restriction-floor}
Let $F\colon\Xcal\times\Ycal\to\{-1,1,*\}$ be a communication problem. Then
for every integer $n\geq1$ and every $\epsilon\geq0,$
\begin{equation}\label{eq:restriction-floor}
  \IC_\epsilon(F^{\oplus n})\geq\IC_\epsilon(F).
\end{equation}
\end{proposition}

\begin{proof}
Let $\pi$ be a protocol for $F^{\oplus n},$ and fix legal
inputs $(x_2,y_2),\ldots,(x_n,y_n)\in\dom F.$ Let $\pi'$ be the protocol
that, on input $(x,y),$ runs $\pi$ on
$((x,x_2,\ldots,x_n),(y,y_2,\ldots,y_n))$ and outputs the XOR of the
output of $\pi$ with $F(x_2,y_2),\ldots,F(x_n,y_n).$ For every legal
input $(x,y),$ the input to $\pi$ is legal, and $\pi'$ errs exactly when
$\pi$ does. Thus $\err(\pi',F)\leq\err(\pi,F^{\oplus n}).$ Since $\pi'$
has the same public randomness and transcript as $\pi,$ its information
cost under any $\mu\in\mathcal P(\dom F)$ equals that of $\pi$ under the
distribution of $((X,x_2,\ldots,x_n),(Y,y_2,\ldots,y_n))$ with
$(X,Y)\sim\mu,$ which lies in $\mathcal P(\dom F^{\oplus n}).$ Thus
$\IC(\pi')\leq\IC(\pi).$ If $\err(\pi,F^{\oplus n})\leq\epsilon,$ it follows
that $\IC_\epsilon(F)\leq\IC(\pi);$ taking the infimum over $\pi$
proves~\eqref{eq:restriction-floor}.
\end{proof}

Conversely, independent execution gives the following upper bounds.

\begin{proposition}\label{prop:general-upper}
Let $F\colon\Xcal\times\Ycal\to\{-1,1,*\}$ be a communication problem. Then
for every integer $n\geq1$ and every $\epsilon\geq0,$
\begin{equation}\label{eq:general-upper}
  \IC_\epsilon(F^{\oplus n})\leq\IC_\epsilon(F^n)
                         \leq n\IC_{\epsilon/n}(F).
\end{equation}
\end{proposition}

\begin{proof}
A protocol for $F^n$ can output the XOR of its $n$ answers without any
additional information. Whenever all $n$ answers are correct, their XOR
is correct. This proves the first inequality.

For the second, fix $\eta>0$ and choose a protocol $\pi$ for $F$ with
worst-case error at most $\epsilon/n$ and
\[
  \IC(\pi)\leq\IC_{\epsilon/n}(F)+\eta.
\]
Run independent copies of $\pi$ on the $n$ coordinate pairs and output
the tuple of answers. The union bound gives whole-vector error at most
$\epsilon$ on every legal input. For any joint distribution
$\nu\in\mathcal P(\dom F^n),$ let $\nu_i$ be the marginal law of
$(X_i,Y_i).$ Each marginal lies in $\mathcal P(\dom F).$
Lemma~\ref{lem:independent-executions} gives
\[
  \IC(\pi^{\otimes n},\nu)
    \leq\sum_{i=1}^n\IC(\pi,\nu_i)
    \leq n\bigl(\IC_{\epsilon/n}(F)+\eta\bigr).
\]
Taking the supremum over $\nu$ and letting $\eta\to0$ proves the
claim. In particular, no independence assumption is made on the input
coordinates in this upper bound.
\end{proof}

\subsection{Margins and discrepancy}\label{subsec:margin}

The \emph{margin} of a partial real matrix
$A\in(\mathbb R\cup\{*\})^{\Xcal\times\Ycal}$ is
\begin{equation}\label{eq:margin-definition}
  \margin(A)=\max_{u,v}\,\min_{(x,y)\in\dom A}A_{xy}\langle u_x,v_y\rangle,
\end{equation}
where the maximum ranges over all dimensions $d\geq1$ and all families
$u=(u_x)_{x\in\Xcal}$ and $v=(v_y)_{y\in\Ycal}$ of unit vectors in
$\mathbb R^d.$ Since such families span a subspace of dimension at most
$|\Xcal|+|\Ycal|,$ we may take $d=|\Xcal|+|\Ycal|,$ and the maximum is
then attained by compactness. Taking all $u_x$ equal to one unit vector
and all $v_y$ equal to an orthogonal one shows that $\margin(A)\geq0.$
The margin of a communication problem
$F\colon\Xcal\times\Ycal\to\{-1,1,*\}$ is that of its characteristic
matrix:
\begin{equation}\label{eq:margin-vector}
  \margin(F)=\margin(A_F)
  =\max_{u,v}\,\min_{(x,y)\in\dom F}F(x,y)\langle u_x,v_y\rangle.
\end{equation}
Clearly $\margin(F)\leq1.$ Moreover, $\margin(F)>0,$ as one sees by
taking $v_y$ to be the standard basis vectors of $\mathbb R^{\Ycal}$ and
$u_x$ to be the $x$th row of $A_F,$ with its $*$ entries replaced
by zeros, normalized to unit length (or any unit vector, if the row has
only $*$ entries). Margins of Euclidean embeddings were
studied, e.g., in~\cite{BES02,FSSS03}. The factorization norm formulation, which we use
in Appendix~\ref{app:norms}, was developed in~\cite{LMSS07,LS09a,LS09b}.

Margin is multiplicative under the following tensor product. For partial
real matrices $A\in(\mathbb R\cup\{*\})^{\Xcal\times\Ycal}$ and
$B\in(\mathbb R\cup\{*\})^{\Xcal'\times\Ycal'},$ the \emph{tensor
product} $A\otimes B$ has rows indexed by $\Xcal\times\Xcal',$ columns
indexed by $\Ycal\times\Ycal',$ domain
\[
  \dom(A\otimes B)
   =\{((x,x'),(y,y')):(x,y)\in\dom A,\ (x',y')\in\dom B\},
\]
and entries $(A\otimes B)_{(x,x'),(y,y')}=A_{xy}B_{x'y'}$ there. For real
matrices, this is the usual Kronecker product. For every communication
problem $F\colon\Xcal\times\Ycal\to\{-1,1,*\}$ and every integer
$n\geq1,$ the characteristic matrix of $F^{\oplus n}$ is the tensor power
$A_F^{\otimes n}.$

\begin{proposition}[Multiplicativity of margin]\label{prop:margin-tensor}
For all partial real matrices $A\in(\mathbb R\cup\{*\})^{\Xcal\times\Ycal}$
and $B\in(\mathbb R\cup\{*\})^{\Xcal'\times\Ycal'},$
\[
  \margin(A\otimes B)=\margin(A)\margin(B).
\]
In particular, $\margin(F^{\oplus n})=\margin(F)^n$ for every
communication problem $F\colon\Xcal\times\Ycal\to\{-1,1,*\}$ and every
integer $n\geq1.$
\end{proposition}

For total sign matrices, the multiplicativity of margin was
established by Lee, Shraibman, and \v{S}palek~\cite{LSS08}; see also Lee and
Mittal~\cite{LM08} and the survey~\cite{LS09c}. We provide a proof for partial
matrices in Appendix~\ref{app:norms}.

For a communication problem $F\colon\Xcal\times\Ycal\to\{-1,1,*\}$ and
$\mu\in\mathcal P(\dom F),$ the \emph{discrepancy} is
\begin{align*}
  \disc_\mu(F)&=\max_{\substack{S\subseteq\Xcal\\T\subseteq\Ycal}}
      \abs*{\sum_{(x,y)\in(S\times T)\cap\dom F}\mu(x,y)F(x,y)},\\[6pt]
  \disc(F)&=\min_{\mu\in\mathcal P(\dom F)}\disc_\mu(F).
\end{align*}
The relation between discrepancy and margin for full sign
matrices was formalized in~\cite{LS09a,LS09b}. We will need 
the following one-sided bound, whose short proof for partial
functions can be found in Appendix~\ref{app:norms}.

\begin{proposition}\label{prop:disc-margin}
For every communication problem $F\colon\Xcal\times\Ycal\to\{-1,1,*\},$
\[
  \disc(F)\leq\margin(F).
\]
\end{proposition}

We use the discrepancy lower bound for information
complexity, due to
Braverman and Weinstein~\cite{BW}.

\begin{fact}[Discrepancy bound]\label{fact:discrepancy}
There are absolute constants $c'>0$ and $c''\geq0$ such that, for every
communication problem $G\colon\Xcal\times\Ycal\to\{-1,1,*\}$ and every
$\mu\in\mathcal P(\dom G),$
\begin{equation}\label{eq:discrepancy-information}
  \IC_{1/3}(G)\geq c'\log\frac1{\disc_\mu(G)}-c''.
\end{equation}
\end{fact}

Braverman and Weinstein prove that every protocol computing a total
function $F$ with error at most $1/10$ under $\mu$ has information cost
$\Omega(\log(1/\disc_\mu(F)))$ under $\mu.$ Fact~\ref{fact:discrepancy}
is a weaker, worst-case form of their result needed in our paper. To
derive it, extend $G$ arbitrarily to a total function $G',$ which changes
neither $\disc_\mu(G)$ nor the error and information cost of any
protocol under $\mu,$ to obtain
\[
\IC_{1/10}(G)\geq
\IC_{1/10}(G,\mu)=
\IC_{1/10}(G',\mu)=\Omega\Bigl(\log\frac1{\disc_\mu(G)}\Bigr).
\]
Finally, observe that $\IC_{1/10}(G)=\Theta(\IC_{1/3}(G)),$ as one
can reduce the error from $1/3$ to below $1/10$ by a
majority vote over constantly many independent runs, which
multiplies the information cost by a constant
(Lemma~\ref{lem:independent-executions}).

The following consequence is an information lower bound in terms
of margin.

\begin{theorem}\label{thm:margin-ic}
There is an absolute constant $c>0$ such that, for every communication
problem $F\colon\Xcal\times\Ycal\to\{-1,1,*\}$ and every integer
$n\geq1,$
\begin{equation}\label{eq:margin-ic}
  \IC_{1/3}(F^{\oplus n})\geq cn\log\frac1{\margin(F)}.
\end{equation}
\end{theorem}

\begin{proof}
Put $I=\IC_{1/3}(F^{\oplus n}).$ If $\IC_{1/3}(F)=0,$ then by
Proposition~\ref{prop:local-variation} every row of $F$ is constant on
its legal inputs, say $F(x,y)=a(x)$ for $(x,y)\in\dom F.$ Taking
$u_x=a(x)e$ and $v_y=e$ in~\eqref{eq:margin-vector}, for a fixed unit
vector $e,$ shows that $\margin(F)=1,$ and there is nothing to prove.
Suppose now that $\IC_{1/3}(F)>0.$ By Propositions~\ref{prop:disc-margin}
and~\ref{prop:margin-tensor},
\[
  \disc(F^{\oplus n})\leq\margin(F^{\oplus n})=\margin(F)^n.
\]
Hence
\begin{align*}
  I&\geq c'n\log\frac1{\margin(F)}-c'',\\
  I&\geq1-h_2\Bigl(\frac13\Bigr),
\end{align*}
where the first inequality is the result of applying
Fact~\ref{fact:discrepancy} to $G=F^{\oplus n}$ and a
distribution $\mu$ attaining $\disc(F^{\oplus n}),$ and
the second is justified by Propositions~\ref{prop:local-variation}
and~\ref{prop:restriction-floor}.
The claimed lower bound on $I$ now follows by
taking a weighted arithmetic mean of these inequalities.
\end{proof}

\section{Total functions}\label{sec:total}

This section proves our main results for total functions.
Here, our key technical contribution is
an XOR lemma with constant additive loss, Theorem~\ref{thm:additive-conversion}.
We prove it in Sections~\ref{subsec:error-preserving}--\ref{subsec:conversion}, on
arbitrary promise domains, so that Section~\ref{sec:partial} can use
it without repeating the argument. Section~\ref{subsec:total-proofs}
then completes the proofs of the direct sum theorem and
XOR lemma for total functions. 

In outline, the proof of Theorem~\ref{thm:additive-conversion} starts
with a protocol for an XOR of $n$ instances of $F$ and
splits it into protocols on half as many instances. On
transcripts with a large posterior bias, a player announces
the more likely answer. On the other transcripts, the
players run a constant-error fallback protocol. A bound on
the fallback under every input distribution permits its cost
to be charged under any posterior distribution. A careful
analysis reveals that the sum of these charges over the
recursion is geometric. All of these steps work with constant
error. Independent executions on disjoint blocks of coordinates then
extend the result to every error $\epsilon\in(0,1/3],$ and a direct
embedding argument handles $\epsilon=0.$ 

\subsection{The standard embedding and fallbacks}\label{subsec:error-preserving}

The following classic construction~\cite{BYJKS04,BBCR10,SY} embeds a single input pair into one coordinate
of a protocol on several coordinates.

\begin{definition}[Standard embedding]\label{def:standard-embedding}
Let $\Dcal\subseteq\Xcal\times\Ycal,$ let $\nu\in\mathcal P(\Dcal),$ and let
$t\geq1$ be an integer. Let $\pi$ be a protocol with input
$(X,Y)=((X_1,\ldots,X_t),(Y_1,\ldots,Y_t))\in\Xcal^t\times\Ycal^t$ and public
randomness $R.$ For $j\in[t],$ the \emph{standard embedding into
coordinate~$j$ of~$\pi$} with respect to $\nu$ is the protocol $\pi^{(j)}$
that operates as follows on input $(x,y)\in\Xcal\times\Ycal.$
\begin{enumerate}[label=(\alph*)]
\item The parties publicly sample $(X_{<j},Y_{>j})$ from its distribution
under $\nu^t,$ and set $(X_j,Y_j)=(x,y).$
\item Alice privately samples $X_{>j}$ from its conditional distribution
under $\nu^t$ given $Y_{>j}.$ Bob privately samples $Y_{<j}$ from its
conditional distribution under $\nu^t$ given $X_{<j}.$
\item The parties run $\pi$ on $(X,Y),$ and each party outputs what 
he or she outputs in this run of $\pi.$
\end{enumerate}
The public randomness of $\pi^{(j)}$ is $(R,X_{<j},Y_{>j}),$ and its
transcript is the transcript of $\pi.$
By construction, if $(x,y)\sim\nu,$ then the completed input
$(X,Y)$ is supported on $\Dcal^t$ and has law $\nu^t.$
\end{definition}

The information cost of $\pi$ under $\nu^t$ splits exactly among its $t$
standard embeddings, as the following identity shows. This identity is
folklore: the embedding argument goes back to Bar-Yossef, Jayram, Kumar,
and Sivakumar~\cite{BYJKS04}, and its form for internal information cost
was made explicit by Barak, Braverman, Chen, and Rao~\cite{BBCR10}; see
also~\cite[Sec.~2.1]{SY}. We include the short proof for completeness.

\begin{fact}[Information cost of the standard embeddings]\label{fact:embedding-cost}
Let $\Dcal\subseteq\Xcal\times\Ycal,$ let $\nu\in\mathcal P(\Dcal),$ and let
$t\geq1$ be an integer. For every protocol $\pi$ with input in
$\Xcal^t\times\Ycal^t,$ the standard embeddings $\pi^{(1)},\ldots,\pi^{(t)}$
into $\pi$ with respect to $\nu$ satisfy
\begin{equation}\label{eq:embedding-cost}
  \sum_{j=1}^t\IC(\pi^{(j)},\nu)=\IC(\pi,\nu^t).
\end{equation}
\end{fact}

\begin{proof}
Let $(X,Y)\sim\nu^t,$ and let $R$ and $M$ be the public randomness and
transcript of $\pi$ on input $(X,Y).$ Fix $j\in[t].$ By construction, an
execution of $\pi^{(j)}$ on an input drawn from $\nu$ induces this same joint
law of $(X,Y,R,M);$ the input of $\pi^{(j)}$ is $(X_j,Y_j),$ and its public
randomness is $(R,X_{<j},Y_{>j}).$ Hence
\begin{align}
  \IC(\pi^{(j)},\nu)
  =I(M;X_j\mid X_{<j},Y_{\geq j},R)+I(M;Y_j\mid X_{\leq j},Y_{>j},R).
  \label{eq:pi-nu}
\end{align}
We claim that $X_{\geq j}$ and $Y_{<j}$ are independent conditional on
$(X_{<j},Y_{\geq j},R),$ and also conditional on
$(X_{<j},Y_{\geq j},R,M).$ The former holds because the coordinate pairs
$(X_i,Y_i)$ are independent and $R$ is independent of the input. For the
latter, condition on values $x_{<j},y_{\geq j},r,m$ for which the posterior
is defined. By~\eqref{eq:rectangle}, the conditional probability of
$(X_{\geq j},Y_{<j})=(x_{\geq j},y_{<j})$ is proportional to
\[
  \prod_{i\geq j}\nu(x_i,y_i)\,a_{r,m}(x)
  \cdot\prod_{i<j}\nu(x_i,y_i)\,b_{r,m}(y),
\]
where $x=(x_{<j},x_{\geq j})$ and $y=(y_{<j},y_{\geq j}).$ The first factor
depends only on $x_{\geq j},$ and the second only on $y_{<j},$ which proves
the claim. Hence, conditioning on $Y_{<j}$ changes neither
$H(X_j\mid X_{<j},Y_{\geq j},R)$ nor $H(X_j\mid X_{<j},Y_{\geq j},R,M),$
thereby forcing $I(M;X_j\mid X_{<j},Y_{\geq j},R) =I(M;X_j\mid X_{<j},Y,R).$
Likewise, for $j<t,$ the claim with $j+1$ in place of $j$ shows that
conditioning on $X_{>j}$ changes neither $H(Y_j\mid X_{\leq j},Y_{>j},R)$
nor $H(Y_j\mid X_{\leq j},Y_{>j},R,M),$ which results in 
$I(M;Y_j\mid X_{\leq j},Y_{>j},R)= I(M;Y_j\mid X,Y_{>j},R).$
Making these substitutions in~\eqref{eq:pi-nu} yields
\[
  \IC(\pi^{(j)},\nu)=I(M;X_j\mid X_{<j},Y,R)+I(M;Y_j\mid X,Y_{>j},R).
\]
Summing over $j$ and applying the chain rule, in forward order for $X$ and in
reverse order for $Y,$ we obtain
\[
  \sum_{j=1}^t\IC(\pi^{(j)},\nu)=I(M;X\mid Y,R)+I(M;Y\mid X,R)
  =\IC(\pi,\nu^t).\qedhere
\]
\end{proof}

For an XOR, the standard embeddings yield protocols for a single copy
with the same error, at the price of two additional one-bit messages
that reveal the XOR of the function values on the other coordinates. This
fact is also folklore: Sawettamalya and Yu~\cite[Thm.~6]{SY} state it
informally, with an additive term $O(1)$ in place of $2,$ and attribute
it to~\cite{BBCR10}. We include the proof for completeness.

\begin{fact}[Error-preserving embedding]\label{fact:ordinary-embedding}
Let $G\colon\Xcal\times\Ycal\to\{-1,1,*\}$ be a communication problem. Let
$\nu\in\mathcal P(\dom G),$ let $t\geq1$ be an integer, and let $\pi$ be a
protocol with
$\err_{\nu^t}(\pi,G^{\oplus t})\leq\epsilon$ and $\IC(\pi,\nu^t)\leq I.$
Then
\[
  \IC_\epsilon(G,\nu)\leq\frac It+2.
\]
\end{fact}

\begin{proof}
Let $\pi^{(1)},\ldots,\pi^{(t)}$ be the standard embeddings into $\pi$ with
respect to $\nu,$ and let $\sigma$ be the protocol that publicly chooses a
uniformly random $j\in[t]$ and runs $\pi^{(j)}.$
By~\eqref{eq:public-mixture-cost} and Fact~\ref{fact:embedding-cost},
\[
  \IC(\sigma,\nu)=\frac1t\sum_{j=1}^t\IC(\pi^{(j)},\nu)
  =\frac{\IC(\pi,\nu^t)}t\leq\frac It.
\]
We extend $\sigma$ to a protocol $\sigma'$ for $G$ as follows. Let $(X,Y)$
be the completed input on which $\pi$ is run inside $\pi^{(j)}.$ After
$\sigma$ ends, Alice sends $\prod_{i>j}G(X_i,Y_i),$ which she can compute
from $X_{>j}$ and the publicly sampled $Y_{>j},$ and Bob sends
$\prod_{i<j}G(X_i,Y_i),$ which he can compute from $Y_{<j}$ and the publicly
sampled $X_{<j}.$ Both parties then output the XOR of these two signs and
the output of $\sigma.$ The two additional messages increase the
information cost by at most two, so that $\IC(\sigma',\nu)\leq I/t+2.$

By Definition~\ref{def:standard-embedding}, the output of $\sigma$ is the
output of $\pi$ on the completed input $(X,Y),$ and $(X_j,Y_j)=(x,y)$ is the
actual input. Since $G^{\oplus t}(X,Y)=G(x,y)\prod_{i\ne j}G(X_i,Y_i),$ the
protocol $\sigma'$ outputs $G(x,y)$ exactly when $\pi$ outputs
$G^{\oplus t}(X,Y).$ If $(x,y)\sim\nu,$ then $(X,Y)\sim\nu^t,$ and therefore
\[
  \err_\nu(\sigma',G)=\err_{\nu^t}(\pi,G^{\oplus t})\leq\epsilon.
\]
Hence $\IC_\epsilon(G,\nu)\leq\IC(\sigma',\nu)\leq I/t+2.$
\end{proof}

Taking the supremum over input distributions and applying the minimax
theorem (Fact~\ref{fact:minimax}), we obtain protocols for smaller XORs whose information cost is
bounded under every input distribution. Moreover, this bound decreases in
proportion to the number of copies. Thus, passing from an XOR of $N$ copies to an
XOR of $k$ copies multiplies the worst-case information cost by $2k/N,$ up
to an additive constant.

\begin{corollary}\label{cor:fallback}
Let $F\colon\Xcal\times\Ycal\to\{-1,1,*\}$ be a communication problem, and
let $k,N\geq1$ be integers such that $k$ divides $N.$ Then
\[
  \IC_{1/5}(F^{\oplus k})
  \leq\frac{2k}{N}\IC_{1/10}(F^{\oplus N})+4.
\]
\end{corollary}

\begin{proof}
Put $G=F^{\oplus k}$ and $t=N/k,$ so that $G^{\oplus t}=F^{\oplus N}.$
Let $\pi$ be any protocol for $F^{\oplus N}$ with worst-case error at most
$1/10,$ and put $I=\IC(\pi).$ For every $\nu\in\mathcal P(\dom G),$
Fact~\ref{fact:ordinary-embedding} gives
\[
  \IC_{1/10}(G,\nu)\leq\frac It+2.
\]
Here $\nu$ is an arbitrary distribution on the legal pairs of $k$-coordinate
blocks; its coordinates need not be independent. Taking the supremum over
$\nu$ and applying Fact~\ref{fact:minimax} yields
\[
  \IC_{1/5}(G)\leq\frac{2I}{t}+4.
\]
Taking the infimum over $\pi$ completes the proof.
\end{proof}

\subsection{The error reduction step}\label{subsec:error-reduction}

Our error reduction step splits a
protocol for an XOR of $2k$ copies into protocols for XORs of $k$ copies,
using the standard embeddings with $t=2.$ In this case we index the two
blocks by $0$ and $1$ rather than by $1$ and $2.$ Thus, for a set
$\Dcal\subseteq\Xcal\times\Ycal,$ a distribution $\nu\in\mathcal P(\Dcal),$
and a protocol $\pi$ with input
$((X_0,X_1),(Y_0,Y_1))\in\Xcal^2\times\Ycal^2,$ the standard embeddings
$\pi^{(0)}$ and $\pi^{(1)}$ into $\pi$ with respect to $\nu$ operate as
follows. In $\pi^{(0)},$ the actual input is placed in block $0,$ the parties
publicly sample $Y_1,$ and Alice privately samples $X_1$ given $Y_1.$ In
$\pi^{(1)},$ the actual input is placed in block $1,$ the parties publicly
sample $X_0,$ and Bob privately samples $Y_0$ given $X_0.$ By
Fact~\ref{fact:embedding-cost},
\begin{equation}\label{eq:binary-cost}
  \IC(\pi^{(0)},\nu)+\IC(\pi^{(1)},\nu)=\IC(\pi,\nu^2).
\end{equation}
Neither the embeddings nor the following lemma requires the promise domain
$\Dcal$ to be a rectangle. In particular, no function value on a crossed
pair such as $(X_0,Y_1)$ is used.

\begin{lemma}\label{lem:cross-independence}
Let $\Dcal\subseteq\Xcal\times\Ycal$ and $\nu\in\mathcal P(\Dcal).$ Let $\pi$
be a protocol with input $((X_0,X_1),(Y_0,Y_1))\in\Xcal^2\times\Ycal^2,$
where $(X_0,Y_0)$ and $(X_1,Y_1)$ are independent with common law $\nu,$ and
let $R$ and $M$ be the public randomness and transcript of $\pi.$ Then
\begin{equation}\label{eq:cross-after}
  X_1\perp Y_0\mid X_0,Y_1,R,M.
\end{equation}
\end{lemma}

\begin{proof}
Condition on values $x_0,y_1,r,m$ for which the posterior is defined.
By~\eqref{eq:rectangle}, the conditional probability of $(X_1,Y_0)=(x_1,y_0)$
is proportional to
\[
  \nu(x_1,y_1)a_{r,m}(x_0,x_1)
  \cdot\nu(x_0,y_0)b_{r,m}(y_0,y_1).
\]
The first factor depends only on $x_1,$ and the second only on $y_0,$
which proves~\eqref{eq:cross-after}.
\end{proof}

The next lemma splits a protocol for $F^{\oplus2k}$ into protocols
$\pi_0$ and $\pi_1$ for $F^{\oplus k}$ whose errors sum to at most $4/3$
times the original error and whose information costs sum to at most the
original cost plus an additive term. Our proof builds on the construction 
of Sawettamalya and Yu~\cite[Sec.~7.1]{SY}. In a critical departure from
their approach, which conditions on the transcripts with sufficiently
large posterior bias, we retain every transcript and run an additional
protocol on the transcripts with low posterior bias. We refer to this
additional protocol as the \emph{fallback protocol}; its information cost
will be controlled uniformly over all input distributions by
Corollary~\ref{cor:fallback}. Since no transcript is discarded, both
$\pi_0$ and $\pi_1$ retain the original product input distribution.

\begin{lemma}\label{lem:one-step}
Let $F\colon\Xcal\times\Ycal\to\{-1,1,*\}$ be a communication problem. Let
$k\geq1$ be an integer and $\mu\in\mathcal P(\dom F).$ Let $\pi$ be a
protocol for $F^{\oplus2k}$ with
\[
  \epsilon=\err_{\mu^{2k}}(\pi,F^{\oplus2k}),
  \qquad I=\IC(\pi,\mu^{2k}).
\]
Suppose that a protocol $\rho$ for $F^{\oplus k}$ satisfies
$\err(\rho,F^{\oplus k})\leq1/5$ and $\IC(\rho)\leq K.$ Then there are
protocols $\pi_0,\pi_1$ for $F^{\oplus k}$ such that, writing
\[
  \epsilon_j=\err_{\mu^k}(\pi_j,F^{\oplus k}),
  \qquad I_j=\IC(\pi_j,\mu^k),
\]
one has
\begin{align}
  \epsilon_0+\epsilon_1&\leq\frac43\epsilon,\label{eq:one-step-error}\\
  I_0+I_1&\leq I+4+\frac{16}3\epsilon K.\label{eq:one-step-cost}
\end{align}
\end{lemma}

\begin{proof}
Regard the input of $\pi$ as a pair of $k$-coordinate blocks, and let
$\pi^{(0)}$ and $\pi^{(1)}$ be the standard embeddings into $\pi$ with
respect to $\mu^k$ (Definition~\ref{def:standard-embedding}), with the
blocks indexed by $0$ and $1.$ Let
$(X,Y)=((X_0,X_1),(Y_0,Y_1))\sim\mu^{2k},$ so that the blocks $(X_0,Y_0)$
and $(X_1,Y_1)$ are independent with common law $\mu^k,$ and let $R$ and
$M$ be the public randomness and transcript of $\pi$ on input $(X,Y).$ Put
$V=(X_0,Y_1,R,M)$ and, for $j\in\{0,1\},$
\[
  B_j=F^{\oplus k}(X_j,Y_j),
  \qquad \Gamma_j=\abs{\Exp[B_j\mid V]}.
\]
Then $B_0B_1=F^{\oplus2k}(X,Y),$ so $\pi$ errs exactly when its output
differs from $B_0B_1.$

We first relate $\Gamma_0\Gamma_1$ to the error of $\pi.$
Lemma~\ref{lem:cross-independence}, applied with $\Xcal^k,$ $\Ycal^k,$
$(\dom F)^k,$ $\mu^k$ in place of $\Xcal,$ $\Ycal,$ $\Dcal,$ $\nu,$ gives
$X_1\perp Y_0\mid V.$ Since $X_0$ and $Y_1$ are components of $V,$ the sign
$B_0$ is a function of $(V,Y_0),$ and the sign $B_1$ is a function of
$(V,X_1).$ Hence $B_0\perp B_1\mid V,$ and therefore
\begin{equation}\label{eq:bias-product}
  \abs{\Exp[B_0B_1\mid V]}=\abs{\Exp[B_0\mid V]\Exp[B_1\mid V]}
  =\Gamma_0\Gamma_1.
\end{equation}
The output of $\pi$ is a function of $(R,M),$ and hence of $V.$
By~\eqref{eq:prediction-error} and~\eqref{eq:bias-product}, conditional on
$V,$ this output differs from $B_0B_1$ with probability at least
$(1-\Gamma_0\Gamma_1)/2.$ Averaging over $V,$ we obtain
\begin{equation}\label{eq:bias-error}
  \Exp[1-\Gamma_0\Gamma_1]\leq2\epsilon.
\end{equation}

We now construct $\pi_0$ and $\pi_1.$ Let $G$ be the event that
$\Gamma_0\Gamma_1\geq1/4.$ We call $G$ the
\emph{good event}: by~\eqref{eq:bias-product}, the posterior bias
$\abs{\Exp[B_0B_1\mid V]}$ is at least $1/4$ on $G.$ For $j\in\{0,1\},$ let
$\widehat B_j$ be the more likely value of $B_j$ given $V.$ Both $\1_G$ and
$\widehat B_j$ are functions of $V.$ In $\pi^{(0)},$ Alice knows $V,$
because $X_0$ is her input, $Y_1$ is publicly sampled, and $R$ and $M$ are
the public randomness and transcript of the run of $\pi.$ In $\pi^{(1)},$
Bob knows $V,$ because $Y_1$ is his input and $X_0$ is publicly sampled. For
$j\in\{0,1\},$ let $\sigma_j$ be the protocol that runs $\pi^{(j)},$ after
which the party that knows $V$ sends the bit $\1_G$ and the sign
$\widehat B_j;$ the output of $\sigma_j$ is $\widehat B_j.$ Let $\pi_j$ be
the protocol that runs $\sigma_j$ and then, if $\1_G=0,$ runs $\rho$ on the
same input with fresh public and private randomness. The output of $\pi_j$
is that of $\rho$ if $\rho$ is run, and $\widehat B_j$ otherwise.

By construction, $\pi_j$ is the protocol $\pi'$ of
Proposition~\ref{prop:continuation} for the function $F^{\oplus k},$ the
protocols $\sigma_j$ and $\rho,$ and the set of transcripts of $\sigma_j$ in
which $\1_G=0.$ When the input of
$\sigma_j$ is drawn from $\mu^k,$ the completed input on which $\pi$ is run
inside $\sigma_j$ has law $\mu^{2k}$ (Definition~\ref{def:standard-embedding}),
so this completed input, together with the public randomness and transcript
of that run of $\pi,$ has the joint law of $(X,Y,R,M).$ In particular, $\rho$
is run with probability $\Prob[\overline G].$ Moreover, the input of $\sigma_j$ is
$(X_j,Y_j),$ so $\sigma_j$ errs exactly when $\widehat B_j\ne B_j.$ Since
$\err(\rho,F^{\oplus k})\leq1/5$ and $\IC(\rho,\nu)\leq K$ for every
$\nu\in\mathcal P(\dom F^{\oplus k}),$ Proposition~\ref{prop:continuation}
gives
\begin{align}
  \epsilon_j&\leq\Prob[G,\ \widehat B_j\ne B_j]+\frac15\Prob[\overline G],
    \label{eq:child-error}\\
  I_j&\leq\IC(\sigma_j,\mu^k)+\Prob[\overline G]K.
    \label{eq:child-cost}
\end{align}

We now prove~\eqref{eq:one-step-error}. By~\eqref{eq:prediction-error},
conditional on $V,$ the more likely value $\widehat B_j$ differs from $B_j$
with probability $(1-\Gamma_j)/2.$ Since $G$ is determined by $V,$ it
follows that $\Prob[G,\ \widehat B_j\ne B_j]=\Exp[(1-\Gamma_j)\1_G]/2.$
Summing~\eqref{eq:child-error} over $j\in\{0,1\},$ we obtain
\begin{equation}\label{eq:error-split}
  \epsilon_0+\epsilon_1
  \leq\Exp\Bigl[\Bigl(1-\frac{\Gamma_0+\Gamma_1}2\Bigr)\1_G
                +\frac25\1_{\overline G}\Bigr].
\end{equation}
We bound the quantity inside this expectation by
$(2/3)(1-\Gamma_0\Gamma_1),$ separately on $G$ and on $\overline G.$ On $G,$ the
product $\Gamma_0\Gamma_1$ lies in $[1/4,1].$ The square root function is
concave, so on $[1/4,1]$ it lies above its chord from $(1/4,1/2)$ to
$(1,1);$ that is, $\sqrt z\geq(2z+1)/3$ for $1/4\leq z\leq1.$ Together with
the inequality of arithmetic and geometric means, this gives
\[
  \Gamma_0+\Gamma_1\geq2\sqrt{\Gamma_0\Gamma_1}
  \geq\frac43\Gamma_0\Gamma_1+\frac23
  \qquad\text{on }G,
\]
which rearranges to
$1-(\Gamma_0+\Gamma_1)/2\leq(2/3)(1-\Gamma_0\Gamma_1).$ On $\overline G,$ we have
$1-\Gamma_0\Gamma_1>3/4,$ and hence $2/5<(2/3)(1-\Gamma_0\Gamma_1).$
Using these bounds in~\eqref{eq:error-split} and then
applying~\eqref{eq:bias-error}, we obtain
\[
  \epsilon_0+\epsilon_1\leq\frac23\Exp[1-\Gamma_0\Gamma_1]
  \leq\frac43\epsilon,
\]
which is~\eqref{eq:one-step-error}.

Finally, we prove~\eqref{eq:one-step-cost}. Sending an additional bit
increases information cost by at most one. As $\sigma_j$ appends the two
one-bit messages $\1_G$ and $\widehat B_j$ to $\pi^{(j)},$ it follows that
$\IC(\sigma_j,\mu^k)\leq\IC(\pi^{(j)},\mu^k)+2.$ Since
$1-\Gamma_0\Gamma_1>3/4$ on $\overline G,$ inequality~\eqref{eq:bias-error} also
gives $(3/4)\Prob[\overline G]\leq\Exp[1-\Gamma_0\Gamma_1]\leq2\epsilon,$ that is,
$\Prob[\overline G]\leq8\epsilon/3.$ Summing~\eqref{eq:child-cost} over
$j\in\{0,1\}$ and applying~\eqref{eq:binary-cost} with $\nu=\mu^k,$ we
conclude that
\begin{align*}
  I_0+I_1&\leq\IC(\pi^{(0)},\mu^k)+\IC(\pi^{(1)},\mu^k)+4+2\Prob[\overline G]K\\
  &\leq I+4+\frac{16}3\epsilon K.\qedhere
\end{align*}
\end{proof}

\subsection{Recursive decomposition}

We use Lemma~\ref{lem:one-step} recursively, with the fallback protocols at
all scales supplied by the original XOR protocol. The following statement is
distributional in its conclusion but uses worst-case information complexity
on its right-hand side.

\begin{lemma}\label{lem:recursive-bound}
Let $F\colon\Xcal\times\Ycal\to\{-1,1,*\}$ be a communication problem, and
let $N\geq1$ be a power of two. Then, with $\alpha=\log(3/2)=0.5849\ldots,$
every input distribution $\mu$ satisfies
\begin{equation}\label{eq:recursive-bound}
  \IC_{N^{-\alpha}/10}(F,\mu)
  \leq\frac{13}{5N}\IC_{1/10}(F^{\oplus N})+4+\frac{32}{5}N^{-\alpha}.
\end{equation}
\end{lemma}

\begin{proof}
If $N=1,$ then~\eqref{eq:recursive-bound} holds because
$\IC_{1/10}(F,\mu)\leq\IC_{1/10}(F).$ Assume therefore that $N\geq2,$ and
let $d=\log N,$ so that $N=2^d$ and $d\geq1.$ Fix a protocol $\pi$ for
$F^{\oplus N}$ with worst-case error at most $1/10,$ and write
$I=\IC(\pi).$ Fix $\eta>0.$ By Corollary~\ref{cor:fallback}, for each
$k\in\{1,2,4,\ldots,N/2\}$ there is a protocol $\rho_k$ for
$F^{\oplus k}$ with worst-case error at most $1/5$ and
\begin{equation}\label{eq:fallback-scale}
  \IC(\rho_k)\leq\frac{2k}{N}I+4+\eta.
\end{equation}
Recall that we refer to the protocol $\rho$ in Lemma~\ref{lem:one-step} as
a fallback protocol; each $\rho_k$ will play this role. As usual, the slack
$\eta$ is needed because the infimum defining information complexity may
not be attained.

We now apply Lemma~\ref{lem:one-step} recursively to build a complete binary
tree of protocols of depth $d$ with root $\pi.$ For $1\leq i\leq d,$ the
nodes at depth $i$ are indexed by the strings $v\in\{0,1\}^i,$ and the
protocol at node $v$ is denoted $\pi_v.$ Lemma~\ref{lem:one-step}, applied
to $\pi$ with the fallback protocol $\rho_{N/2},$ gives the protocols
$\pi_0$ and $\pi_1.$ For $1\leq i<d$ and $v\in\{0,1\}^i,$ the same lemma,
applied to $\pi_v$ with the fallback protocol $\rho_{N/2^{i+1}},$ gives the
protocols $\pi_{v0}$ and $\pi_{v1}.$ All these applications use the
same distribution $\mu.$ By induction, $\pi_v$ is a protocol for
$F^{\oplus N/2^i}$ for every $v\in\{0,1\}^i.$

Let $E_0$ and $L_0$ be the error and information cost of $\pi$ under
$\mu^N,$ so that $E_0\leq1/10$ and $L_0\leq I.$ For $1\leq i\leq d$ and
$v\in\{0,1\}^i,$ let $\epsilon_v$ and $I_v$ be the error and information
cost of $\pi_v$ under $\mu^{N/2^i},$ and define
\[
  E_i=\sum_{v\in\{0,1\}^i}\epsilon_v,
  \qquad L_i=\sum_{v\in\{0,1\}^i}I_v.
\]
Summing~\eqref{eq:one-step-error} over the nodes at depth $i$ gives
$E_{i+1}\leq(4/3)E_i$ for $0\leq i<d,$ and therefore
\begin{equation}\label{eq:level-errors}
  E_i\leq\frac1{10}\left(\frac43\right)^i,
  \qquad i=0,1,\ldots,d.
\end{equation}
Only this sum of errors is used; an individual node is not required to have
small error.

The fallback protocol used at depth $i$ is $\rho_{N/2^{i+1}},$ and
\eqref{eq:fallback-scale} bounds its information cost by $I/2^i+4+\eta.$
Therefore, summing~\eqref{eq:one-step-cost} over the nodes at depth $i$ and
using~\eqref{eq:level-errors}, we obtain
\begin{align}
  L_{i+1}
    &\leq L_i+4\cdot2^i
      +\frac{16}3\left(\frac I{2^i}+4+\eta\right)E_i\nonumber\\
    &\leq L_i+4\cdot2^i
      +\frac8{15}\left(\frac I{2^i}+4+\eta\right)
                    \left(\frac43\right)^i.
      \label{eq:level-information}
\end{align}
Summing over $i=0,1,\ldots,d-1$ yields
\begin{align}
  L_d
    &\leq I+4(N-1)
      +\frac8{15}I\sum_{i=0}^{d-1}\left(\frac23\right)^i
      +\frac8{15}(4+\eta)\sum_{i=0}^{d-1}\left(\frac43\right)^i
      \nonumber\\
    &\leq\frac{13}{5}I+4(N-1)
      +\frac85(4+\eta)\left(\left(\frac43\right)^d-1\right).
      \label{eq:leaf-information}
\end{align}
The coefficient of $I$ is bounded independently of $d$ because its
contribution at depth $i$ is proportional to $(2/3)^i.$

The leaves of our constructed tree are the $N$ protocols $\pi_v$ with $v\in\{0,1\}^d.$
They are protocols for $F,$ with errors $\epsilon_v$ and information costs
$I_v$ under $\mu.$ The final protocol publicly chooses a uniformly random
$v\in\{0,1\}^d$ and runs $\pi_v.$ Since $(2/3)^d=N^{-\alpha},$
equations~\eqref{eq:public-mixture-cost}, \eqref{eq:public-mixture-error},
\eqref{eq:level-errors}, and~\eqref{eq:leaf-information} imply that this
protocol has error at most
\[
  \frac{E_d}{N}\leq\frac1{10}N^{-\alpha}
\]
and information cost at most
\[
  \frac{L_d}{N}
  \leq\frac{13}{5}\frac I N+4+\frac85(4+\eta)N^{-\alpha}.
\]
Letting $\eta\to0$ and then taking the infimum over the initial
protocol $\pi$ proves~\eqref{eq:recursive-bound}.
\end{proof}

\subsection{Completing the reduction}\label{subsec:conversion}

We now convert a protocol for $F^{\oplus n}$ with
constant error into a protocol for $F$ with error $1/(6n).$ The conversion
applies to arbitrary communication problems and has no dependence on the
input domain.

\begin{lemma}\label{lem:xor-conversion}
Let $F\colon\Xcal\times\Ycal\to\{-1,1,*\}$ be a communication problem. Then
for every integer $n\geq1,$
\begin{align}
  \IC_{1/(6n)}(F)
  &\leq\frac{32}{n}\IC_{1/10}(F^{\oplus n})+64,\label{eq:xor-conversion}\\
  \IC_{1/(6n)}(F)
  &\leq\frac{480}{n}\IC_{1/3}(F^{\oplus n})+64.\label{eq:xor-conversion-third}
\end{align}
\end{lemma}

\begin{proof}
Let $\pi$ be any protocol for $F^{\oplus n}$ with worst-case error at most
$1/10,$ and put $I=\IC(\pi).$ Let $N$ be the largest power of two that does
not exceed $n,$ so that $N>n/2.$ Fix legal input pairs
$(x_i,y_i)\in\dom F$ for $N<i\leq n.$ Let $\pi'$ be the protocol for
$F^{\oplus N}$ that runs $\pi$ on its input extended by these pairs and
outputs the XOR of the output of $\pi$ with
$\prod_{i=N+1}^nF(x_i,y_i).$ As in the proof of
Proposition~\ref{prop:restriction-floor}, the protocol $\pi'$ has
worst-case error at most $1/10$ and satisfies $\IC(\pi')\leq I.$ In
particular, $\IC_{1/10}(F^{\oplus N})\leq I.$

Let $\alpha=\log(3/2)=0.5849\ldots,$ as in Lemma~\ref{lem:recursive-bound},
and put
\[
  I_N=\frac{13}{5}\frac I N+4+\frac{32}{5}N^{-\alpha}.
\]
By Lemma~\ref{lem:recursive-bound} and the bound
$\IC_{1/10}(F^{\oplus N})\leq I,$ every $\mu\in\mathcal P(\dom F)$
satisfies $\IC_{N^{-\alpha}/10}(F,\mu)\leq I_N.$ Since $I_N$ does not
depend on $\mu,$ taking the supremum over $\mu$ and applying
Fact~\ref{fact:minimax} gives
\begin{equation}\label{eq:final-minimax}
  \IC_{N^{-\alpha}/5}(F)\leq2I_N.
\end{equation}

We now reduce the error by a majority vote. Put $p=N^{-\alpha}/5,$ and fix
$\eta>0.$ By~\eqref{eq:final-minimax}, there is a protocol $\sigma$ for $F$
with $\err(\sigma,F)\leq p$ and $\IC(\sigma)\leq2I_N+\eta.$ Let $\sigma'$ be
the protocol that runs three copies of $\sigma$ on its input with
independent randomness and outputs the majority of their answers. On a
given legal input, each copy errs with probability at most $p,$
independently of the other copies, and $\sigma'$ errs only if at least two
copies err. The union bound over the three pairs of copies therefore gives
\begin{equation}\label{eq:three-repetitions}
  \err(\sigma',F)\leq3p^2=\frac3{25}N^{-2\alpha}.
\end{equation}
Here it matters that the majority vote
comes after the application of Fact~\ref{fact:minimax}: the bound $3p^2$ uses the guarantee
$\err(\sigma,F)\leq p$ on every legal input, and a bound on the error
averaged over an input distribution would not suffice. To
compare~\eqref{eq:three-repetitions} with $1/(6n),$ write $N=2^d,$ so that
$N^{-2\alpha}=(4/9)^d,$ and recall that $n\leq2N-1.$ Then
\[
  n\err(\sigma',F)
  \leq\frac3{25}\,(2^{d+1}-1)\Bigl(\frac49\Bigr)^d
  \leq\frac{112}{675}<\frac16.
\]
For the middle step, the quantity $(2^{d+1}-1)(4/9)^d$ equals $1,$ $4/3,$
and $112/81$ for $d=0,1,2,$ and it decreases from then on, because the
ratio of its consecutive values is at most $20/21$ for $d\geq2.$

By Lemma~\ref{lem:independent-executions} and the remarks following it,
$\IC(\sigma')\leq3\IC(\sigma)\leq6I_N+3\eta.$ Since $N>n/2$ and
$N^{-\alpha}\leq1,$
\[
  6I_N=\frac{78}{5}\frac I N+24+\frac{192}{5}N^{-\alpha}
  \leq\frac{156}{5}\frac I n+\frac{312}{5}
  \leq\frac{32I}{n}+64.
\]
Hence $\IC_{1/(6n)}(F)\leq\IC(\sigma')\leq32I/n+64+3\eta.$ Letting
$\eta\to0$ and then taking the infimum over $\pi$
proves~\eqref{eq:xor-conversion}.

Finally, for every communication problem $G$ with output set $\{-1,1\},$ the
majority of $15$ independent
runs of a protocol with error at most $1/3$ has error at most
\[
  \sum_{j=8}^{15}\binom{15}{j}
        \left(\frac13\right)^j\left(\frac23\right)^{15-j}
  =\frac{422009}{4782969}<\frac1{10}.
\]
By Lemma~\ref{lem:independent-executions}, its information cost is at most
$15$ times the original cost. As a result, $\IC_{1/10}(G)\leq15\IC_{1/3}(G).$
Taking $G=F^{\oplus n}$ in~\eqref{eq:xor-conversion}
proves~\eqref{eq:xor-conversion-third}.
\end{proof}

The recursion, the minimax theorem, and the majority votes in this
proof all work with constant error. Independent executions on disjoint
blocks of coordinates extend~\eqref{eq:xor-conversion-third} to every
error $\epsilon\in(0,1/3],$ and a direct embedding argument handles
$\epsilon=0.$ This gives the main technical result of this section.

\begin{theorem}[XOR lemma with constant additive loss]\label{thm:additive-conversion}
Let $F\colon\Xcal\times\Ycal\to\{-1,1,*\}$ be a communication problem. Then
for every integer $n\geq1$ and every $\epsilon\in[0,1/3],$
\begin{equation}\label{eq:additive-conversion}
  \IC_{\epsilon/n}(F)
  \leq\frac{480}{n}\IC_\epsilon(F^{\oplus n})+64.
\end{equation}
\end{theorem}

\begin{proof}
Suppose first that $\epsilon>0,$ and put $r=\lfloor1/(3\epsilon)\rfloor.$
Then $r\epsilon\leq1/3,$ and $r\geq1/(6\epsilon)$ because
$\lfloor t\rfloor\geq t/2$ for $t\geq1.$ Let $\pi$ be a protocol for
$F^{\oplus n}$ with worst-case error at most $\epsilon.$ By the remarks
following Lemma~\ref{lem:independent-executions}, independent executions
of $\pi$ on $r$ disjoint blocks of $n$ coordinate pairs give a protocol for
$F^{\oplus rn}$ with worst-case error at most $r\epsilon\leq1/3$ and
worst-case information cost at most $r\IC(\pi).$
Hence~\eqref{eq:xor-conversion-third}, applied with $rn$ in place of $n,$
gives
\[
  \IC_{1/(6rn)}(F)\leq\frac{480}{rn}\,r\IC(\pi)+64
  =\frac{480}{n}\IC(\pi)+64.
\]
Since $1/(6rn)\leq\epsilon/n,$ taking the infimum over $\pi$
proves~\eqref{eq:additive-conversion}.

It remains to treat $\epsilon=0.$ Let $\pi$ be a protocol for
$F^{\oplus n}$ that never errs on a legal input, and fix
$\mu\in\mathcal P(\dom F).$ The protocol $\sigma'$ in the proof of
Fact~\ref{fact:ordinary-embedding}, constructed from $\pi$ with $F,$ $n,$
$\mu$ in place of $G,$ $t,$ $\nu,$ satisfies
$\IC(\sigma',\mu)\leq\IC(\pi)/n+2.$ Moreover, $\sigma'$ errs on a legal
input exactly when $\pi$ errs on the completed input, which is legal.
Thus $\sigma'$ never errs on a legal input, even outside the support of
$\mu.$ The class of protocols for $F$ that never err on a legal input is
closed under finite public mixtures, and
Proposition~\ref{prop:class-minimax} gives
$\IC_0(F)\leq\IC(\pi)/n+2.$ Taking the infimum over $\pi,$ we
obtain $\IC_0(F)\leq\IC_0(F^{\oplus n})/n+2,$ which
implies~\eqref{eq:additive-conversion} for $\epsilon=0.$
\end{proof}

\subsection{Main results on total functions}\label{subsec:total-proofs}

We now prove the direct sum theorem and XOR lemma for total functions.
Both proofs combine Theorem~\ref{thm:additive-conversion} with a linear
lower bound for the multiple-copy problem, $F^n$ or $F^{\oplus n}.$
In the case of the direct sum
theorem, the linear lower bound is supplied by
Proposition~\ref{prop:local-variation}, so we can give the proof right
away. We restate the theorem for convenience.

\begin{restated}{Theorem~\ref*{main:ds}}{Direct sum theorem, restated}
There is a universal constant $c>0$ such that for every
communication problem $F\colon\Xcal\times\Ycal\to\{-1,1\},$
every integer $n\geq1,$ and every $\epsilon\in[0,1/3],$
\begin{equation*}
  cn\IC_{\epsilon/n}(F)
    \leq\IC_\epsilon(F^n)
    \leq n\IC_{\epsilon/n}(F).\tag{\ref*{eq:total-direct-sum}}
\end{equation*}
\end{restated}

\begin{proof}
If $F$ is constant, then all information complexities in the
theorem statement are zero, and the result holds trivially.
For nonconstant $F,$ the upper bound is
Proposition~\ref{prop:general-upper}. Moreover,
\begin{align*}
1-h_2\Bigl(\frac13\Bigr)&\leq\frac{\IC_\epsilon(F^n)}n,\\
\IC_{\epsilon/n}(F)&\leq\frac{480\IC_\epsilon(F^n)}n+64,
\end{align*}
where the first inequality uses Proposition~\ref{prop:local-variation}
and $\epsilon\leq1/3,$ and the second is valid by Theorem~\ref{thm:additive-conversion} and
Proposition~\ref{prop:general-upper}. Taking a weighted arithmetic mean
of these inequalities gives the lower bound
in~\eqref{eq:total-direct-sum}, with
$c=1/(480+64/(1-h_2(1/3)))>1/1264.$
\end{proof}

The XOR lemma requires more care, because $\IC_\epsilon(F^{\oplus n})$ may
stay bounded as $n$ grows. This is the case for the exceptional class
of XOR-reducible functions, whose definition we recall.

\begin{definition}\label{def:xor-reducible}
A function $F\colon\Xcal\times\Ycal\to\{-1,1\}$ is \emph{XOR-reducible}
if there are functions $a\colon\Xcal\to\{-1,1\}$ and
$b\colon\Ycal\to\{-1,1\}$ such that $F(x,y)=a(x)b(y)$ for all
$x\in\Xcal$ and $y\in\Ycal.$
\end{definition}

We now settle this class by proving
Proposition~\ref{main:xor-reducible} from the introduction,
which we restate for convenience.

\begin{restated}{Proposition~\ref*{main:xor-reducible}}{XOR-reducible
functions, restated}
Let $F\colon\Xcal\times\Ycal\to\{-1,1\}$ be XOR-reducible, and let
$n\geq1$ be an integer.
\begin{enumerate}
\item If $F$ is constant, then $\IC_0(F^{\oplus n})=\IC_0(F)=0.$
\item If $F$ is not constant, then
$\IC_\epsilon(F^{\oplus n})$ and $\IC_{\epsilon/n}(F)$ lie in
$[1-h_2(1/3),2]$ for every $\epsilon\in[0,1/3].$
\end{enumerate}
\end{restated}

\begin{proof}
If $F$ is constant, then so is $F^{\oplus n},$ and the parties can
output the answer without communication; hence
$\IC_0(F^{\oplus n})=\IC_0(F)=0.$ Suppose now that $F$ is not constant,
and write $F(x,y)=a(x)b(y).$ To compute $F^{\oplus n},$ Alice and Bob
announce the signs $\prod_ia(x_i)$ and $\prod_ib(y_i),$ respectively,
and both output the XOR of these signs. This protocol never errs. It
consists of two one-bit messages, so its information cost is at most $2$
under every input distribution. Hence
$\IC_\epsilon(F^{\oplus n})\leq\IC_0(F^{\oplus n})\leq2,$ and the case
$n=1$ gives $\IC_{\epsilon/n}(F)\leq\IC_0(F)\leq2.$

For the lower bounds, recall from Proposition~\ref{prop:local-variation}
that a nonconstant total function has a row or a column with two
different values. That proposition, applied with a single copy, therefore
gives $\IC_{\epsilon/n}(F)\geq1-h_2(\epsilon/n)\geq1-h_2(1/3)$ and
$\IC_\epsilon(F)\geq1-h_2(\epsilon)\geq1-h_2(1/3).$ Finally,
Proposition~\ref{prop:restriction-floor} gives
$\IC_\epsilon(F^{\oplus n})\geq\IC_\epsilon(F).$
\end{proof}

It remains to establish a perfect XOR lemma for every function
outside this class. As with the direct sum theorem, we
will need a linear lower bound for $F^{\oplus n},$ supplied
by the following lemma.

\begin{lemma}\label{lem:xor-linear}
Let $F\colon\Xcal\times\Ycal\to\{-1,1\}$ be a communication problem that
is not XOR-reducible. Then for every integer $n\geq1,$
\[
  \IC_{1/3}(F^{\oplus n})\geq\frac n{48}\Bigl(1-h_2\Bigl(\frac13\Bigr)\Bigr).
\]
\end{lemma}

\begin{proof}
Fix $x_0\in\Xcal$ and $y_0\in\Ycal.$ Since $F$ is not XOR-reducible, it
differs from the XOR-reducible function
$(x,y)\mapsto F(x,y_0)F(x_0,y_0)F(x_0,y)$ on some input $(x_1,y_1),$
that is,
\[
  F(x_0,y_0)F(x_0,y_1)F(x_1,y_0)F(x_1,y_1)=-1.
\]
This forces $x_0\ne x_1$ and $y_0\ne y_1.$ Moreover, three of the four
factors on the left are equal, and the fourth differs from them. Swapping
the names $x_0$ and $x_1$ if necessary, and likewise $y_0$ and $y_1,$ 
we may assume that the fourth factor is $F(x_1,y_1).$ Then
\[
  F(x_u,y_v)=s(-1)^{uv},\qquad u,v\in\{0,1\},
\]
for some sign $s.$ In other words, the restriction of $F$ to the rows
$x_0,x_1$ and the columns $y_0,y_1$ is the two-bit AND function
$\AND_2(u,v)=(-1)^{uv}$ if $s=1,$ and its negation $\neg\AND_2$ if
$s=-1.$

The XOR of $n$ copies of $\AND_2$ is the inner product function
$\IP_n(u,v)=(-1)^{\sum_{i=1}^nu_iv_i}$ on $\{0,1\}^n\times\{0,1\}^n.$
Corollary~\ref{cor:ip-lower} in Appendix~\ref{app:inner-product} gives
$\IC_{1/3}(\IP_n)\geq(1-h_2(1/3))n/48;$ up to the constant factor, this
is a standard consequence of the discrepancy lower bound for information
complexity~\cite{BW}. It therefore suffices to observe that
$\IC_{1/3}(\IP_n)\leq\IC_{1/3}(F^{\oplus n}),$ due to the
identity $\IP_n(u,v)=
s^nF^{\oplus n}((x_{u_1},\ldots,x_{u_n}),(y_{v_1},\ldots,y_{v_n}))$ 
for all $u,v\in\{0,1\}^n.$ 
\end{proof}

We are now in a position to prove the XOR lemma.

\begin{restated}{Theorem~\ref*{main:xor}}{XOR lemma, restated}
There is a universal constant $c>0$ such that for every
communication problem $F\colon\Xcal\times\Ycal\to\{-1,1\}$
that is not XOR-reducible, and for every integer $n\geq1$ and every
$\epsilon\in[0,1/3],$
\begin{equation*}
  cn\IC_{\epsilon/n}(F)
    \leq\IC_\epsilon(F^{\oplus n})
    \leq n\IC_{\epsilon/n}(F).\tag{\ref*{eq:total-xor}}
\end{equation*}
\end{restated}

\begin{proof}
The upper bound is again Proposition~\ref{prop:general-upper}. For the
lower bound, we follow the proof of Theorem~\ref{main:ds}, with
Lemma~\ref{lem:xor-linear} in place of
Proposition~\ref{prop:local-variation}. That lemma, together with
$\IC_\epsilon(F^{\oplus n})\geq\IC_{1/3}(F^{\oplus n}),$ and
Theorem~\ref{thm:additive-conversion} give
\begin{align*}
  &\frac1{48}\Bigl(1-h_2\Bigl(\frac13\Bigr)\Bigr)
     \leq\frac{\IC_\epsilon(F^{\oplus n})}n,\\
  &\IC_{\epsilon/n}(F)\leq\frac{480\IC_\epsilon(F^{\oplus n})}n+64.
\end{align*}
Taking a weighted arithmetic mean
of these inequalities gives the lower bound
in~\eqref{eq:total-xor}, with
$c=1/(480+3072/(1-h_2(1/3)))>1/38080.$
\end{proof}

\section{Partial functions}\label{sec:partial}

This section proves our results for partial functions,
Theorems~\ref{main:ds-partial}--\ref{main:characterization}. We start
with the direct sum theorem. Since
Theorem~\ref{thm:additive-conversion} holds on arbitrary promise domains,
the proof of Theorem~\ref{main:ds} carries over to the
setting of arbitrary partial functions, and the only difference
is that the role of
the constant functions is now played by the communication problems in
which every row and every column is constant on its legal inputs. We
restate the theorem for convenience.

\begin{restated}{Theorem~\ref*{main:ds-partial}}{Direct sum theorem, restated}
There is a universal constant $c>0$ such that for every
communication problem $F\colon\Xcal\times\Ycal\to\{-1,1,*\},$
every integer $n\geq1,$ and every $\epsilon\in[0,1/3],$
\begin{equation*}
  cn\IC_{\epsilon/n}(F)
    \leq\IC_\epsilon(F^n)
    \leq n\IC_{\epsilon/n}(F).\tag{\ref*{eq:partial-direct-sum}}
\end{equation*}
\end{restated}

\begin{proof}
The upper bound is Proposition~\ref{prop:general-upper}. For the lower
bound, suppose first that every row and every column of $F$ is constant on
its legal inputs. Then Proposition~\ref{prop:local-variation} gives
$\IC_{\epsilon/n}(F)\leq\IC_0(F)=0,$ and there is nothing to prove. Otherwise,
the same proposition gives
$\IC_\epsilon(F^n)\geq(1-h_2(\epsilon))n\geq(1-h_2(1/3))n,$ and the rest
of the proof of Theorem~\ref{main:ds} applies verbatim. In particular, one
can again take $c=1/1264.$
\end{proof}

The rest of this section is devoted to the XOR question,
which requires considerable additional ideas.

\Needspace{15\baselineskip}
\subsection{From announced to split outputs}
\label{subsec:partial-obstruction}

For total functions, Proposition~\ref{main:xor-reducible} and
Theorem~\ref{main:xor} show that the XOR lemma
$\IC_\epsilon(F^{\oplus n})\geq cn\IC_{\epsilon/n}(F)$
holds precisely for those functions $F$ that are constant or not
XOR-reducible. We will show that for partial functions, it does not suffice
for $F$ to disagree on its domain with every XOR-reducible
function (Theorem~\ref{main:counterexample} from the
introduction). The protocol in its proof motivates the split-output
measure, which we define at the end of this subsection. 

\Needspace{13\baselineskip}
\begin{restated}{Theorem~\ref*{main:counterexample}}{Counterexample, restated}
For every integer $n\geq1,$ there is a communication problem
$F\colon\Xcal\times\Ycal\to\{-1,1,*\}$ that disagrees on
$\dom F$ with every XOR-reducible function and yet satisfies
\begin{align*}
  &\IC_0(F^{\oplus n})\leq3,\tag{\ref*{eq:counterexample-xor}}\\
  &\IC_{1/3}(F)\geq1-h_2\Bigl(\frac13\Bigr).
     \tag{\ref*{eq:counterexample-single}}
\end{align*}
\end{restated}

\begin{proof}
Let $m>n$ be an integer to be set later. 
Let $F_m\colon\mathbb Z_m\times\mathbb Z_m\to\{-1,1,*\}$ be the communication
problem whose legal inputs are the pairs $(x,y)$ with $y\in\{x,x+1\},$
and whose value is $-1$ on $(m-1,0)$ and $1$ on all other
legal inputs.

Every row and every column of the characteristic matrix of $F_m$
contains exactly two legal inputs. Hence, if $F_m(x,y)=a(x)b(y)$ held on
$\dom F_m$ for some $\pm1$-valued functions $a$ and $b,$ the product of the values of
$F_m$ over all legal inputs would be $1,$ since each factor $a(x)$ and
$b(y)$ would occur exactly twice. However, this product is $-1,$ because
exactly one legal input has value $-1.$ Therefore $F_m$ disagrees on its domain 
with every XOR-reducible function.

Order the elements of $\mathbb Z_m$ in the natural way as $0<1<2<\cdots<m-1.$ For $t\in\mathbb Z_m,$ put
\[
  a_t(x)=(-1)^{\1_{\{x\geq t\}}},\qquad
  b_t(y)=(-1)^{\1_{\{y>t\}}},\qquad
  G_t(x,y)=a_t(x)b_t(y).
\]
For $x\ne m-1,$ we have $G_t(x,x+1)=1,$ because $x\geq t$ if and only if
$x+1>t.$ Moreover, $G_t(m-1,0)=-1,$ and $G_t(x,x)=-1$ precisely when
$x=t.$ Thus $G_t$ agrees with $F_m$ on every legal input except $(t,t).$ A
uniformly random $t\in\mathbb Z_m$ therefore errs on any given legal input
with probability at most $1/m.$

To compute $F_m^{\oplus n}$ without error, the parties instead use a value
of $t$ whose exceptional input $(t,t)$ occurs in no coordinate. Since
Alice's inputs take at most $n<m$ distinct values, such values exist, and
Alice can find one. Specifically, the parties publicly choose a uniformly
random ordering $t_0,t_1,\ldots,t_{m-1}$ of $\mathbb Z_m,$ and Alice
announces the least $J$ such that $t_J\notin\{x_1,\ldots,x_n\}.$ Alice
then announces $\prod_{i=1}^n a_{t_J}(x_i),$ Bob announces
$\prod_{i=1}^n b_{t_J}(y_i),$ and both output the XOR of these two signs.
As $G_{t_J}(x_i,y_i)=F_m(x_i,y_i)$ for every $i,$ the output is always
correct.

Now fix an input distribution, let $(X,Y)$ be the input, and let $R$ be
the public ordering. The index $J$ is a function of $(X,R),$ so it reveals
nothing about $Y$ to Alice and at most $H(J)$ bits about $X$ to Bob, and
the two announced signs add at most $2.$ For $j\geq1,$ we have $J\geq j$
exactly when $t_0,\ldots,t_{j-1}$ all lie in $\{X_1,\ldots,X_n\},$ a
set of at most $n$ elements, so that $\Prob[J\geq j]\leq(n/m)^j.$ Now set
$m=8n.$ Then $\Exp J=\sum_{j\geq1}\Prob[J\geq j]\leq1/7.$ The
nonnegativity of relative entropy, applied to the law of $J$ and the
geometric distribution $Q(j)=(7/8)\,8^{-j}$ on $\{0,1,2,\ldots\},$ gives
\[
  H(J)\leq\Exp\log\frac1{Q(J)}=\log\frac87+3\Exp J
  \leq\log\frac87+\frac37<1.
\]
Hence the protocol has information cost less than $3$ under every input
distribution, which proves~\eqref{eq:counterexample-xor}.
For~\eqref{eq:counterexample-single},
use $F_m(m-1,m-1)\ne F_m(m-1,0)$ along with
Proposition~\ref{prop:local-variation} to get
$\IC_{1/3}(F_m)\geq1-h_2(1/3).$
\end{proof}

The function $F_m$ in the proof of Theorem~\ref{main:counterexample}
disagrees on its domain with every XOR-reducible function yet can be
computed to high accuracy by a probability distribution on XOR-reducible
functions. What makes the XOR of multiple copies of $F_m$ inexpensive is
that each party can combine his or her local signs over all coordinates and
announce the result only once. In any protocol of this form, the final
announcement costs at most $2$ bits, however many copies there are, so
only the cost of producing the local signs grows with the number of
copies. To measure this cost, we keep the communication model and the
definition of information cost but change the output convention:
each party now ends with a sign, the answer is the product of the two signs,
and neither sign is announced.

\Needspace{11\baselineskip}
\begin{definition}\label{def:split-output}
Let $F\colon\Xcal\times\Ycal\to\{-1,1,*\}$ be a communication problem. A
\emph{split-output protocol}
ends with local signs $A$ and $B$ produced by Alice and Bob, respectively,
which we call their \emph{shares}. Each share is a function of that
player's local view. The protocol is correct when $AB=F(x,y).$ Neither
share is appended to the transcript.
The protocol's information cost is still~\eqref{eq:information-cost}. For this
output convention, write $\err(\pi,F)$ for the maximum, over legal input
pairs, of $\Prob[AB\ne F(x,y)].$ Define
\begin{equation}\label{eq:shared-ic}
  \SIC_\epsilon(F)
   =\inf_{\pi:\,\err(\pi,F)\leq\epsilon}\IC(\pi).
\end{equation}
The supremum defining $\IC(\pi)$ is over all distributions in
$\mathcal P(\dom F),$ as before.
\end{definition}

To relate this output convention to the standard model, we have
$\SIC_\epsilon(F)\leq\IC_\epsilon(F)\leq\SIC_\epsilon(F)+2.$
For the first inequality, use the announced answer as Alice's share and
set Bob's share to $1.$ For the second, append both local shares to a
split-output protocol; these messages cost at most two bits of internal
information and make the answer public.

\subsection{Extraction via grouping}\label{subsec:shared-embedding}

The lower bound in Theorem~\ref{main:characterization} requires
extracting, from a protocol for $F^{\oplus n}$ with error $\epsilon,$ a
split-output protocol for a single copy of $F$ with error
$\epsilon/n.$ The main work is the case $\epsilon=1/3,$ which we carry out
in this subsection and the next. As a first step,
Lemma~\ref{lem:tensor} extracts one copy from a split-output protocol
with a factor-$n$ decrease in information, no increase in
error, and no additive loss. Recall that its counterpart for announced
outputs, Fact~\ref{fact:ordinary-embedding}, incurred a
constant additive loss due to the use of the standard
embedding followed by two one-bit messages, in which Alice announces the XOR of
the function values on the coordinates after the embedded
coordinate and Bob
the XOR on those before it. With split outputs, each party instead
multiplies his or her share by that XOR.

\Needspace{7\baselineskip}
\begin{lemma}\label{lem:tensor}
For every communication problem $F\colon\Xcal\times\Ycal\to\{-1,1,*\},$
every $k\ge1,$ and every $\epsilon\ge0,$
\[
 \SIC_\epsilon(F) \le \frac{\SIC_\epsilon(F^{\oplus k})}k.
\]
\end{lemma}

\begin{proof}
Let $\pi$ be a split-output protocol for $F^{\oplus k}$ with worst-case
error at most $\epsilon$ and worst-case information cost $I.$ Fix
$\mu\in\mathcal P(\dom F),$ and let $\pi^{(1)},\ldots,\pi^{(k)}$ be the standard
embeddings into $\pi$ with respect to $\mu;$ Definition~\ref{def:standard-embedding}
and Fact~\ref{fact:embedding-cost} apply verbatim to split-output
protocols, since they concern only the inputs, the randomness, and the
transcript. The new protocol publicly chooses a uniformly random $j\in[k]$
and runs $\pi^{(j)},$ in which Alice and Bob obtain shares $A$ and $B.$
The new shares are
\[
 A'=A\prod_{i>j}F(X_i,Y_i),\qquad
 B'=B\prod_{i<j}F(X_i,Y_i),
\]
which the parties compute without further communication: Alice knows
$X_{>j}$ and $Y_{>j},$ and Bob knows $X_{<j}$ and $Y_{<j}.$ As
$A'B'=AB\prod_{i\ne j}F(X_i,Y_i),$ the new protocol is correct exactly
when $\pi$ is correct on the completed input, which is legal 
whenever the actual input is legal. Hence the new protocol has error at
most $\epsilon$ on every legal input, even outside the support of $\mu.$
By~\eqref{eq:public-mixture-cost} and Fact~\ref{fact:embedding-cost}, its
information cost under $\mu$ is $\IC(\pi,\mu^k)/k\leq I/k.$

For every $\mu,$ we have therefore produced a split-output protocol for
$F$ with worst-case error at most $\epsilon$ and information cost at most
$I/k$ under $\mu.$ Let $\Ccal$ be the class of all split-output protocols
for $F$ with worst-case error at most $\epsilon.$ A public mixture of
protocols in $\Ccal$ again belongs to $\Ccal,$ because its error on each
legal input is a weighted average of their errors on that input.
Proposition~\ref{prop:class-minimax}, like
Fact~\ref{fact:embedding-cost}, concerns only the inputs, the randomness,
and the transcript, so it applies to $\Ccal$ and gives
$\SIC_\epsilon(F)\leq I/k.$ Taking the infimum over $\pi$ proves the result.
\end{proof}

Lemma~\ref{lem:tensor} leaves the error unchanged, so the error reduction must
come from elsewhere. The only mechanism we have developed so far that converts a
protocol for an XOR into a protocol for a single copy with smaller error
is Theorem~\ref{thm:additive-conversion}, and it introduces an additive
loss of $64.$ Our next result uses grouping to partially overcome this loss. We split the copies
into $m$ blocks of $k$ copies each, apply
Lemma~\ref{lem:xor-conversion} to the block problem
$F^{\oplus k}$ with $m$ copies, and reduce the error further by a
majority vote. Lemma~\ref{lem:tensor} then extracts a single copy from a
block, which divides the additive loss by $k.$ The majority vote must
precede the extraction, because it requires announced answers. The
resulting additive loss, of order $1/k,$ is affordable unless
$\IC_{1/3}(F^{\oplus n})$ is very small compared with $n,$ and
Section~\ref{subsec:rounding} treats that case.

\Needspace{7\baselineskip}
\begin{lemma}\label{lem:block}
For every communication problem $F\colon\Xcal\times\Ycal\to\{-1,1,*\}$ and
all integers $k,m\geq1,$
\begin{equation}\label{eq:block}
 \SIC_{1/(16m^3)}(F)\le\frac{2400}{km}\IC_{1/3}(F^{\oplus km})
 +\frac{320}k.
\end{equation}
\end{lemma}

\begin{proof}
Inequality~\eqref{eq:xor-conversion-third} in Lemma~\ref{lem:xor-conversion},
applied to the composition $(F^{\oplus k})^{\oplus m}=F^{\oplus km},$
gives
\[
 \IC_{1/(6m)}(F^{\oplus k})\le\frac{480}m\IC_{1/3}(F^{\oplus km})+64.
\]
Fix $\eta>0$ and a protocol $\sigma$ for $F^{\oplus k}$ with worst-case error at most
$1/(6m)$ and $\IC(\sigma)\le\IC_{1/(6m)}(F^{\oplus k})+\eta.$ Let $\sigma'$ be the
protocol that runs five copies of $\sigma$ with independent randomness
and outputs the majority of their answers. On a given legal input, each
copy errs with probability at most $1/(6m),$ independently of the other
copies, and $\sigma'$ errs only if at least three copies err. The union
bound over the $\binom53$ triples of copies therefore gives
\[
 \err(\sigma',F^{\oplus k})\le\binom53\Bigl(\frac1{6m}\Bigr)^3=\frac5{108m^3}
 <\frac1{16m^3}.
\]
By Lemma~\ref{lem:independent-executions}, $\IC(\sigma')\le5\IC(\sigma).$
Regard the answer of $\sigma'$ as Alice's share, and set Bob's share
to $1.$ Lemma~\ref{lem:tensor} then gives
\[
 \SIC_{1/(16m^3)}(F)\le\frac1k\SIC_{1/(16m^3)}(F^{\oplus k})
 \le\frac5k\Bigl(\frac{480}m\IC_{1/3}(F^{\oplus km})+64+\eta\Bigr).
\]
Letting $\eta\to0$ proves~\eqref{eq:block}.
\end{proof}

\subsection{Extraction via rounding}\label{subsec:rounding}

The additive loss in Lemma~\ref{lem:block} is affordable only when
$\IC_{1/3}(F^{\oplus n})$ is not too small compared with
$n.$ In the complementary case,
Theorem~\ref{thm:margin-ic} forces the margin of $F$ to be close to $1,$
and we convert the vectors that achieve the margin into accurate shares.
For this purpose, we use a one-message variant of random hyperplane
rounding. Weighting the choice of direction by a squared coordinate
reduces the error from first to third order in the angle, at an
information cost of second order.

\Needspace{7\baselineskip}
\begin{lemma}\label{lem:rounding}
Let $F\colon\Xcal\times\Ycal\to\{-1,1,*\}$ be a communication problem,
and put $s=1-\margin(F)^2.$ Then
\begin{equation}\label{eq:rounding}
 \SIC_{s^{3/2}}(F)\le\frac2{\ln2}s.
\end{equation}
\end{lemma}

To explain the key idea behind this result, fix unit vectors
$u_x,v_y\in\mathbb R^d$ that achieve the margin of $F$
in~\eqref{eq:margin-vector}, so that
$F(x,y)\langle u_x,v_y\rangle\ge\margin(F)=\sqrt{1-s}$ for every legal
input $(x,y).$ For a legal input $(x,y),$ let $\theta\in[0,\pi/2]$ be the
angle between $u_x$ and $F(x,y)v_y,$ so that
$\cos\theta=F(x,y)\langle u_x,v_y\rangle\ge\sqrt{1-s}$ and hence
$\sin^2\theta\le s.$ The natural protocol without
communication is \emph{random
hyperplane rounding:} the players use public randomness to choose a
uniformly random unit vector $w$ and output
$A=\sgn\langle u_x,w\rangle$ and $B=\sgn\langle
v_y,w\rangle,$ respectively. This rounding operation is central to the
semidefinite approximation algorithms of Goemans and
Williamson~\cite{GW95}. In our notation, 
$AB\ne F(x,y)$ exactly when the hyperplane orthogonal to $w$ separates
$u_x$ from $F(x,y)v_y,$ which happens with probability
$\theta/\pi.$ The underlying sign-correlation identity goes back to
Sheppard~\cite{She99} and plays a key role in Krivine's
proof~\cite{Kri79} of Grothendieck's inequality (see Alon and
Naor~\cite{AN06} and Pisier~\cite{Pis12}), and in communication
complexity, signs of random projections were used by Kremer, Nisan, and
Ron~\cite{KNR99} to estimate inner products. For the unweighted random
direction, the disagreement probability is linear in $\theta.$ For
Lemma~\ref{lem:rounding} this is too large: on a legal input with
$\sin^2\theta=s,$ the error of hyperplane rounding is at least
$\sqrt s/\pi,$ whereas the lemma promises error of order $s^{3/2}.$

More generally, no communication-free protocol can achieve the guarantee
of Lemma~\ref{lem:rounding} when $s$ is small. Consider any protocol
without communication. Let $R$ denote its public
randomness, which by Proposition~\ref{prop:finite-fixed} we may assume to
take finitely many values, and let $a_x(R)=\Exp[A\mid x,R]$ and
$b_y(R)=\Exp[B\mid y,R],$ where the expectations are over the
players' private randomness. Since the private randomness of the two
players is independent, $\Exp[AB\mid x,y,R]=a_x(R)b_y(R),$
and therefore
\[
 F(x,y)\,\Exp_R\bigl[a_x(R)\,b_y(R)\bigr]
 =1-2\Prob[AB\ne F(x,y)]
\]
for every legal input $(x,y).$ Here the vectors
$(\sqrt{\Prob[R=r]}\,a_x(r))_r$ and $(\sqrt{\Prob[R=r]}\,b_y(r))_r,$ indexed by
the values $r$ of $R,$ have norm at most $1,$ and their inner product is
$\Exp_R[a_x(R)\,b_y(R)].$ If the error is at most $\epsilon<1/2$ on every legal input,
then padding these vectors to unit norm in orthogonal auxiliary
directions yields unit vectors with margin at least
$1-2\epsilon,$ so that $\margin(F)\ge1-2\epsilon.$ Consequently, some legal
input has error at least $(1-\margin(F))/2=(1-\sqrt{1-s})/2\ge s/4,$ which
is much larger than $s^{3/2}$ when $s$ is small.

Our protocol in the proof of Lemma~\ref{lem:rounding} is a weighted form
of hyperplane rounding with a single message. The random direction is
one of the $d$ vectors of a public random orthonormal basis. Alice
selects it with probability equal to the squared coordinate of $u_x$
along it, and she sends its index to Bob. Let $\Phi$ be the angle of the
projection of the selected direction onto a plane containing $u_x$ and
$F(x,y)v_y,$ measured from $u_x.$ The two signs disagree exactly when
$\Phi$ lies in one of two arcs of length $\theta$ that start at
$\pm\pi/2.$ For a uniformly random direction, $\Phi$ is uniformly
distributed on $[0,2\pi),$ and this gives the error $\theta/\pi$ of
hyperplane rounding. Alice's selection multiplies the density of $\Phi$
at each angle $\phi$ by $2\cos^2\phi,$ which is at most $2\sin^2\theta$
on the two arcs. As a result, the error drops from $\theta/\pi$ to
$(\theta-\sin\theta\cos\theta)/\pi\le\sin^3\theta,$ while the index
reveals only information of order $\sin^2\theta\le s$ about $u_x.$
A related mechanism, in which Alice chooses between two shared random
directions and tells Bob her choice, underlies the one-bit simulation of
singlet correlations by Toner and Bacon~\cite{TB03}, in the form given by
Degorre, Laplante, and Roland~\cite{DLR05}.

\Needspace{3\baselineskip}
\begin{proof}
As before, let $u_x,v_y\in\mathbb R^d$ be unit vectors that achieve the
margin of $F$ in~\eqref{eq:margin-vector}, so that
$F(x,y)\langle u_x,v_y\rangle\ge\margin(F)=\sqrt{1-s}$ for every legal
input $(x,y).$ If $d=1,$ then $F(x,y)u_xv_y\ge\margin(F)>0$ forces
$F(x,y)=u_xv_y$ on every legal input,
so the local signs $u_x$ and $v_y$ solve the problem without communication.
Assume henceforth that $d\ge2.$
Publicly choose a
uniformly random (Haar distributed) orthonormal basis
$R=(e_1,\ldots,e_d)$ of $\mathbb R^d.$ For a unit vector $u,$ let $p_u^R$
be the probability distribution on $[d]$ given by
\begin{equation}\label{eq:square-coordinates}
 p_u^R(i)=\langle u,e_i\rangle^2.
\end{equation}
On input $x,$ Alice privately samples $Z\sim p_{u_x}^R$ and sends $Z$ to
Bob. The shares are
\[
 A=\sgn\langle u_x,e_Z\rangle,\qquad
 B=\sgn\langle v_y,e_Z\rangle.
\]
(Ties occur with probability zero and may be resolved arbitrarily.)

To analyze the protocol, fix unit vectors $u,v$ with
$\langle u,v\rangle=\cos\theta,$ where $\theta\in[0,\pi/2],$ and let
$Z\sim p_u^R.$ We claim that there are random variables $r>0$ and
$\Phi\in[0,2\pi)$ with
\begin{equation}\label{eq:angle}
 \langle u,e_Z\rangle=r\cos\Phi,\qquad
 \langle v,e_Z\rangle=r\cos(\Phi-\theta),
\end{equation}
such that $\Phi$ has density $\cos^2\phi/\pi$ on $[0,2\pi).$
Let $O$ be an orthogonal matrix with $Ou=(1,0,0,\ldots,0)$ and
$Ov=(\cos\theta,\sin\theta,0,\ldots,0).$
Since $\langle u,e_i\rangle=\langle Ou,Oe_i\rangle,$
$\langle v,e_i\rangle=\langle Ov,Oe_i\rangle,$ and $(Oe_i)_{i\le d}$ is
again a uniformly random orthonormal basis, we may assume that
$u=(1,0,0,\ldots,0)$ and $v=(\cos\theta,\sin\theta,0,\ldots,0).$
Write the first two coordinates of $e_i$ in polar form as
$r_i(\cos\phi_i,\sin\phi_i).$ Then
\[
 \langle u,e_i\rangle=r_i\cos\phi_i,\qquad
 \langle v,e_i\rangle
 =r_i(\cos\theta\cos\phi_i+\sin\theta\sin\phi_i)
 =r_i\cos(\phi_i-\theta),
\]
so~\eqref{eq:angle} holds with $(r,\Phi)=(r_Z,\phi_Z).$ Each $e_i$ is
uniformly distributed on the unit sphere. By invariance under rotations
of the plane of the first two coordinates, $\phi_i$ is uniform on
$[0,2\pi)$ and independent of $r_i,$ and by exchangeability of the
coordinates $\Exp r_i^2=\Exp[e_i(1)^2+e_i(2)^2]=2/d.$ Hence, for every
nonnegative measurable $g,$
\begin{align*}
 \Exp\,g(\Phi)
 &=\Exp_R\,\Exp\bigl[g(\phi_Z)\bigm|R\bigr]
  =\Exp_R\Bigl[\,\sum_{i=1}^d p_u^R(i)\,g(\phi_i)\Bigr]\\
 &=\Exp_R\Bigl[\,\sum_{i=1}^d r_i^2\cos^2\phi_i\,g(\phi_i)\Bigr]\\
 &=\sum_{i=1}^d\Exp\bigl[r_i^2\bigr]\,\Exp\bigl[\cos^2\phi_i\,g(\phi_i)\bigr]\\
 &=d\cdot\frac2d\cdot\frac1{2\pi}
   \int_0^{2\pi}\cos^2\phi\,g(\phi)\,d\phi.
\end{align*}
The second equality holds because, given $R,$ the index $Z$ has
distribution $p_u^R.$ The third uses
$p_u^R(i)=\langle u,e_i\rangle^2=r_i^2\cos^2\phi_i,$ the fourth uses the
linearity of expectation and the independence of $r_i$ and $\phi_i,$ and
the fifth uses $\Exp r_i^2=2/d$ and the uniformity of $\phi_i.$ This is the
claimed density. Finally, $r>0$ almost surely: conditioning
on $R$ as above,
\[
 \Prob[r_Z=0]
 =\Exp_R\left[\sum_{i=1}^d r_i^2\cos^2\phi_i\,\1_{\{r_i=0\}}\right]=0.
\]

By~\eqref{eq:angle}, the signs of $\langle u,e_Z\rangle$ and
$\langle v,e_Z\rangle$ differ exactly when $\cos\Phi$ and
$\cos(\Phi-\theta)$ have opposite signs, that is, up to a null set, when
$\Phi$ lies in one of the two arcs $(\pi/2,\pi/2+\theta)$ and
$(3\pi/2,3\pi/2+\theta).$ The two arcs contribute equally, so
\begin{equation}\label{eq:angular-error}
 \Prob\bigl[\sgn\langle u,e_Z\rangle\ne\sgn\langle v,e_Z\rangle\bigr]
 =\frac2\pi\int_{\pi/2}^{\pi/2+\theta}\cos^2\phi\,d\phi
 =\frac{\theta-\sin\theta\cos\theta}{\pi}.
\end{equation}

We now compute the expected divergence between $p_u^R$ and $p_v^R.$
Given $R,$ the index $Z$ has distribution $p_u^R,$ so by the definition
of relative entropy,
\[
 \klnat{p_u^R}{p_v^R}
 =\sum_{i:\,p_u^R(i)>0}p_u^R(i)\ln\frac{p_u^R(i)}{p_v^R(i)}
 =\Exp\Bigl[\ln\frac{p_u^R(Z)}{p_v^R(Z)}\Bigm|R\Bigr].
\]
By~\eqref{eq:square-coordinates} and~\eqref{eq:angle},
\[
 \frac{p_u^R(Z)}{p_v^R(Z)}
 =\frac{\langle u,e_Z\rangle^2}{\langle v,e_Z\rangle^2}
 =\frac{r^2\cos^2\Phi}{r^2\cos^2(\Phi-\theta)}
 =\frac{\cos^2\Phi}{\cos^2(\Phi-\theta)}
\]
almost surely. Here the factor $r^2$ cancels because $r>0,$ and the
denominator is nonzero almost surely because $\Phi$ has a density.
Taking logarithms and averaging over $R$ and $Z,$ we obtain
\begin{equation}\label{eq:angular-divergence}
\begin{split}
 \Exp_R \klnat{p_u^R}{p_v^R}
 &=\Exp\bigl[\ln\cos^2\Phi-\ln\cos^2(\Phi-\theta)\bigr]\\
 &=\frac1\pi\int_0^{2\pi}\cos^2\phi\,
   \bigl(\ln\cos^2\phi-\ln\cos^2(\phi-\theta)\bigr)\,d\phi
 =2\sin^2\theta,
\end{split}
\end{equation}
where the second step uses the density of $\Phi,$ and the last step is
proved in Claim~\ref{claim:calculus} below. The first two steps are
legitimate because, by the same claim, the function
$\phi\mapsto\ln\cos^2\phi$ is integrable over a period. Indeed, by the
density of $\Phi$ and periodicity, both $\Exp\abs{\ln\cos^2\Phi}$ and
$\Exp\abs{\ln\cos^2(\Phi-\theta)}$ are at most
$\frac1\pi\int_0^{2\pi}\abs{\ln\cos^2\phi}\,d\phi.$

Now fix a legal input $(x,y),$ and apply~\eqref{eq:angular-error}
and~\eqref{eq:angular-divergence} to $u=u_x$ and $v=F(x,y)v_y,$ so that
$\cos\theta=F(x,y)\langle u_x,v_y\rangle.$ The protocol uses $v_y$
rather than $v,$ since Bob does not know $F(x,y),$ but the sign
$F(x,y)=\pm1$ is harmless. Indeed, almost surely
$\sgn\langle v,e_Z\rangle=F(x,y)\sgn\langle v_y,e_Z\rangle,$ so the
error event $AB\ne F(x,y)$ is the event
$\sgn\langle u,e_Z\rangle\ne\sgn\langle v,e_Z\rangle$
of~\eqref{eq:angular-error}. Moreover, $p_v^R=p_{v_y}^R,$ because
$p_v^R$ depends on $v$ only through the squares $\langle v,e_i\rangle^2.$

We now proceed to calculate the error on $(x,y).$
By the choice of $u_x$ and $v_y,$ we have $\cos\theta\ge\sqrt{1-s},$
that is,
$\sin^2\theta\le s.$ By~\eqref{eq:angular-error}, it follows that
\begin{align*}
 \Prob[AB\ne F(x,y)]
 &=\frac{\theta-\sin\theta\cos\theta}{\pi}
  =\frac{\theta\sin^2\theta-\cos\theta\,(\sin\theta-\theta\cos\theta)}{\pi}\\
 &\le\frac{\theta\sin^2\theta}{\pi}
  \le\frac{\sin^3\theta}{2}
  \le s^{3/2}
\end{align*}
on every legal input. The first inequality holds because
$\cos\theta\ge0$ and $\sin\theta-\theta\cos\theta\ge0,$ where the latter
holds because $\sin\theta-\theta\cos\theta$ vanishes at $\theta=0$ and
has derivative $\theta\sin\theta\ge0$ on $[0,\pi/2].$ 
The second inequality amounts to
$\theta\le(\pi/2)\sin\theta,$ which holds because the sine is concave on
$[0,\pi/2]$ and hence lies above its chord from $(0,0)$ to $(\pi/2,1).$
The last inequality uses $\sin^2\theta\le s.$ 

Finally, we analyze the information cost of the proposed
protocol. For an arbitrary distribution $\mu\in\mathcal P(\dom F),$ use
$p_{v_y}^R$ as a reference distribution for the message
conditionally on $Y=y$ and $R$:
\[
 I(Z;X\mid Y,R)
 \le\Exp_{(x,y)\sim\mu,\,R}\,\kl{p_{u_x}^R}{p_{v_y}^R}
 \le\frac2{\ln2}\,s,
\]
with relative entropy in bits. The first inequality is the reference-distribution identity
\eqref{eq:kl-mixture}, applied conditionally on $Y=y$ and $R$ with
reference law $p_{v_y}^R.$ Finally, $I(Z;Y\mid X,R)=0$ because the message depends
only on $X,$ $R,$ and Alice's private randomness. The bound is uniform in
$\mu,$ proving~\eqref{eq:rounding}.
\end{proof}

\Needspace{6\baselineskip}
\begin{claim}\label{claim:calculus}
The function $\phi\mapsto\ln\cos^2\phi,$ which equals $-\infty$ at the
zeros of the cosine, is integrable over $[0,2\pi).$ Moreover, for every
$\theta\in\mathbb R,$
\[
 \frac1\pi\int_0^{2\pi}\cos^2\phi\,
 \bigl(\ln\cos^2\phi-\ln\cos^2(\phi-\theta)\bigr)\,d\phi=2\sin^2\theta.
\]
\end{claim}

\begin{proof}
The function $\ln\cos^2\phi$ is even, $\pi$-periodic and nonpositive.
By concavity, $\cos\phi\ge1-2\phi/\pi$ on $[0,\pi/2],$ so
\[
 \int_0^{2\pi}\abs{\ln\cos^2\phi}\,d\phi
 =4\int_0^{\pi/2}\bigl(-\ln\cos^2\phi\bigr)\,d\phi
 \le-8\int_0^{\pi/2}\ln\Bigl(1-\frac{2\phi}{\pi}\Bigr)\,d\phi
 =4\pi.
\]
This proves integrability, and it also shows that every integral below
converges absolutely, so the integral in the claim may be split and each
piece may be computed as an improper integral at the zeros of the cosine.

Denote by $J$ the left-hand side of the claim. We first move the shift
by $\theta$ from the logarithm to the weight. Substituting
$\phi=\psi+\theta$ in the second piece and using $2\pi$-periodicity,
\[
 \int_0^{2\pi}\cos^2\phi\,\ln\cos^2(\phi-\theta)\,d\phi
 =\int_0^{2\pi}\cos^2(\psi+\theta)\,\ln\cos^2\psi\,d\psi,
\]
and therefore
\begin{equation}\label{eq:claim-shifted}
 J=\frac1\pi\int_0^{2\pi}\bigl(\cos^2\phi-\cos^2(\phi+\theta)\bigr)
   \ln\cos^2\phi\,d\phi.
\end{equation}
Next we expand the weight. By the double-angle formula
$\cos^2t=\frac12(1+\cos2t)$ and the addition formula for the
cosine,
\[
 \cos^2\phi-\cos^2(\phi+\theta)
 =\tfrac12\bigl(\cos2\phi-\cos(2\phi+2\theta)\bigr)
 =\sin^2\theta\,\cos2\phi+\tfrac12\sin2\theta\,\sin2\phi.
\]
The second term contributes nothing to~\eqref{eq:claim-shifted}: by
periodicity, $\int_0^{2\pi}\sin2\phi\,\ln\cos^2\phi\,d\phi$ equals the
integral over $[-\pi,\pi),$ and there the integrand is odd. Hence
\[
 J=\frac{\sin^2\theta}{\pi}\int_0^{2\pi}\cos2\phi\,\ln\cos^2\phi\,d\phi.
\]

For the remaining integral, note that $\cos2\phi\ln\cos^2\phi$ is even
and $\pi$-periodic, so its integral over $[0,2\pi)$ is four times its
integral over $[0,\pi/2].$ On $[0,\pi/2),$ the function $\ln\cos^2\phi$
has derivative $-2\tan\phi.$ Integrating by parts, with
$\frac12\sin2\phi$ as an antiderivative of $\cos2\phi,$ and using
$\sin2\phi\tan\phi=2\sin^2\phi,$ we get
\[
 \int_0^{\pi/2}\cos2\phi\,\ln\cos^2\phi\,d\phi
 =\Bigl[\tfrac12\sin2\phi\,\ln\cos^2\phi\Bigr]_0^{\pi/2}
 +\int_0^{\pi/2}2\sin^2\phi\,d\phi
 =\frac\pi2.
\]
Here the integral is improper at $\pi/2,$ and the boundary term
vanishes at both endpoints: at $0$ because $\sin0=0,$ and as
$\phi\to\pi/2$ because
$\frac12\sin2\phi\ln\cos^2\phi=2\sin\phi\,(\cos\phi\ln\cos\phi)\to0,$
as $t\ln t\to0$ for $t\to0^+.$ Hence
$\int_0^{2\pi}\cos2\phi\ln\cos^2\phi\,d\phi=4\cdot\pi/2=2\pi$ and
$J=(\sin^2\theta/\pi)\cdot2\pi=2\sin^2\theta.$
\end{proof}

\subsection{Completing extraction and the characterization}\label{subsec:completion}

The lower bound in Theorem~\ref{main:characterization}
crucially rests on an additive-free extraction theorem,
which we now prove. Analogous to the extraction in
Theorem~\ref{thm:additive-conversion}, the new result
bounds the information complexity of a single copy of
$F$ with error $\epsilon/n$ in terms of the information
complexity of $F^{\oplus n}$ with error $\epsilon,$ but involves
no additive loss due to a subtle use of the split-output
measure.

\begin{theorem}[Split-output
extraction]\label{thm:extraction}
There is an absolute constant $C>0$ such that every communication problem
$F\colon\Xcal\times\Ycal\to\{-1,1,*\}$ satisfies, for every integer $n\geq1$
and every $\epsilon\in[0,1/3],$
\begin{equation}\label{eq:extraction}
  \SIC_{\epsilon/n}(F)\leq\frac C n\IC_\epsilon(F^{\oplus n}).
\end{equation}
\end{theorem}

\begin{proof}
We first prove the case of error $1/3$ in a sharper form:
\begin{equation}\label{eq:extraction-third}
  \SIC_{1/(6n)}(F)\leq\frac C n\IC_{1/3}(F^{\oplus n})
  \qquad(n\geq1).
\end{equation}
Put $I=\IC_{1/3}(F^{\oplus n}).$ We derive two upper bounds on split-output
information complexity, one from each of Lemmas~\ref{lem:block}
and~\ref{lem:rounding}, and then show that one of them always
gives~\eqref{eq:extraction-third}.

The first bound holds for every integer $m$ with $1\leq m\leq n.$ Put
$k=\lfloor n/m\rfloor.$ Since $\lfloor t\rfloor\geq t/2$ for $t\geq1,$
we have $n/2\leq km\leq n$ and $k\geq n/(2m).$ Fixing $n-km$ legal input
pairs, as in the proof of Proposition~\ref{prop:restriction-floor},
shows that $\IC_{1/3}(F^{\oplus km})\leq I.$ Lemma~\ref{lem:block}
therefore gives
\begin{equation}\label{eq:grouping-bound}
  \SIC_{1/(16m^3)}(F)\leq\frac{4800\,I}n+\frac{640\,m}n.
\end{equation}
For the second bound, put $s=1-\margin(F)^2,$ and let $c$ be the constant
of Theorem~\ref{thm:margin-ic}. Since $1-t^2\leq2\ln(1/t)$ for
$0<t\leq1,$ Theorem~\ref{thm:margin-ic} implies that
\begin{equation}\label{eq:margin-deficit}
  s\leq(2\ln2)\log\frac1{\margin(F)}\leq\frac{2\ln2}c\cdot\frac In.
\end{equation}
Applying Lemma~\ref{lem:rounding}, we obtain
\begin{equation}\label{eq:rounding-bound}
  \SIC_{s^{3/2}}(F)\leq\frac2{\ln2}\,s\leq\frac4c\cdot\frac In.
\end{equation}

It remains to choose between the two bounds. If $s^{3/2}\leq1/(6n),$
then~\eqref{eq:rounding-bound} proves~\eqref{eq:extraction-third} for every
$C\geq4/c.$ In the complementary case, $s>(6n)^{-2/3},$
and~\eqref{eq:margin-deficit} gives $n^{-2/3}<6^{2/3}s\leq(5/c)\,I/n,$
because $2\cdot6^{2/3}\ln2<5.$ We then apply~\eqref{eq:grouping-bound}
with $m=\lceil(3n/8)^{1/3}\rceil,$ the least integer such that
$1/(16m^3)\leq1/(6n).$ This choice is admissible because $3n\leq8n^3,$
so that $1\leq m\leq n.$ Since $\lceil t\rceil\leq2t$ for $t\geq1/2,$ we
also have $m\leq2(3n/8)^{1/3}=(3n)^{1/3},$ and~\eqref{eq:grouping-bound}
gives
\[
  \SIC_{1/(6n)}(F)\leq\frac{4800\,I}n+640\cdot3^{1/3}\,n^{-2/3}
  <4800\Bigl(1+\frac1c\Bigr)\frac In.
\]
Thus $C=4800(1+1/c)$ suffices in every case.

Now let $0<\epsilon\leq1/3,$ and put $r=\lfloor1/(3\epsilon)\rfloor,$ so that
$r\epsilon\leq1/3$ and $r\geq1/(6\epsilon),$ as in the proof of
Theorem~\ref{thm:additive-conversion}. Let $\pi$ be a protocol for
$F^{\oplus n}$ with worst-case error at most $\epsilon.$ By the remarks
following Lemma~\ref{lem:independent-executions}, independent executions
of $\pi$ on $r$ disjoint blocks of $n$ coordinate pairs give a protocol for
$F^{\oplus rn}$ with worst-case error at most $1/3$ and worst-case
information cost at most $r\IC(\pi).$ Hence~\eqref{eq:extraction-third},
applied with $rn$ in place of $n,$ gives
$\SIC_{1/(6rn)}(F)\leq C\IC(\pi)/n.$ Since $1/(6rn)\leq\epsilon/n,$ taking
the infimum over $\pi$ proves~\eqref{eq:extraction}. Finally, for
$\epsilon=0,$ Lemma~\ref{lem:tensor} and the comparison of the two output
conventions in Section~\ref{subsec:partial-obstruction} give
$\SIC_0(F)\leq\SIC_0(F^{\oplus n})/n\leq\IC_0(F^{\oplus n})/n,$ which
implies~\eqref{eq:extraction} because $C\geq1.$
\end{proof}

We now restate our main result for partial functions,
Theorem~\ref{main:characterization}, and complete its proof. The
function, its domain, and the number of copies may all vary; the implicit
constants do not.

\begin{restated}{Theorem~\ref*{main:characterization}}{XOR characterization
for partial functions, restated}
For every communication problem $F\colon\Xcal\times\Ycal\to\{-1,1,*\},$
every integer $n\geq1,$ and every $\epsilon\in[0,1/3],$
\begin{equation*}
  \IC_\epsilon(F^{\oplus n})
    =\Theta\!\left(\IC_\epsilon(F)+n\SIC_{\epsilon/n}(F)\right),
    \tag{\ref*{eq:shared-characterization}}
\end{equation*}
with absolute implicit constants.
\end{restated}

The baseline term $\IC_\epsilon(F)$ is necessary: a nonconstant XOR-reducible
function has zero split-output cost but, by
Proposition~\ref{main:xor-reducible}, positive announced-output
cost. The accuracy $\epsilon/n$ in the other term is also
retained, since there is no reason to expect that 
the single-copy information cost remains stable as the error decreases.

\begin{proof}[Proof of Theorem~\ref{main:characterization}]
Write $I=\IC_\epsilon(F^{\oplus n}),$ $I_1=\IC_\epsilon(F),$ and
$S=\SIC_{\epsilon/n}(F).$ Proposition~\ref{prop:restriction-floor} and
Theorem~\ref{thm:extraction} give $I\geq I_1$ and $I\geq nS/C.$ A
weighted average of these bounds yields
\[
  I\geq\frac{I_1+nS}{1+C}.
\]
For the reverse bound, run split-output protocols for $F$ with worst-case
error at most $\epsilon/n$ independently on the $n$ coordinate pairs, and let
each party output the XOR of his or her $n$ shares. As in the proof of
Proposition~\ref{prop:general-upper}, this shows that
$\SIC_\epsilon(F^{\oplus n})\leq nS,$ and the comparison of the two output
conventions in Section~\ref{subsec:partial-obstruction} gives
\[
  I\leq nS+2.
\]
By Proposition~\ref{prop:local-variation}, either
$I_1\geq1-h_2(\epsilon)\geq1-h_2(1/3)>2/25,$
in which case $I\leq25(I_1+nS),$ or every row and every column of $F,$ and
hence of $F^{\oplus n},$ is constant on its legal inputs, in which case
$I=0.$ This proves the theorem.
\end{proof}

\section{The distributional direct sum theorem}\label{sec:dist-direct-sum}

This section and the next establish our results in the distributional setting,
where errors are averaged over the input distribution.  The proofs of
Sections~\ref{sec:total} and~\ref{sec:partial} do not carry over because
their main tools, namely the minimax theorem (Fact~\ref{fact:minimax}),
majority votes, and the margin bound (Theorem~\ref{thm:margin-ic}), rely on
guarantees for every input or every input distribution. As a result, this
section and the next develop substantial additional machinery.

The standard convention for communication protocols is that the
output is a function of the public randomness and the transcript. In this
formalism, the output is ``announced,'' in the sense that both parties, and
anyone who sees the transcript, know the answer. 
We prove a multiplicative direct sum theorem under every
input distribution, with single-copy error of order $\epsilon/n,$
for \emph{local} outputs (Theorem~\ref{main:dist-direct-sum}). In this output
convention, each party at the end of the protocol computes the answer but 
need not announce it.  Our result settles the standard model of announced 
outputs as well, because announced information complexity is
within an additive constant of local-output information complexity at the
same error (Theorem~\ref{thm:local-comparison}). In particular, we obtain
a direct sum theorem for announced outputs in which the single copy is
measured with local outputs (Corollary~\ref{main:ds-announced}).

\subsection{Pricing errors and choosing a block}\label{subsec:block-choice}

As the first step of our proof, we establish a far-reaching generalization of the error
reduction step of Section~\ref{subsec:error-reduction} to the distributional
setting.  We show that, for some block of coordinates, the direct sum over
that block has a protocol whose information cost and error are proportional
to the size of the block. For this purpose, we charge errors in units of
information. Formally, for a communication problem $F$ with a finite output set, a
distribution $\mu\in\mathcal P(\dom F),$ and $\beta\geq0,$ define
\begin{equation}\label{eq:priced}
  W_\beta(F,\mu)=\inf_\pi\bigl\{\IC(\pi,\mu)
     +\beta\err_\mu(\pi,F)\bigr\},
\end{equation}
where the infimum ranges over all protocols $\pi$ for $F,$ with outputs
announced as in Section~\ref{subsec:protocols}. Each unit of error thus
costs $\beta$ bits, and we refer to $W_\beta(F,\mu)$ as the
\emph{priced information complexity} of $F.$ A protocol that announces a
fixed answer without any communication has information cost $0$
and error at most $1,$ so that $W_\beta(F,\mu)$ lies in $[0,\beta].$

In the error reduction step of Section~\ref{subsec:error-reduction}, a
fallback protocol runs after another protocol, on the conditional
distribution of the input given the public randomness and transcript of the
first protocol, and the fallback's information cost is bounded under every input
distribution by Corollary~\ref{cor:fallback}. In the distributional setting, we bound the
fallback's priced cost under
such a conditional distribution through the following proposition. Recall
from Fact~\ref{fact:concavity} that the information cost of a fixed
protocol $\pi$ is a concave function of the input distribution. The
proposition exhibits an affine function of the input distribution $\mu'$
that lies above $\IC(\pi,\mu')+\beta\err_{\mu'}(\pi,F)$ and coincides with
it at $\mu'=\mu.$ For a protocol with finite public randomness, this affine
function is given by a function on inputs, defined through relative
entropies between distributions of the pair $(R,M).$

\begin{proposition}[Cost profile]\label{prop:cost-profile}
Let $F$ be a communication problem, let $\mu\in\mathcal P(\dom F)$ and
$\beta\geq0,$ and let $\pi$ be a protocol for $F$ with finite public
randomness $R$ and transcript $M.$
For $(x,y)\in\supp\mu,$ let $Q_{xy},$ $Q_x,$ and $Q_y$ be the conditional
distributions of $(R,M)$ given $(X,Y)=(x,y),$ given $X=x,$ and given $Y=y,$
respectively, where $(X,Y)\sim\mu,$ and put
\[
  c(x,y)=\kl{Q_{xy}}{Q_x}+\kl{Q_{xy}}{Q_y}
    +\beta\Prob[\pi(x,y)\ne F(x,y)],
\]
where $\pi(x,y)$ denotes the output of $\pi$ on input $(x,y).$ Then
\begin{align*}
  \Exp_\mu c&=\IC(\pi,\mu)+\beta\err_\mu(\pi,F),\\
  \Exp_{\mu'}c&\geq\IC(\pi,\mu')+\beta\err_{\mu'}(\pi,F)
  \qquad\text{for every }\mu'\in\mathcal P(\supp\mu).
\end{align*}
We call $c$ the \emph{cost profile} of $\pi$ with respect to $\mu$ and
$\beta.$
\end{proposition}

\begin{proof}
Let $\mu'\in\mathcal P(\supp\mu),$ and let $(X,Y)\sim\mu'.$ Given
$(X,Y)=(x,y),$ the pair $(R,M)$ has distribution $Q_{xy},$ whatever the
input distribution. Fix $y$ with $\Prob[Y=y]>0;$ then $Q_y$ is defined,
because $\supp\mu'\subseteq\supp\mu.$ Conditional on $Y=y,$ the
distribution of $(R,M)$ is the average of $Q_{Xy}$ over the conditional
distribution of $X.$ The reference-distribution
identity~\eqref{eq:kl-mixture}, with reference distribution $Q_y,$ and the
nonnegativity of relative entropy therefore give
\[
  I(R,M;X\mid Y=y)\leq\Exp\bigl[\kl{Q_{XY}}{Q_Y}\bigm|Y=y\bigr],
\]
with equality if $\mu'=\mu,$ since the average of $Q_{Xy}$ is then $Q_y$
itself. Averaging over $Y$ and adding the analogous bound for
$I(R,M;Y\mid X),$ we obtain from~\eqref{eq:information-cost-R} that
\[
  \IC(\pi,\mu')\leq\Exp\bigl[\kl{Q_{XY}}{Q_X}+\kl{Q_{XY}}{Q_Y}\bigr],
\]
with equality if $\mu'=\mu.$ Adding $\beta\err_{\mu'}(\pi,F),$ the
average of $\beta\Prob[\pi(x,y)\ne F(x,y)]$ over $(x,y)\sim\mu',$ to both
sides proves both claims.
\end{proof}

We develop the error reduction mechanism in
the generality of direct sums of communication problems that may differ.
For communication problems $F_0\colon\Xcal_0\times\Ycal_0\to\Zcal_0\cup\{*\}$
and $F_1\colon\Xcal_1\times\Ycal_1\to\Zcal_1\cup\{*\},$ let $F_0\times F_1$
be the communication problem on
$(\Xcal_0\times\Xcal_1)\times(\Ycal_0\times\Ycal_1)$ whose value on
$((x_0,x_1),(y_0,y_1))$ is $(F_0(x_0,y_0),F_1(x_1,y_1))$ if
$(x_0,y_0)\in\dom F_0$ and $(x_1,y_1)\in\dom F_1,$ and $*$ otherwise. For
$\mu_0\in\mathcal P(\dom F_0)$ and $\mu_1\in\mathcal P(\dom F_1),$ let
$\mu_0\otimes\mu_1$ be the distribution of $((X_0,X_1),(Y_0,Y_1))$ for
independent $(X_0,Y_0)\sim\mu_0$ and $(X_1,Y_1)\sim\mu_1.$ After grouping
coordinates, $F^{k+k'}=F^k\times F^{k'}$ and
$\mu^{k+k'}=\mu^k\otimes\mu^{k'}$ for all integers $k,k'\geq1.$ As in
Section~\ref{subsec:error-reduction}, we index the two blocks by $0$ and
$1.$ Definition~\ref{def:standard-embedding} extends verbatim to the
distribution $\mu_0\otimes\mu_1,$ whose two blocks may have different input
sets, in place of $\nu^2.$ In the standard embedding $\pi^{(0)},$ the actual
input is placed in block $0,$ the parties publicly sample $Y_1$ from its
distribution under $\mu_1,$ and Alice privately samples $X_1$ from its
conditional distribution given $Y_1;$ in $\pi^{(1)},$ the actual input is
placed in block $1,$ the parties publicly sample $X_0,$ and Bob privately
samples $Y_0$ given $X_0.$ The only property of $\nu^2$ used in the proofs
of Fact~\ref{fact:embedding-cost} and Lemma~\ref{lem:cross-independence} is
the independence of the two blocks, so these proofs apply unchanged. Thus,
for every protocol $\pi$ with input
$((X_0,X_1),(Y_0,Y_1))\sim\mu_0\otimes\mu_1,$ public randomness $R,$ and
transcript $M,$
\begin{align}
  &\IC(\pi^{(0)},\mu_0)+\IC(\pi^{(1)},\mu_1)
     =\IC(\pi,\mu_0\otimes\mu_1),\label{eq:product-cost}\\
  &X_1\perp Y_0\mid X_0,Y_1,R,M.\label{eq:product-cross}
\end{align}

We are now in a position to prove the promised counterpart of the error reduction step,
Lemma~\ref{lem:one-step}, for priced information complexity. It bounds the
priced information complexity of $F_0\times F_1$ from below by nearly the
sum of those of $F_0$ and $F_1,$ provided that these are small
compared with $\beta.$

\begin{lemma}\label{lem:priced-product}
Let $F_0\colon\Xcal_0\times\Ycal_0\to\Zcal_0\cup\{*\}$ and
$F_1\colon\Xcal_1\times\Ycal_1\to\Zcal_1\cup\{*\}$ be communication
problems with finite output sets, let $\mu_0\in\mathcal P(\dom F_0)$ and
$\mu_1\in\mathcal P(\dom F_1),$ and let $\beta>0.$ Then, with
$W_j=W_\beta(F_j,\mu_j),$
\begin{equation}\label{eq:priced-product}
  W_\beta(F_0\times F_1,\mu_0\otimes\mu_1)
  \geq W_0+W_1-\frac{W_0W_1}\beta-2.
\end{equation}
\end{lemma}

It is helpful to recall that priced information complexity $W_\beta$ ranges
in $[0,\beta].$ The lemma states that the normalized slack $1-W_\beta/\beta$
is submultiplicative up to an additive term $2/\beta$:
\begin{equation}\label{eq:priced-slack}
  1-\frac{W_\beta(F_0\times F_1,\mu_0\otimes\mu_1)}\beta
  \leq\Bigl(1-\frac{W_0}\beta\Bigr)\Bigl(1-\frac{W_1}\beta\Bigr)
     +\frac2\beta.
\end{equation}

\begin{proof}
Fix $\eta>0.$ For $j\in\{0,1\},$ fix a protocol $\rho_j$ for $F_j$ such
that $\IC(\rho_j,\mu_j)+\beta\err_{\mu_j}(\rho_j,F_j)\leq
W_j+\eta;$ by Proposition~\ref{prop:finite-fixed}, we may take $\rho_j$ to
have finite public randomness. Let $c_j$ be its cost profile with respect
to $\mu_j$ and $\beta.$ By Proposition~\ref{prop:cost-profile},
$\Exp_{\mu_j}c_j\leq W_j+\eta;$ in particular, $c_j$ is finite on
$\supp\mu_j.$ Let $\pi$ be an arbitrary protocol for $F_0\times F_1,$ and
put $\epsilon=\err_{\mu_0\otimes\mu_1}(\pi,F_0\times F_1)$ and
$I=\IC(\pi,\mu_0\otimes\mu_1).$ We will show that
\begin{equation}\label{eq:priced-goal}
  W_0+W_1\leq I+\beta\epsilon+2+\frac{(W_0+\eta)(W_1+\eta)}\beta.
\end{equation}
Letting $\eta\to0$ and then taking the infimum over $\pi$
proves~\eqref{eq:priced-product}.

Let $((X_0,X_1),(Y_0,Y_1))\sim\mu_0\otimes\mu_1,$ let $R$ and $M$ be the public
randomness and transcript of $\pi$ on this input, and put
$V=(X_0,Y_1,R,M).$ Let $Z_0$ and $Z_1$ be the components for blocks $0$
and $1$ of the answer that $\pi$ announces, which are functions of
$(R,M).$ For $j\in\{0,1\},$ put
\[
  B_j=F_j(X_j,Y_j),\qquad
  E_j=\Prob[Z_j\ne B_j\mid V],\qquad
  C_j=\Exp[c_j(X_j,Y_j)\mid V].
\]
By~\eqref{eq:product-cross}, the random variables $X_1$ and $Y_0$ are
independent conditional on $V.$ Since $X_0$ and $Y_1$ are components of
$V,$ it follows that functions of $(V,Y_0)$ and functions of $(V,X_1)$ are
independent conditional on $V.$ In particular, $B_0$ and $B_1$ are
independent conditional on $V.$ As $Z_0$ and $Z_1$ are functions of $V,$
the protocol $\pi$ answers correctly with conditional probability
$(1-E_0)(1-E_1)$ given $V,$ and averaging over $V$ gives
\begin{equation}\label{eq:default-error}
  \Exp\bigl[1-(1-E_0)(1-E_1)\bigr]=\epsilon.
\end{equation}
Likewise,
\begin{align}
  \Exp[C_0C_1]
  &=\Exp_V\bigl[\Exp[c_0(X_0,Y_0)\mid V]\,\Exp[c_1(X_1,Y_1)\mid V]\bigr]
     \nonumber\\
  &=\Exp_V\Exp[c_0(X_0,Y_0)\,c_1(X_1,Y_1)\mid V]\nonumber\\
  &=\Exp\bigl[c_0(X_0,Y_0)\,c_1(X_1,Y_1)\bigr]\nonumber\\
  &=\Exp_{\mu_0}c_0\cdot\Exp_{\mu_1}c_1,
\label{eq:priced-cross}
\end{align}
where the second step uses the conditional independence, and the final step 
uses the independence of $(X_0,Y_0)$ and $(X_1,Y_1).$

We now construct protocols $\pi_0$ and $\pi_1$ for $F_0$ and
$F_1,$ respectively. Let $\pi^{(0)}$ and $\pi^{(1)}$ be the standard
embeddings into $\pi$ described above. As in the proof of
Lemma~\ref{lem:one-step}, Alice knows $V$ in $\pi^{(0)},$ and Bob knows $V$
in $\pi^{(1)}.$ For $j\in\{0,1\},$ let $G_j$ be the event that
$\beta E_j\leq C_j,$ which is determined by $V.$ We call $G_j$ the
\emph{good event} for block $j$: on $G_j,$ the default output $Z_j$ is
kept. Let $\sigma_j$ be the protocol that runs $\pi^{(j)},$ after which the
party that knows $V$ sends the bit $\1_{G_j}.$ Let $\pi_j$ be the protocol
that runs $\sigma_j$ and then, if $\1_{G_j}=0,$ runs $\rho_j$ on the same
input with fresh public and private randomness. The output of $\pi_j$ is
that of $\rho_j$ if $\rho_j$ is run, and $Z_j$ otherwise. As $Z_j$ is a
function of the public randomness and transcript of $\sigma_j,$ the
protocol $\pi_j$ announces its output.

We claim that, for $j\in\{0,1\},$
\begin{equation}\label{eq:priced-child}
  \IC(\pi_j,\mu_j)+\beta\err_{\mu_j}(\pi_j,F_j)
  \leq\IC(\pi^{(j)},\mu_j)+1+\Exp\bigl[\min\{\beta E_j,C_j\}\bigr].
\end{equation}
Apart from its output, $\pi_j$ is the protocol $\pi'$ of
Proposition~\ref{prop:continuation} for the protocols $\sigma_j$ and
$\rho_j$ and the set of transcripts of $\sigma_j$ in which $\1_{G_j}=0.$
When the input of $\sigma_j$ is drawn from $\mu_j,$ the completed input
on which $\pi$ is run inside $\sigma_j$ has distribution $\mu_0\otimes\mu_1,$
so this completed input, together with the public randomness and
transcript of that run of $\pi,$ has the joint distribution of
$((X_0,X_1),(Y_0,Y_1),R,M),$ and the input of $\sigma_j$ is $(X_j,Y_j).$
Let $T_j$ be the public randomness and transcript of $\sigma_j,$ and let
$\mu_{j,T_j}\in\mathcal P(\supp\mu_j)$ be the conditional distribution of
$(X_j,Y_j)$ given $T_j.$ The proof of Proposition~\ref{prop:continuation}
shows that
\[
  \IC(\pi_j,\mu_j)
  =\IC(\sigma_j,\mu_j)+\Exp\bigl[\1_{\overline G_j}\IC(\rho_j,\mu_{j,T_j})\bigr],
\]
and the error of $\pi_j$ under $\mu_j$ is the sum of
$\Exp[\1_{G_j}E_j]$ and
$\Exp[\1_{\overline G_j}\err_{\mu_{j,T_j}}(\rho_j,F_j)].$ By
Proposition~\ref{prop:cost-profile}, applied to $\rho_j$ and the
distributions $\mu_{j,T_j},$ and since the event $G_j$ is determined both by
$T_j$ and by $V,$
\begin{align*}
  \Exp\bigl[\1_{\overline G_j}\bigl(\IC(\rho_j,\mu_{j,T_j})
     +\beta\err_{\mu_{j,T_j}}(\rho_j,F_j)\bigr)\bigr]
  &\leq\Exp\bigl[\1_{\overline G_j}\Exp[c_j(X_j,Y_j)\mid T_j]\bigr]\\
  &=\Exp\bigl[\1_{\overline G_j}c_j(X_j,Y_j)\bigr]
   =\Exp\bigl[\1_{\overline G_j}C_j\bigr].
\end{align*}
Finally, $\sigma_j$ appends a one-bit message to $\pi^{(j)},$ so that
$\IC(\sigma_j,\mu_j)\leq\IC(\pi^{(j)},\mu_j)+1.$ Combining these
bounds with the identity
$\1_{G_j}\beta E_j+\1_{\overline G_j}C_j=\min\{\beta E_j,C_j\}$ proves the
claim.

To complete the proof, we use the elementary inequality
\[
  \min\{\beta e,s\}+\min\{\beta e',s'\}
  \leq\beta\bigl(1-(1-e)(1-e')\bigr)+\frac{ss'}\beta,
  \qquad e,e'\in[0,1],\quad s,s'\geq0.
\]
To verify it, put $z=\min\{e,s/\beta\}$ and $z'=\min\{e',s'/\beta\}.$
The function $(z,z')\mapsto z+z'-zz'$ is nondecreasing in each argument on
$[0,1]^2,$ so that $z+z'-zz'\leq e+e'-ee'=1-(1-e)(1-e');$ moreover,
$zz'\leq ss'/\beta^2.$ Multiplying by $\beta$ gives the inequality.
Since $W_j$ is at most the left-hand side of~\eqref{eq:priced-child},
summing over $j\in\{0,1\}$ and applying~\eqref{eq:product-cost} and the
elementary inequality with $e=E_0,$ $e'=E_1,$ $s=C_0,$ $s'=C_1,$ we obtain
\begin{align*}
  W_0+W_1
  &\leq I+2+\beta\Exp\bigl[1-(1-E_0)(1-E_1)\bigr]
     +\frac{\Exp[C_0C_1]}\beta\\
  &\leq I+2+\beta\epsilon+\frac{(W_0+\eta)(W_1+\eta)}\beta,
\end{align*}
where the second step uses~\eqref{eq:default-error},
\eqref{eq:priced-cross}, and $\Exp_{\mu_j}c_j\leq W_j+\eta.$ This
proves~\eqref{eq:priced-goal}.
\end{proof}

Iterating Lemma~\ref{lem:priced-product} over a partition
$n=k_1+k_2+\cdots+k_m$ of the $n$
coordinates into $m$ blocks, much as Lemma~\ref{lem:recursive-bound} iterates
the error reduction step, we now show that the normalized slack of $F^n$ is
at most that of some block of $k_i$ coordinates, raised to the power
$n/k_i,$ up to an additive term $2/\beta$ for each of the $m-1$ splits.

\begin{lemma}\label{lem:block-choice}
Let $F\colon\Xcal\times\Ycal\to\{-1,1,*\}$ be a communication problem, let
$\mu\in\mathcal P(\dom F)$ and $\beta>0,$ and let $n=k_1+k_2+\cdots+k_m$
for positive integers $k_1,k_2,\ldots,k_m.$ Then for some $i\in[m],$
\begin{equation}\label{eq:block-choice}
  \Bigl(1-\frac{W_\beta(F^{k_i},\mu^{k_i})}\beta\Bigr)^{n/k_i}
  \geq1-\frac{W_\beta(F^n,\mu^n)+2(m-1)}\beta.
\end{equation}
\end{lemma}

\begin{proof}
For $j\in[m],$ put $s_j=1-W_\beta(F^{k_j},\mu^{k_j})/\beta,$ so that
$0\leq s_j\leq1.$ We first show, by induction on $m,$ that
\begin{equation}\label{eq:slack-product}
  1-\frac{W_\beta(F^n,\mu^n)}\beta\leq s_1s_2\cdots s_m+\frac{2(m-1)}\beta.
\end{equation}
For $m=1,$ this holds with equality. For $m\geq2,$ grouping coordinates
gives $F^n=F^{k_1}\times F^{n-k_1}$ and $\mu^n=\mu^{k_1}\otimes\mu^{n-k_1},$
so that~\eqref{eq:priced-slack} and the inductive hypothesis for the
partition $n-k_1=k_2+\cdots+k_m$ give
\[
  1-\frac{W_\beta(F^n,\mu^n)}\beta
  \leq s_1\Bigl(1-\frac{W_\beta(F^{n-k_1},\mu^{n-k_1})}\beta\Bigr)
     +\frac2\beta
  \leq s_1\Bigl(s_2\cdots s_m+\frac{2(m-2)}\beta\Bigr)+\frac2\beta,
\]
which implies~\eqref{eq:slack-product} because $0\leq s_1\leq1.$ Now, the
product $s_1s_2\cdots s_m$ is a weighted
geometric mean of the numbers $s_j^{n/k_j},$ with weights $k_j/n$ that sum
to $1,$ and is therefore at most the largest of these numbers. Together
with~\eqref{eq:slack-product}, this proves the lemma.
\end{proof}

\subsection{Direct sum with constant additive loss}\label{subsec:additive-loss}

When a protocol for $F^n$ reveals at least a constant number of bits per copy,
Lemma~\ref{lem:block-choice} alone, applied with blocks of size $1,$ implies
the direct sum theorem. The result below is Theorem~\ref{main:ds-additive}
from the introduction, which we restate for convenience. Its constant
additive loss comes from the two bits per split in
Lemma~\ref{lem:priced-product} and is negligible in this regime.

\begin{restated}{Theorem~\ref*{main:ds-additive}}{Distributional direct sum with
constant additive loss, restated}
For every communication problem $F\colon\Xcal\times\Ycal\to\{-1,1,*\},$
every probability distribution $\mu$ on the domain of $F,$ every integer
$n\geq1,$ and every $\epsilon\in[0,1/3],$
\begin{equation*}
  \frac n4\bigl(\IC_{4\epsilon/n}(F,\mu)-8\bigr)
  \leq\IC_\epsilon(F^n,\mu^n)
  \leq n\IC_{\epsilon/n}(F,\mu).\tag{\ref*{eq:ds-additive}}
\end{equation*}
\end{restated}

\begin{proof}
For the upper bound, fix $\eta>0$ and a protocol $\pi$ for $F$ with error
at most $\epsilon/n$ under $\mu$ and
$\IC(\pi,\mu)\leq\IC_{\epsilon/n}(F,\mu)+\eta.$ Run independent copies of
$\pi$ on the $n$ coordinate pairs, and announce the tuple of the $n$
answers. Under $\mu^n,$ each coordinate pair has distribution $\mu,$ so the
union bound gives error at most $\epsilon.$ By
Lemma~\ref{lem:independent-executions}, the information cost is at most
$n(\IC_{\epsilon/n}(F,\mu)+\eta).$ Letting $\eta\to0$ proves the upper
bound.

For the lower bound, let $\pi$ be a protocol for $F^n$ with error at most
$\epsilon$ under $\mu^n,$ and put $I=\IC(\pi,\mu^n).$ It suffices to show
that $\IC_{4\epsilon/n}(F,\mu)\leq4I/n+8,$ since taking the infimum over
$\pi$ then proves the lower bound. Suppose first that $\epsilon=0.$
Let $\rho$ be the
protocol that publicly chooses a uniformly random $i\in[n],$ runs the
standard embedding $\pi^{(i)}$ into coordinate $i$ of $\pi$ with respect to
$\mu$ (Definition~\ref{def:standard-embedding}), and outputs the component
for coordinate $i$ of the answer that $\pi$ announces. When the input of
$\rho$ is drawn from $\mu,$ the completed input has distribution $\mu^n,$
so $\rho$ has error $0$ under $\mu,$ and its information cost is $I/n$
by~\eqref{eq:public-mixture-cost} and Fact~\ref{fact:embedding-cost}.
Hence $\IC_0(F,\mu)\leq I/n.$

Now let $\epsilon>0,$ and put $\beta=(I+2n)/\epsilon.$ The protocol $\pi$
shows that $W_\beta(F^n,\mu^n)\leq I+\beta\epsilon,$ so that
Lemma~\ref{lem:block-choice}, applied with $n$ blocks of size $1,$ gives
\[
  \Bigl(1-\frac{W_\beta(F,\mu)}\beta\Bigr)^n
  \geq1-\epsilon-\frac{I+2(n-1)}\beta
  >1-2\epsilon.
\]
Since $(1-t)^n\leq e^{-tn}$ for $0\leq t\leq1,$ it follows that
\[
  \frac{W_\beta(F,\mu)}\beta
  <\frac1n\ln\frac1{1-2\epsilon}
  \leq\frac{3\ln3}n\,\epsilon
  <\frac{4\epsilon}n,
\]
where the middle step holds because the function
$t\mapsto\ln(1/(1-2t))$ is convex on $[0,1/3],$ vanishes at $0,$ and equals
$\ln3$ at $1/3.$ Hence
there is a protocol $\sigma$ for $F$ such that
\[
  \IC(\sigma,\mu)+\beta\err_\mu(\sigma,F)
  <\frac{4\epsilon\beta}n=\frac{4I}n+8.
\]
In particular, $\err_\mu(\sigma,F)<4\epsilon/n$ and $\IC(\sigma,\mu)<4I/n+8,$
which proves our claim.
\end{proof}

The additive term in~\eqref{eq:ds-additive} is negligible when a single
copy requires a constant number of bits. For example, if
$\IC_{4\epsilon/n}(F,\mu)\geq16,$ then~\eqref{eq:ds-additive} gives
$
  n\IC_{4\epsilon/n}(F,\mu)/8
  \leq\IC_\epsilon(F^n,\mu^n)
  \leq n\IC_{\epsilon/n}(F,\mu).
$
As we will see shortly, the additive term cannot be removed as long as the 
single copy is measured with announced outputs (Theorem~\ref{thm:announced-counterexample}).
The rest of this section removes the additive term altogether by
working in the model with local outputs
(Theorem~\ref{main:dist-direct-sum}).

\subsection{Local outputs}\label{subsec:local-outputs}

The protocols in Lemma~\ref{lem:block-choice} announce their outputs. To
efficiently extract a single copy from such a protocol, we will instead let each
party settle on an answer of his or her own, based on checks that only that
party can perform. Thus, we still require each party to compute the answer,
but we do not require the answer to be a function of the public randomness
and transcript. Let $F\colon\Xcal\times\Ycal\to\Zcal\cup\{*\}$ be a
communication problem with a finite output set $\Zcal.$ A
\emph{local-output protocol} $\pi$ for $F$ ends with Alice holding an
answer $A\in\Zcal$ and Bob holding an answer $B\in\Zcal,$ each a function
of that party's input, private randomness, and $T=(R,M),$ the public
randomness and transcript of $\pi.$ Its error under $\mu\in\mathcal P(\dom
F)$ is $\err_\mu(\pi,F)=1-\Prob[A=B=F(X,Y)],$ where $(X,Y)\sim\mu.$ The
\emph{local-output information complexity} is
\begin{equation}\label{eq:lic}
  \LIC_\epsilon(F,\mu)
    =\inf_{\pi:\,\err_\mu(\pi,F)\leq\epsilon}\IC(\pi,\mu),
\end{equation}
where the infimum ranges over local-output protocols for $F.$ Local-output protocols are a
natural formalism and have appeared in previous work. For example, Barak,
Braverman, Chen, and Rao~\cite{BBCR10} consider protocols that end with each
party outputting a value computed from his or her input and the transcript,
and that succeed if both values are correct. Bauer, Moran, and
Yehudayoff~\cite{BMY15} call this convention internal computation, as
opposed to external computation in which the transcript determines the
value, and they observe that the two notions coincide for deterministic protocols that compute a total function. Fontes, Laplante, Lauri\`ere, 
and Nolin~\cite{FLLN23} call it the local model.

\begin{remark}\label{rem:local-outputs}
Individual outputs by Alice and Bob, such as local answers and the shares
of Definition~\ref{def:split-output}, may be assumed to be functions of the
respective party's input and of $T=(R,M),$ the public randomness and
transcript.  Indeed, let $A$ be an output of Alice that depends on her
input, her private randomness, and $T,$ and let $B$ be an analogous output
of Bob.  Recall from Section~\ref{subsec:protocols} that conditional on
$(X,Y,T)=(x,y,t),$ the outputs $A$ and $B$ are independent, with
conditional distributions that depend only on $(x,t)$ and $(y,t),$
respectively. Let the parties choose two additional random seeds publicly,
and let Alice sample her output from its conditional distribution given
$(x,t)$ using the first seed, and Bob from his given $(y,t)$ using the
second. This preserves the joint distribution of $(X,Y,T,A,B),$ and since
the seeds are independent of everything else, it does not change the
information cost either. We include such seeds in $T$ whenever they are
used.  
\end{remark}

An announced answer can serve as a local answer for both Alice and Bob, so that
$\LIC_\epsilon(F,\mu)\leq\IC_\epsilon(F,\mu).$ Conversely, local answers
that agree with high probability can be announced at a small cost, no
matter how large the set of possible answers. For this, the parties compare
their answers bit by bit after a public random relabeling, and the
comparison reveals only the position of the first disagreement.  This is
the idea behind Braverman's zero-error protocol for the equality function
with constant information cost~\cite[Prop.~3.21]{Braverman15}, which
compares random linear hashes of the two inputs one at a time.

\begin{lemma}\label{lem:announcement}
Let $\pi$ be a protocol at the end of which Alice holds $A(X,T)$ and Bob
holds $B(Y,T),$ where $A$ and $B$ take values in a finite set $\Zcal$ and
$T=(R,M)$ consists of the public randomness and transcript of $\pi.$ Let
$p=\Prob[A\ne B]$ under a given input distribution. Then $\pi$ can be
extended by a continuation that announces $A$ whenever $A=B,$ at an
additional information cost of at most $2h_2(p)+4p$ under that
distribution.
\end{lemma}

\begin{proof}
If $|\Zcal|=1,$ there is nothing to prove. Otherwise, fix an injective
encoding of $\Zcal$ by strings in $\{0,1\}^b,$ where
$b=\lceil\log|\Zcal|\rceil.$ The parties publicly choose a uniformly random
permutation $R'$ of $\{0,1\}^b$ and compare the strings $R'(A)$ and $R'(B)$
bit by bit: at each step, Alice sends her next bit, and Bob replies whether
his bit agrees. At the first disagreement, the parties stop and announce a
fixed element of $\Zcal.$ If all $b$ bits agree, Alice has sent all of
$R'(A),$ and the parties announce $A.$

Let $J$ be the first position at which $R'(A)$ and $R'(B)$ differ, with
$J=\infty$ if $A=B,$ and let $M'$ be the transcript of the continuation.
Before position $J$ the two strings agree, and at position $J$ they differ,
so $M'$ is a function of $(A,R',J)$ and also of $(B,R',J).$ Since $A$ is a
function of $(X,T)$ and $B$ is a function of $(Y,T),$ it follows that $M'$
is determined by $J$ given $(X,T,R'),$ and also given $(Y,T,R').$ Here it
matters that the outputs do not depend on the parties' private
randomness; for a local-output protocol, Remark~\ref{rem:local-outputs}
removes this dependence. As $R'$ is independent of everything else, the
continuation increases the information cost by
\[
  I(M';X\mid Y,T,R')+I(M';Y\mid X,T,R')
  \leq H(J\mid Y,T,R')+H(J\mid X,T,R')
  \leq2H(J\mid T).
\]
It remains to bound $H(J\mid T).$ Write $p_T=\Prob[A\ne B\mid T].$ On the
event $A=B,$ we have $J=\infty,$ so the conditional entropy of $J$ given
$T$ is $h_2(p_T)$ plus $p_T$ times the entropy of $J$ given $T$ and the
event $A\ne B.$ Conditional on $T$ and $A\ne B,$ the images of two distinct
strings under $R'$ form a uniformly random pair of distinct strings, so
that $\Prob[J=j]=2^{-j}/(1-2^{-b})$ for $1\leq j\leq b.$ This distribution
has entropy $\Exp J+\log(1-2^{-b})<\Exp J<2.$ Hence
$H(J\mid T)\leq\Exp[h_2(p_T)+2p_T]\leq h_2(p)+2p$ by the concavity of
$h_2.$
\end{proof}

We now use Lemma~\ref{lem:announcement} to prove~\eqref{eq:local-comparison}
from the introduction, which compares local and announced outputs.

\begin{theorem}[Local versus announced outputs]\label{thm:local-comparison}
Let $F\colon\Xcal\times\Ycal\to\Zcal\cup\{*\}$ be a communication problem
with a finite output set, let $\mu\in\mathcal P(\dom F),$ and let
$0\leq\epsilon\leq1/2.$ Then
\begin{equation*}
  \LIC_\epsilon(F,\mu)\leq\IC_\epsilon(F,\mu)
  \leq\LIC_\epsilon(F,\mu)+2h_2(\epsilon)+4\epsilon.
\end{equation*}
\end{theorem}

\begin{proof}
The first inequality was noted above. For the
second, assume, as we may by Remark~\ref{rem:local-outputs}, that the
answers of a local-output protocol are functions of the respective party's
input and $T,$ and apply
Lemma~\ref{lem:announcement}. The announced answer is correct whenever
$A=B=F(X,Y),$ so the error does not increase. The two answers differ with
probability at most $\epsilon,$ and the function $t\mapsto2h_2(t)+4t$ is
increasing on $[0,1/2],$ so the additional cost is at most
$2h_2(\epsilon)+4\epsilon.$
\end{proof}

We conclude this subsection with an example showing that, with announced
outputs, a truly multiplicative direct sum theorem fails even for a
function on two bits. In this example, the input bits of the two parties
differ with probability $\epsilon/n.$ For $n$ copies, Alice knows all $n$
answers, and Bob guesses them correctly from his own inputs with
probability at least $1-\epsilon,$ so that, by
Theorem~\ref{thm:local-comparison}, the answers can be announced at a cost
of less than $4.$ A single copy with error of order $\epsilon/n,$ on the
other hand, has information cost larger than $\epsilon/n$ by a factor
logarithmic in $n.$

\begin{theorem}[Counterexample for announced outputs]
\label{thm:announced-counterexample}
Let $F\colon\{0,1\}\times\{0,1\}\to\{-1,1\}$ be given by $F(x,y)=(-1)^x.$
Let $\epsilon\in(0,1/3],$ let $n\geq1$ be an integer, and let $\mu$ be the
probability distribution on $\{0,1\}\times\{0,1\}$ that assigns
probability $(1-\epsilon/n)/2$ to each of $(0,0)$ and $(1,1),$ and
probability $\epsilon/(2n)$ to each of $(0,1)$ and $(1,0).$ Then
\begin{align}
  &\IC_\epsilon(F^n,\mu^n)<4,\label{eq:announced-many}\\
  &n\IC_{C\epsilon/n}(F,\mu)\geq\epsilon\log\frac n{12C\epsilon}
    \qquad\text{for every }C\geq1.\label{eq:announced-single}
\end{align}
\end{theorem}

\begin{proof}
Let $(X,Y)\sim\mu.$ Then $X$ and $Y$ are uniformly random bits, and
$\Prob[X\ne Y]=\epsilon/n.$

For~\eqref{eq:announced-many}, consider the local-output protocol for
$F^n$ without communication in which Alice outputs
$((-1)^{x_1},\ldots,(-1)^{x_n})$ and Bob outputs
$((-1)^{y_1},\ldots,(-1)^{y_n}).$ Alice's answer is always correct, and
Bob's answer is correct unless $x_i\ne y_i$ for some $i.$ Under $\mu^n,$
the union bound shows that this happens with probability at most
$\epsilon,$ so that $\LIC_\epsilon(F^n,\mu^n)=0.$
Theorem~\ref{thm:local-comparison}, applied to $F^n$ and $\mu^n,$ now
gives $\IC_\epsilon(F^n,\mu^n)\leq2h_2(\epsilon)+4\epsilon,$ which is less
than $4$ because the function $t\mapsto2h_2(t)+4t$ is increasing on
$[0,1/2]$ and $2h_2(1/3)+4/3<4.$

For~\eqref{eq:announced-single}, fix $C\geq1,$ put $\delta=C\epsilon/n,$
and let $\pi$ be a protocol for $F$ with $\err_\mu(\pi,F)\leq\delta.$ By
Proposition~\ref{prop:finite-fixed}, we may assume that $\pi$ has finite
public randomness. For $x,y\in\{0,1\},$ let $p_{xy},$ $p_x,$ and $q_y$ be
the conditional distributions of the output of $\pi$ given $(X,Y)=(x,y),$
given $X=x,$ and given $Y=y,$ respectively. As the output is a function of
the public randomness and transcript, these distributions are the images
of the distributions $Q_{xy},$ $Q_x,$ and $Q_y$ in
Proposition~\ref{prop:cost-profile} under a common map. That proposition,
with $\beta=0,$ and the data-processing inequality therefore give
\begin{equation}\label{eq:announced-output-cost}
  \IC(\pi,\mu)\geq\Exp\bigl[\kl{p_{XY}}{p_X}+\kl{p_{XY}}{q_Y}\bigr].
\end{equation}

As $X$ is uniformly random, $\pi$ errs with probability at most $2\delta$
given $X=x,$ so that $p_x(-(-1)^x)\leq2\delta.$ Likewise, $\pi$ errs with
probability at most $2\delta$ given $Y=y,$ and given $Y=y,$ we have $X=y$
except with probability $\epsilon/n.$ Hence $q_y(-(-1)^y)\leq s,$ where
$s=\epsilon/n+2\delta.$ Now let $x\ne y.$ Then $F(x,y)=(-1)^x=-(-1)^y,$
and these bounds show that $p_x(z)q_y(z)\leq s$ for both $z\in\{-1,1\}.$
Since $p_{xy}$ is a distribution on two points, it follows that
\[
  \kl{p_{xy}}{p_x}+\kl{p_{xy}}{q_y}
  =\sum_zp_{xy}(z)\Bigl(2\log p_{xy}(z)+\log\frac1{p_x(z)q_y(z)}\Bigr)
  \geq-2+\log\frac1s,
\]
where the sum ranges over the $z$ with $p_{xy}(z)>0.$ As relative entropy
is nonnegative and $\Prob[X\ne Y]=\epsilon/n,$
\eqref{eq:announced-output-cost} now gives
\[
  \IC(\pi,\mu)\geq\frac\epsilon n\log\frac1{4s}
  \geq\frac\epsilon n\log\frac n{12C\epsilon},
\]
where the second step uses $4s=(4+8C)\epsilon/n\leq12C\epsilon/n.$ Taking
the infimum over $\pi$ proves~\eqref{eq:announced-single}.
\end{proof}

For constants $\epsilon\in(0,1/3]$ and $C\geq1,$ the right-hand side
of~\eqref{eq:announced-single} tends to infinity with $n.$ Hence, there is
no constant $c>0$ such that
$\IC_\epsilon(F^n,\mu^n)\geq cn\IC_{C\epsilon/n}(F,\mu)$ for every
$n\geq1$ and every input distribution $\mu.$ For $C=4,$ this shows that
the additive term in Theorem~\ref{main:ds-additive} cannot be removed.

\subsection{Extraction with local outputs}\label{subsec:local-extraction}

Consider a local-output protocol for $F^n$ with error $\epsilon$
under $\mu^n.$ In this subsection, we extract from it a local-output
protocol for a single copy of $F,$ dividing the information cost and the
error by a factor of order $n.$ We need two lemmas, the first of which is a distributional
counterpart of Proposition~\ref{prop:local-variation}: unless each party's
input determines the answer, the information complexity of $F^n$ grows
linearly in $n.$

\begin{lemma}\label{lem:input-only}
Let $F\colon\Xcal\times\Ycal\to\{-1,1,*\}$ be a communication problem,
let $\mu\in\mathcal P(\dom F),$ and let $p_A$ and $p_B$ be the least errors
of any predictions of $F(X,Y)$ based on $X$ and on $Y,$ respectively, where
$(X,Y)\sim\mu.$ Then for every integer $n\geq1,$
\[
  p_A+p_B\leq\frac{3\ln2}{2n}\,\bigl(\LIC_{1/3}(F^n,\mu^n)+2\bigr).
\]
\end{lemma}

\begin{proof}
Let $\pi$ be a local-output protocol for $F^n$ with error at most $1/3$
under $\mu^n,$ and put $I=\IC(\pi,\mu^n).$ Let $(X,Y)\sim\mu^n,$ with
coordinate pairs $(X_i,Y_i),$ and let $T=(R,M)$ consist of the public
randomness and transcript of $\pi$ on input $(X,Y).$ By
Remark~\ref{rem:local-outputs}, we may assume that
Alice's answer is a function $A(X,T)$ of her input and $T,$ and then, by
Proposition~\ref{prop:finite-fixed}, that $\pi$ has finite public
randomness. Let $P$ be the
distribution of $(X,Y,T),$ and let $Q$ be the distribution of $(X,Y,T'),$
where, conditional on $(X,Y),$ the random variable $T'$ has the
conditional distribution of $T$ given $X.$ Then
$\kl PQ=I(T;Y\mid X)=I(M;Y\mid X,R),$ where the second step holds because
$R$ is independent of $(X,Y).$

Let $E$ be the event that Alice's answer equals the value of $F^n$ on the
input, so that $P(E)\geq2/3.$ Under $Q,$ the random variable $T'$ is
conditionally independent of $Y$ given $X,$ so $A(X,T')$ is a randomized
prediction of $F^n(X,Y)$ based on $X.$ Conditional on $X,$ the values
$F(X_1,Y_1),\ldots,F(X_n,Y_n)$ are independent, and the probability that
any prediction of $F(X_i,Y_i)$ is correct is at most $1-e(X_i),$ where
$e(x)$ is the least error of a prediction of $F(X_1,Y_1)$ given $X_1=x.$
Since $X_1,\ldots,X_n$ are independent and $\Exp e(X_1)=p_A,$ it follows
that $Q(E)\leq\Exp\prod_{i=1}^n(1-e(X_i))=(1-p_A)^n.$ By the
data-processing inequality for relative entropy, applied to the indicator
of $E,$
\[
  I(M;Y\mid X,R)
  \geq P(E)\log\frac{P(E)}{Q(E)}
     +P(\overline E)\log\frac{P(\overline E)}{Q(\overline E)}
  \geq P(E)\log\frac1{Q(E)}-h_2(P(E)),
\]
where the last step uses $Q(\overline E)\leq1.$ Since $P(E)\geq2/3,$
$Q(E)\leq(1-p_A)^n,$ and $h_2\leq1,$ we conclude that
\[
  I(M;Y\mid X,R)\geq\frac{2n}3\log\frac1{1-p_A}-1
  \geq\frac{2n}{3\ln2}\,p_A-1.
\]
An analogous argument for Bob's answer gives
$I(M;X\mid Y,R)\geq(2n/(3\ln2))\,p_B-1.$ Adding these bounds, we obtain
$I\geq(2n/(3\ln2))(p_A+p_B)-2.$ Taking the infimum over $\pi$ proves the
lemma.
\end{proof}

The second lemma is an elementary variance bound, which we will apply
again in Section~\ref{subsec:vector-lift}. It concerns the average, over the
coordinates of a product distribution, of the conditional expectation of a
bounded random variable given one coordinate.

\begin{lemma}\label{lem:variance}
Let $\mu\in\mathcal P(\Xcal\times\Ycal),$ let $n\geq1$ be an integer, and
let $(X,Y)\sim\mu^n,$ with coordinate pairs $(X_i,Y_i).$ Let $L$ be a
random variable with values in $[0,1]$ on the same probability space. Put
$\epsilon=\Exp L,$ and define a function $\bar\epsilon$ on $\supp\mu$ by
\[
  \bar\epsilon(x,y)=\frac1n\sum_{i=1}^n\Exp\bigl[L\bigm|(X_i,Y_i)=(x,y)\bigr].
\]
Then $\Exp_\mu\bar\epsilon=\epsilon$ and $\Var_\mu(\bar\epsilon)\leq\epsilon/n.$
\end{lemma}

\begin{proof}
The identity holds because each $(X_i,Y_i)$ has distribution $\mu.$ For
the inequality, put $f_i(x,y)=\Exp[L\mid(X_i,Y_i)=(x,y)]-\epsilon,$ so that
$\bar\epsilon-\epsilon=\frac1n\sum_if_i.$ The random variables $f_i(X_i,Y_i)$ have
mean zero and, by the independence of the coordinate pairs, are pairwise
orthogonal. Moreover, $f_i(X_i,Y_i)=\Exp[L-\epsilon\mid X_i,Y_i],$ so that
$\Exp[(L-\epsilon)\,f_i(X_i,Y_i)]=\Exp[f_i(X_i,Y_i)^2].$ Hence
\[
  0\leq\Exp\Bigl[\Bigl(L-\epsilon-\sum_{i=1}^nf_i(X_i,Y_i)\Bigr)^2\Bigr]
  =\Exp\bigl[(L-\epsilon)^2\bigr]-\sum_{i=1}^n\Exp\bigl[f_i(X_i,Y_i)^2\bigr],
\]
and therefore $\sum_i\Exp[f_i(X_i,Y_i)^2]\leq\Exp[(L-\epsilon)^2]\leq\epsilon,$
where the last step uses $L^2\leq L.$ By the convexity of the square,
$(\bar\epsilon-\epsilon)^2\leq\frac1n\sum_if_i^2$ pointwise. Taking expectations
under $\mu,$ we conclude that
$\Var_\mu(\bar\epsilon)\leq\frac1n\sum_i\Exp[f_i(X_i,Y_i)^2]\leq\epsilon/n.$
\end{proof}

We now carry out the extraction, as follows.  The standard embedding into a
uniformly random coordinate $i$ of a protocol $\sigma$ for $F^n$ divides
the information cost by $n$ (Fact~\ref{fact:embedding-cost}) but does not
reduce the error, because $\sigma$ may err on many coordinates at once. To
reduce the error, we have each party check the answers of $\sigma$ on
the coordinates where he or she knows both inputs, namely, Alice on the
coordinates after $i$ and Bob on those before $i,$ and keep the answer for
coordinate $i$ only if the check passes. If Alice's check passes but her
answer for coordinate $i$ is wrong, then $i$ is the last coordinate on
which her answers are wrong, which has probability at most $\epsilon/n$
because $i$ is uniformly random; the same holds for Bob, with the first
coordinate in place of the last. The parties run two independent copies of
this procedure. Each party keeps the answer from the first copy in which
his or her check passes and, if both checks fail, falls back on the best
prediction of $F(X,Y)$ based on his or her own input. The analysis relies
on the two lemmas just proven.

\Needspace{12\baselineskip}
\begin{lemma}\label{lem:two-checks}
Let $F\colon\Xcal\times\Ycal\to\{-1,1,*\}$ be a communication problem, let
$\mu\in\mathcal P(\dom F),$ and let $p_A$ and $p_B$ be as in
Lemma~\ref{lem:input-only}. Let $n\geq1$ be an integer, and let $\sigma$ be a
local-output protocol for $F^n$ with error $\epsilon$ under $\mu^n.$ Then
there is a local-output protocol $\pi$ for $F$ such that
\begin{align*}
  \IC(\pi,\mu)&\leq\frac{2\IC(\sigma,\mu^n)}n,\\
  \err_\mu(\pi,F)&\leq2\epsilon^2(p_A+p_B)+\frac{8\epsilon}n.
\end{align*}
\end{lemma}

\begin{proof}
Let $(X,Y)\sim\mu^n,$ with coordinate pairs $(X_i,Y_i),$ and let
$A=(A_1,\ldots,A_n)$ and $B=(B_1,\ldots,B_n)$ be the answers of Alice and
Bob in $\sigma$ on input $(X,Y).$ Let $E$ be the event that $A$ or $B$
differs from $F^n(X,Y),$ so that $\Prob[E]=\epsilon.$ For $i\in[n],$
consider the standard embedding $\sigma^{(i)}$ into coordinate $i$ of
$\sigma$ with respect to $\mu$ (Definition~\ref{def:standard-embedding}).
In $\sigma^{(i)},$ Alice knows the coordinate pairs $(X_\ell,Y_\ell)$ with
$\ell>i,$ and Bob knows those with $\ell<i.$ We say that Alice's check
passes if $A_\ell=F(X_\ell,Y_\ell)$ for every $\ell>i,$ and that Bob's check
passes if $B_\ell=F(X_\ell,Y_\ell)$ for every $\ell<i.$ If Alice's check
passes and $A_i\ne F(X_i,Y_i),$ then $i$ is the largest index at which $A$
differs from $F^n(X,Y);$ likewise, if Bob's check passes and
$B_i\ne F(X_i,Y_i),$ then $i$ is the smallest index at which $B$ differs
from $F^n(X,Y).$

Let $\rho$ be the protocol that publicly chooses a uniformly random
$i\in[n]$ and runs $\sigma^{(i)}$ on its input. By~\eqref{eq:public-mixture-cost}
and Fact~\ref{fact:embedding-cost}, $\IC(\rho,\mu)=\IC(\sigma,\mu^n)/n.$
When the input of $\rho$ is drawn from $\mu,$ the completed input and the
run of $\sigma$ have the same joint distribution for every $i,$ so that
they are independent of $i.$ By the previous paragraph, the probability
that Alice's check passes and $A_i\ne F(X_i,Y_i)$ is therefore at most
$\Prob[E]/n=\epsilon/n,$ and the same bound holds for Bob. Moreover, a
check fails only if $E$ occurs. Fix $(x,y)\in\supp\mu,$ and run
$\rho$ on input $(x,y).$ Conditional on $i,$ the completed input, together
with the answers of the run of $\sigma,$ has the conditional distribution
of $(X,Y,A,B)$ given $(X_i,Y_i)=(x,y).$ Hence the probability that at
least one of the two checks fails is at most
\[
  \bar\epsilon(x,y)=\frac1n\sum_{i=1}^n\Prob\bigl[E\bigm|(X_i,Y_i)=(x,y)\bigr].
\]
Lemma~\ref{lem:variance}, applied to the coordinate pairs $(X_i,Y_i)$ and
the indicator of $E,$ gives $\Exp_\mu\bar\epsilon=\epsilon$ and
$\Var_\mu(\bar\epsilon)\leq\epsilon/n.$

Now let $\pi$ be the protocol that runs two independent copies of $\rho$
on its input. Each party outputs his or her answer for the actual
coordinate in the first copy in which his or her check passes, and, if
both checks fail, the best prediction of $F(x,y)$ based on his or her own
input. The two parties may use different copies, and no further
communication takes place. By Lemma~\ref{lem:independent-executions},
$\IC(\pi,\mu)\leq2\IC(\rho,\mu)=2\IC(\sigma,\mu^n)/n.$

It remains to bound the error. Alice's final answer is wrong only if, in
one of the two copies, her check passes and her answer for the actual
coordinate is wrong, or both of her checks fail and her prediction is
wrong. On input $(x,y)\in\supp\mu,$ the two copies are independent, so the
latter event has probability at most $\bar\epsilon(x,y)^2e_A(x,y),$ where
$e_A(x,y)\in\{0,1\}$ indicates that Alice's prediction is wrong on $(x,y).$
Defining $e_B$ analogously for Bob and applying the union bound, we obtain
\[
  \err_\mu(\pi,F)\leq\frac{4\epsilon}n+\Exp_\mu\bigl[\bar\epsilon^2(e_A+e_B)\bigr].
\]
Since $\bar\epsilon^2\leq2\epsilon^2+2(\bar\epsilon-\epsilon)^2,$ $e_A+e_B\leq2,$
$\Exp_\mu e_A=p_A,$ and $\Exp_\mu e_B=p_B,$ it follows that
\[
  \Exp_\mu\bigl[\bar\epsilon^2(e_A+e_B)\bigr]
  \leq2\epsilon^2(p_A+p_B)+4\Var_\mu(\bar\epsilon)
  \leq2\epsilon^2(p_A+p_B)+\frac{4\epsilon}n.
\]
Combining the last two displays proves the lemma.
\end{proof}

\subsection{Optimal distributional direct sum theorem}\label{subsec:dist-direct-sum}

We are now in a position to establish the distributional direct sum
theorem from the introduction, which we restate for convenience. Its proof
distinguishes three regimes. Protocols that reveal at least one bit per copy
are handled by Theorem~\ref{main:ds-additive}, after their answers are
announced via Lemma~\ref{lem:announcement}. For the others, we extract a
single copy by Lemma~\ref{lem:two-checks}, either directly from the given
protocol, if it reveals at most one bit, or from a block chosen by
Lemma~\ref{lem:block-choice}.

\begin{restated}{Theorem~\ref*{main:dist-direct-sum}}{Distributional
direct sum theorem, restated}
For every communication problem $F\colon\Xcal\times\Ycal\to\{-1,1,*\},$
every probability distribution $\mu$ on the domain of $F,$ every integer
$n\geq1,$ and every $\epsilon\in[0,1/100],$
\begin{equation*}
  \frac n{32}\LIC_{32\epsilon/n}(F,\mu)
    \leq\LIC_\epsilon(F^n,\mu^n)
    \leq n\LIC_{\epsilon/n}(F,\mu).\tag{\ref*{eq:dist-direct-sum}}
\end{equation*}
Moreover,
\begin{equation*}
  \LIC_0(F^n,\mu^n)=n\LIC_0(F,\mu).\tag{\ref*{eq:dist-direct-sum-zero}}
\end{equation*}
\end{restated}

\begin{proof}
For the upper bound, fix $\eta>0$ and a local-output protocol $\pi$ for
$F$ with error at most $\epsilon/n$ under $\mu$ and
$\IC(\pi,\mu)\leq\LIC_{\epsilon/n}(F,\mu)+\eta.$ Run independent copies of
$\pi$ on the $n$ coordinate pairs, and let each party's answer be the
tuple of his or her $n$ answers. Under $\mu^n,$ each coordinate pair has
distribution $\mu,$ so the union bound gives error at most $\epsilon.$ By
Lemma~\ref{lem:independent-executions}, the information cost is at most
$n(\LIC_{\epsilon/n}(F,\mu)+\eta).$ Letting $\eta\to0$ proves the upper
bound.

Next, let $\sigma$ be a local-output protocol for $F^n$ with error $0$
under $\mu^n,$ and let $\rho$ be the protocol that publicly chooses a
uniformly random $i\in[n]$ and runs $\sigma^{(i)},$ in which each party
outputs his or her answer for coordinate $i.$ When the input of $\rho$ is
drawn from $\mu,$ the completed input has distribution $\mu^n,$ so $\rho$
has error $0$ under $\mu,$ and its information cost is $\IC(\sigma,\mu^n)/n$
by~\eqref{eq:public-mixture-cost} and Fact~\ref{fact:embedding-cost}. Hence
$\LIC_0(F,\mu)\leq\LIC_0(F^n,\mu^n)/n.$ Together with the upper bound, this
proves~\eqref{eq:dist-direct-sum-zero} and the lower bound
in~\eqref{eq:dist-direct-sum} for $\epsilon=0.$

Now let $0<\epsilon\leq1/100,$ let $\pi$ be a local-output protocol for $F^n$
with error at most $\epsilon$ under $\mu^n,$ and put $I=\IC(\pi,\mu^n).$ It
suffices to show that $\LIC_{32\epsilon/n}(F,\mu)\leq32I/n,$ since taking
the infimum over $\pi$ then proves the lower bound. By
Remark~\ref{rem:local-outputs}, the answers of $\pi$ may be assumed to be
functions of the respective party's input and of the public randomness and
transcript, so Lemma~\ref{lem:announcement} extends $\pi$ to a protocol
$\pi'$ for $F^n$ that announces Alice's answer whenever the two answers
agree. In particular, $\pi'$ is correct whenever both answers are, so its
error under $\mu^n$ is at most $\epsilon.$ The two answers of $\pi$ differ
with probability at most $\epsilon\leq1/100,$ and the function
$t\mapsto2h_2(t)+4t$ is increasing on $[0,1/2],$ so $I'=\IC(\pi',\mu^n)$
satisfies
\[
  I'\leq I+2h_2\Bigl(\frac1{100}\Bigr)+\frac4{100}<I+\frac14.
\]

Suppose first that $I\geq n,$ which is the high-information regime. As
$\pi'$ shows that $\IC_\epsilon(F^n,\mu^n)\leq I',$ the lower bound
in~\eqref{eq:ds-additive} gives
\[
  \LIC_{32\epsilon/n}(F,\mu)\leq\IC_{4\epsilon/n}(F,\mu)
  \leq\frac{4I'}n+8<\frac{5I}n+\frac{8I}n<\frac{32I}n,
\]
where the third step uses $1\leq n\leq I.$ This proves our claim.

It remains to treat the case $I<n.$ Let $p_A$ and $p_B$
be as in Lemma~\ref{lem:input-only}, which gives
\begin{equation}\label{eq:input-only-bound}
  p_A+p_B\leq\frac{3\ln2}{2n}\,(I+2).
\end{equation}
In the low-information regime $I\leq1,$ we have $p_A+p_B<4/n$ by~\eqref{eq:input-only-bound}, because
$(9/2)\ln2<4.$ Lemma~\ref{lem:two-checks}, applied to $\pi,$
gives a local-output protocol for $F$ with information cost at most $2I/n$
and error at most
\[
  2\epsilon^2(p_A+p_B)+\frac{8\epsilon}n<\frac{8\epsilon^2+8\epsilon}n
  <\frac{9\epsilon}n,
\]
which proves our claim.

Finally, in the intermediate regime $1<I<n,$ we have $I'<5I/4.$ Put $m=\lceil I\rceil.$ Then
$2\leq m\leq n$ and $m\leq2I.$ Partition the $n$ coordinates into $m$
blocks of sizes $\lfloor n/m\rfloor$ and $\lceil n/m\rceil,$ each of size at
most $2n/m.$ By~\eqref{eq:input-only-bound} and the inequality $I+2<3I,$ we
have $p_A+p_B<4I/n\leq4m/n.$ Put $\beta=(I'+2m)/\epsilon.$ The protocol
$\pi'$ shows that $W_\beta(F^n,\mu^n)\leq I'+\beta\epsilon,$ so that
Lemma~\ref{lem:block-choice} gives the size $k$ of one of the blocks such
that
\[
  \Bigl(1-\frac{W_\beta(F^k,\mu^k)}\beta\Bigr)^{n/k}
  \geq1-\epsilon-\frac{I'+2(m-1)}\beta>1-2\epsilon.
\]
Since $(1-t)^{n/k}\leq e^{-tn/k}$ for $0\leq t\leq1,$ and
$\ln(1/(1-2\epsilon))\leq2\epsilon/(1-2\epsilon)<3\epsilon,$ it follows
that $W_\beta(F^k,\mu^k)<3\epsilon\beta k/n.$ Hence there is a protocol
$\sigma$ for $F^k$ that announces its output, such that
\[
  \IC(\sigma,\mu^k)+\beta\epsilon'<\frac{3\epsilon\beta k}n,
\]
where $\epsilon'=\err_{\mu^k}(\sigma,F^k).$ In particular,
$\epsilon'<3\epsilon k/n\leq6\epsilon/m,$ and therefore
$2\epsilon'(p_A+p_B)<48\epsilon/n\leq48\epsilon/k.$

An announced answer can serve as the answer of both parties, so $\sigma$
is also a local-output protocol with error $\epsilon'$ under $\mu^k.$
Lemma~\ref{lem:two-checks}, applied to $\sigma$ with $k$ in place of $n,$
gives a local-output
protocol $\pi''$ for $F$ such that
\begin{align*}
  \err_\mu(\pi'',F)&\leq2\epsilon'^2(p_A+p_B)+\frac{8\epsilon'}k
    <\frac{\epsilon'}k\,(8+48\epsilon)<\frac{32\epsilon}n,\\
  \IC(\pi'',\mu)&\leq\frac{2\IC(\sigma,\mu^k)}k<\frac{6\epsilon\beta}n
    =\frac{6(I'+2m)}n<\frac{32I}n,
\end{align*}
where the last step in the first line uses $\epsilon'/k<3\epsilon/n$ and
$3(8+48\epsilon)<32,$ and the last step in the second line uses
$I'<5I/4$ and $m\leq2I.$ This proves our claim.
\end{proof}

For announced outputs, Theorems~\ref{main:dist-direct-sum}
and~\ref{thm:local-comparison} give Corollary~\ref{main:ds-announced} from
the introduction, which we restate for convenience. It is a distributional
counterpart of Theorems~\ref{main:ds} and~\ref{main:ds-partial}, in which
the single copy is measured with local outputs. At error $0,$ its additive
term vanishes.

\begin{restated}{Corollary~\ref*{main:ds-announced}}{Distributional direct
sum for announced outputs, restated}
For every communication problem $F\colon\Xcal\times\Ycal\to\{-1,1,*\},$
every probability distribution $\mu$ on the domain of $F,$ every integer
$n\geq1,$ and every $\epsilon\in[0,1/100],$
\begin{equation*}
  \frac n{32}\LIC_{32\epsilon/n}(F,\mu)
    \leq\IC_\epsilon(F^n,\mu^n)
    \leq n\LIC_{\epsilon/n}(F,\mu)+2h_2(\epsilon)+4\epsilon.
    \tag{\ref*{eq:ds-announced}}
\end{equation*}
\end{restated}

\begin{proof}
By Theorem~\ref{thm:local-comparison}, applied to $F^n$ and $\mu^n,$
\[
  \LIC_\epsilon(F^n,\mu^n)\leq\IC_\epsilon(F^n,\mu^n)
  \leq\LIC_\epsilon(F^n,\mu^n)+2h_2(\epsilon)+4\epsilon.
\]
Together with~\eqref{eq:dist-direct-sum}, these bounds prove the
corollary.
\end{proof}

\section{The distributional XOR lemma}\label{sec:distributional}

Recall from Section~\ref{subsec:results} that in the distributional setting, information complexity with announced
outputs does not obey a truly multiplicative XOR lemma.
We overcome this obstacle by identifying the right
output convention, namely protocols with vector outputs,
for which multiplicativity holds
(Theorem~\ref{main:dist-xor}). This XOR lemma for information
complexity with vector outputs is our main result in this section.
It holds under every input distribution, with single-copy error of order
$\epsilon/n.$ Since announced and vector-output information
complexity differ by at most an additive constant
(Theorem~\ref{thm:vector-comparison}), our work also
yields an optimal XOR lemma for announced outputs
(Theorem~\ref{main:xor-announced}), with the single copy
measured with vector outputs or, alternatively, with split
outputs.

\subsection{XOR lemma with constant additive loss}\label{subsec:xor-additive-loss}

We start by proving an XOR lemma with constant additive
loss, in the original model of information complexity with
announced outputs.  The proof is an adaptation of the priced
information complexity framework of
Section~\ref{sec:dist-direct-sum}.  For communication
problems $F_0\colon\Xcal_0\times\Ycal_0\to\{-1,1,*\}$ and
$F_1\colon\Xcal_1\times\Ycal_1\to\{-1,1,*\},$ let $F_0\oplus
F_1$ be the communication problem on
$(\Xcal_0\times\Xcal_1)\times(\Ycal_0\times\Ycal_1)$ whose
value on $((x_0,x_1),(y_0,y_1))$ is
$F_0(x_0,y_0)F_1(x_1,y_1)$ if $(x_0,y_0)\in\dom F_0$ and
$(x_1,y_1)\in\dom F_1,$ and $*$ otherwise.  After grouping
coordinates, $F^{\oplus(k+k')}=F^{\oplus k}\oplus F^{\oplus
k'}$ for all integers $k,k'\geq1.$ The counterpart of
Lemma~\ref{lem:priced-product} for XORs reads as follows.

\begin{lemma}\label{lem:priced-product-xor}
Let $F_0\colon\Xcal_0\times\Ycal_0\to\{-1,1,*\}$ and
$F_1\colon\Xcal_1\times\Ycal_1\to\{-1,1,*\}$ be communication problems,
let $\mu_0\in\mathcal P(\dom F_0)$ and $\mu_1\in\mathcal P(\dom F_1),$ and
let $\beta>0.$ Then, with $W_j=W_\beta(F_j,\mu_j),$
\[
  W_\beta(F_0\oplus F_1,\mu_0\otimes\mu_1)
  \geq W_0+W_1-\frac{2W_0W_1}\beta-4.
\]
\end{lemma}

For a communication problem $F$ with values in $\{-1,1,*\},$ the priced
information complexity $W_\beta(F,\mu)$ lies in $[0,\beta/2],$ because the
more likely value of $F(X,Y),$ announced without any communication, is
wrong with probability at most $1/2.$ In analogy
with~\eqref{eq:priced-slack}, the lemma states that the normalized slack
$1-2W_\beta/\beta$ is submultiplicative up to an additive term $8/\beta$:
\begin{equation}\label{eq:priced-slack-xor}
  1-\frac{2W_\beta(F_0\oplus F_1,\mu_0\otimes\mu_1)}\beta
  \leq\Bigl(1-\frac{2W_0}\beta\Bigr)\Bigl(1-\frac{2W_1}\beta\Bigr)
     +\frac8\beta.
\end{equation}
The additive term $4$ in the lemma is twice that in
Lemma~\ref{lem:priced-product}, because the default answer for each block
must now be announced, at the cost of one more bit.

\begin{proof}
We follow the proof of Lemma~\ref{lem:priced-product}, in which only the
default outputs and the final computation change. Let $\pi$ be a protocol
for $F_0\oplus F_1,$ let $Z$ be the answer that it announces, and put
$\epsilon=\err_{\mu_0\otimes\mu_1}(\pi,F_0\oplus F_1)$ and
$I=\IC(\pi,\mu_0\otimes\mu_1).$ With $V=(X_0,Y_1,R,M)$ as before and
$B_j=F_j(X_j,Y_j),$ let $\widehat B_j$ be the more likely value of $B_j$
given $V,$ and put
\[
  E_j=\Prob[\widehat B_j\ne B_j\mid V]=\frac{1-\abs{\Exp[B_j\mid V]}}2,
\]
so that $0\leq E_j\leq1/2.$ The default output for block $j$ is
$\widehat B_j.$ As in the proof of Lemma~\ref{lem:priced-product}, the
signs $B_0$ and $B_1$ are independent conditional on $V.$ Since $Z$ is a
function of $V,$ it follows that $\pi$ errs with conditional probability
\[
  \frac{1-Z\,\Exp[B_0\mid V]\,\Exp[B_1\mid V]}2
  \geq\frac{1-(1-2E_0)(1-2E_1)}2=E_0+E_1-2E_0E_1
\]
given $V,$ and averaging over $V$ gives, in place
of~\eqref{eq:default-error},
\[
  \Exp[E_0+E_1-2E_0E_1]\leq\epsilon.
\]
As before, $G_j$ is the event that $\beta E_j\leq C_j.$ The protocol
$\sigma_j$ runs $\pi^{(j)},$ after which the party that knows $V$ sends the
bit $\1_{G_j}$ and the sign $\widehat B_j,$ and the output of $\pi_j$ is
that of $\rho_j$ if $\rho_j$ is run, and $\widehat B_j$ otherwise. Since
$\sigma_j$ appends two one-bit messages to $\pi^{(j)}$ rather than one, the
term $1$ in~\eqref{eq:priced-child} becomes $2.$ In place of the elementary
inequality in the proof of Lemma~\ref{lem:priced-product}, we use
\[
  \min\{\beta e,s\}+\min\{\beta e',s'\}
  \leq\beta(e+e'-2ee')+\frac{2ss'}\beta
\]
for $e,e'\in[0,1/2]$ and $s,s'\geq0,$ which follows in the same way,
because the function
$(z,z')\mapsto z+z'-2zz'$ is nondecreasing in each argument on
$[0,1/2]^2.$ Summing~\eqref{eq:priced-child} over $j\in\{0,1\}$ as before,
we obtain
\begin{align*}
  W_0+W_1
  &\leq I+4+\beta\Exp[E_0+E_1-2E_0E_1]+\frac{2\Exp[C_0C_1]}\beta\\
  &\leq I+4+\beta\epsilon+\frac{2(W_0+\eta)(W_1+\eta)}\beta,
\end{align*}
where the second step uses~\eqref{eq:priced-cross} and
$\Exp_{\mu_j}c_j\leq W_j+\eta.$ Letting $\eta\to0$ and then taking the
infimum over $\pi$ proves the lemma.
\end{proof}

Iterating Lemma~\ref{lem:priced-product-xor} gives the
counterpart of Lemma~\ref{lem:block-choice} for XORs.

\begin{lemma}\label{lem:block-choice-xor}
Let $F\colon\Xcal\times\Ycal\to\{-1,1,*\}$ be a communication problem, let
$\mu\in\mathcal P(\dom F)$ and $\beta>0,$ and let $n=k_1+k_2+\cdots+k_m$
for positive integers $k_1,k_2,\ldots,k_m.$ Then for some $i\in[m],$
\[
  \Bigl(1-\frac{2W_\beta(F^{\oplus k_i},\mu^{k_i})}\beta\Bigr)^{n/k_i}
  \geq1-\frac{2W_\beta(F^{\oplus n},\mu^n)+8(m-1)}\beta.
\]
\end{lemma}

\begin{proof}
The proof of Lemma~\ref{lem:block-choice} applies verbatim, with XORs,
\eqref{eq:priced-slack-xor}, $2W_\beta,$ and $8/\beta$ in place of direct
sums, \eqref{eq:priced-slack}, $W_\beta,$ and $2/\beta.$
\end{proof}

Taking blocks of size $1$ in Lemma~\ref{lem:block-choice-xor}, we now
obtain the XOR lemma with constant additive loss from the introduction
(Theorem~\ref{main:xor-additive}), which we restate for convenience. Its
additive term reflects the four bits sent in each split of
Lemma~\ref{lem:priced-product-xor}, and it matters only when the
information cost per copy is $O(1).$

\begin{restated}{Theorem~\ref*{main:xor-additive}}{Distributional XOR
lemma with constant additive loss, restated}
For every communication problem $F\colon\Xcal\times\Ycal\to\{-1,1,*\},$
every probability distribution $\mu$ on the domain of $F,$ every integer
$n\geq1,$ and every $\epsilon\in[0,1/6],$
\begin{equation*}
  \frac n4\bigl(\IC_{4\epsilon/n}(F,\mu)-16\bigr)
  \leq\IC_\epsilon(F^{\oplus n},\mu^n)
  \leq n\IC_{\epsilon/n}(F,\mu).\tag{\ref*{eq:xor-additive}}
\end{equation*}
\end{restated}

\begin{proof}
For the upper bound, fix $\eta>0$ and a protocol $\pi$ for $F$ with error
at most $\epsilon/n$ under $\mu$ and
$\IC(\pi,\mu)\leq\IC_{\epsilon/n}(F,\mu)+\eta.$ Run independent copies of
$\pi$ on the $n$ coordinate pairs, and output the product of the $n$
announced answers. Under $\mu^n,$ each coordinate pair has distribution
$\mu,$ so the union bound gives error at most $\epsilon.$ By
Lemma~\ref{lem:independent-executions}, the information cost is at most
$n(\IC_{\epsilon/n}(F,\mu)+\eta).$ Letting $\eta\to0$ proves the upper
bound.

For the lower bound, let $\pi$ be a protocol for $F^{\oplus n}$ with error
at most $\epsilon$ under $\mu^n,$ and put $I=\IC(\pi,\mu^n).$ It suffices
to show that $\IC_{4\epsilon/n}(F,\mu)\leq4I/n+16,$ since taking the
infimum over $\pi$ then proves the lower bound. For $\epsilon=0,$ this
follows from Fact~\ref{fact:ordinary-embedding}, which gives
$\IC_0(F,\mu)\leq I/n+2.$

Now let $\epsilon>0,$ and put $\beta=(I+4n)/\epsilon.$ The protocol $\pi$
shows that $W_\beta(F^{\oplus n},\mu^n)\leq I+\beta\epsilon,$ so that
Lemma~\ref{lem:block-choice-xor}, applied with $n$ blocks of size $1,$
gives
\[
  \Bigl(1-\frac{2W_\beta(F,\mu)}\beta\Bigr)^n
  \geq1-2\epsilon-\frac{2I+8(n-1)}\beta
  >1-4\epsilon.
\]
Since $(1-t)^n\leq e^{-tn}$ for $0\leq t\leq1,$ it follows that
\[
  \frac{2W_\beta(F,\mu)}\beta
  <\frac1n\ln\frac1{1-4\epsilon}
  \leq\frac{6\ln3}n\,\epsilon
  <\frac{8\epsilon}n,
\]
where the middle step holds because the function
$t\mapsto\ln(1/(1-4t))$ is convex on $[0,1/6],$ vanishes at $0,$ and
equals $\ln3$ at $1/6.$ Hence there is a protocol $\sigma$ for $F$ such
that
\[
  \IC(\sigma,\mu)+\beta\err_\mu(\sigma,F)
  <\frac{4\epsilon\beta}n=\frac{4I}n+16.
\]
In particular, $\err_\mu(\sigma,F)<4\epsilon/n$ and $\IC(\sigma,\mu)<4I/n+16,$
which proves our claim.
\end{proof}

The lower bound of Theorem~\ref{main:xor-additive} subsumes and substantially
strengthens the distributional XOR lemma~\eqref{eq:sy-distributional} of
Sawettamalya and Yu~\cite{SY}, which concerns the error
$\epsilon=1/10.$ Specifically, our lower bound 
has no term that depends on $|\Xcal||\Ycal|,$ and its single copy has error
$4\epsilon/n$ rather than $n^{-0.41}.$

\subsection{Vector outputs}\label{subsec:vector-outputs}

The constant additive loss in Theorem~\ref{main:xor-additive}
cannot be removed as long as the single copy is measured
with announced outputs, as the example in Section~\ref{subsec:results}
shows. This example highlights the need
for a revised output convention, one that would support a truly multiplicative
XOR lemma.
Accordingly, the output convention that we adopt for the XOR problem
requires no announcement. As with local outputs in
Section~\ref{subsec:local-outputs}, each party ends with an output of his or
her own, here a unit vector.
A protocol $\pi$ has \emph{vector outputs} if, at the end of $\pi,$ Alice
holds a unit vector $u(X,T)\in\mathbb R^d$ and Bob holds a unit vector
$v(Y,T)\in\mathbb R^d,$ where $(X,Y)$ is the input, $T=(R,M)$ consists of
the public randomness and transcript of $\pi,$ and $u$ and $v$ are fixed
functions with values in the unit sphere of $\mathbb R^d.$ These vectors
are not communicated, so they do not affect information cost. For a
communication problem $F\colon\Xcal\times\Ycal\to\{-1,1,*\}$ and
$\mu\in\mathcal P(\dom F),$ the \emph{deficit} of $\pi$ under $\mu$ is
\[
  \Delta_\mu(\pi,F)=\Exp\bigl[1-F(X,Y)\langle u(X,T),v(Y,T)\rangle\bigr],
\]
where the expectation is over $(X,Y)\sim\mu.$
We refer to $\Delta_\mu(\pi,F)/2$ as the \emph{vector error} of $\pi$
under $\mu.$ The \emph{vector-output information complexity} is
\begin{equation}\label{eq:vic}
  \VIC_\epsilon(F,\mu)
    =\inf_{\pi:\,\Delta_\mu(\pi,F)/2\,\leq\,\epsilon}\IC(\pi,\mu),
\end{equation}
where the infimum ranges over protocols with vector outputs
of arbitrary finite dimension.

For a split-output protocol $\pi$ for $F$ and $\mu\in\mathcal P(\dom F),$ we write
$\err_\mu(\pi,F)=\Prob[AB\ne F(X,Y)],$ where $(X,Y)\sim\mu$ and $A$ and $B$
are the shares of $\pi.$ Define
\begin{equation}\label{eq:dist-sic}
  \SIC_\epsilon(F,\mu)
    =\inf_{\pi:\,\err_\mu(\pi,F)\leq\epsilon}\IC(\pi,\mu).
\end{equation}
A split-output protocol whose shares are functions of the respective
party's input and of $T$ has vector outputs in dimension $1,$ and its vector
error is then $\err_\mu(\pi,F),$ because $1-F(X,Y)AB$ is twice the
indicator of the event $AB\ne F(X,Y).$
Conversely, we now show that vector outputs can be converted
into shares with no increase in error.
Regev and Toner~\cite{RT09} showed that public randomness and two bits of
communication from Alice to Bob suffice to turn unit vectors $u$ and $v,$
held by Alice and Bob, respectively, into signs $A$ and $B$ with
$\Exp[AB]=\langle u,v\rangle.$ The information cost of our conversion,
by contrast, vanishes as the error tends to zero.

\begin{lemma}\label{lem:vector-rounding}
Let $F\colon\Xcal\times\Ycal\to\{-1,1,*\}$ be a communication problem, and
let $\mu\in\mathcal P(\dom F).$ Then for every $0\leq\epsilon\leq1,$
\[
  \SIC_\epsilon(F,\mu)\leq\VIC_\epsilon(F,\mu)+\frac8{\ln2}\,\epsilon(1-\epsilon).
\]
\end{lemma}

\begin{proof}
If $\epsilon\geq1/2,$ the lemma holds trivially: the split-output protocol
without communication in which Alice outputs a uniformly random sign and
Bob outputs $1$ has error $1/2,$ so that $\SIC_\epsilon(F,\mu)=0.$ Now let
$\epsilon<1/2,$ and let $\pi$ be a protocol with vector outputs and vector
error at most $\epsilon$ under $\mu.$ It suffices to show that
$\SIC_\epsilon(F,\mu)\leq\IC(\pi,\mu)+(8/\ln2)\,\epsilon(1-\epsilon),$
since taking the infimum over $\pi$ then proves the lemma.
By Proposition~\ref{prop:finite-fixed}, we may assume that $\pi$ has finite
public randomness. For each value $t$ of $T$ with $\Prob[T=t]>0,$ let
$\mu_t$ be the conditional distribution of the input given $T=t,$ and
choose families $(u^t_x)$ and $(v^t_y)$ of unit vectors that maximize
$\Gamma_t=\Exp_{\mu_t}[F(X,Y)\langle u^t_X,v^t_Y\rangle];$ as with the margin in
Section~\ref{subsec:margin}, the maximum is attained in
$\mathbb R^{|\Xcal|+|\Ycal|}.$
Comparing $\Gamma_t$ with the values attained by the output vectors of $\pi$
given $T=t$ and by the tensor powers $(u^t_x)^{\otimes k}$ and
$(v^t_y)^{\otimes k}$ for odd $k,$ whose signed inner products are
$(F(x,y)\langle u^t_x,v^t_y\rangle)^k,$ we obtain
\begin{equation}\label{eq:odd-moments}
  \begin{aligned}
    &\Exp\Gamma_T\geq1-2\epsilon,\\
    &\Exp_{\mu_t}\bigl[(F(X,Y)\langle u^t_X,v^t_Y\rangle)^k\bigr]\leq\Gamma_t
      \quad(k=1,3,5,\ldots).
  \end{aligned}
\end{equation}

Let $\pi'$ be the split-output protocol that runs $\pi$ and then the
protocol from the proof of Lemma~\ref{lem:rounding}, with $u^T_x$ and
$v^T_y$ in place of $u_x$ and $v_y,$ and let $A$ and $B$ be its shares. For
unit vectors $u,v$ with $\langle u,v\rangle=\cos\theta,$ where
$0\leq\theta\leq\pi,$ the rounding produces signs whose product has
expectation
\[
  \rho(\cos\theta)=1-\frac{2(\theta-\sin\theta\cos\theta)}\pi
  =1-\frac{2\theta-\sin2\theta}\pi,
\]
because the signs differ with probability
$(\theta-\sin\theta\cos\theta)/\pi$: by~\eqref{eq:angular-error} if
$\theta\leq\pi/2,$ and by~\eqref{eq:angular-error} applied to $u$ and $-v$
otherwise, since negating $v$ negates Bob's sign and replaces $\theta$ by
$\pi-\theta.$ We have
$\rho(0)=0$ and, as differentiation with respect to $\theta$ shows,
$\rho'(z)=(4/\pi)\sqrt{1-z^2}.$ Integrating the binomial series of
$\sqrt{1-z^2},$ whose coefficients after the constant term are negative, we
obtain
\[
  \rho(z)=\frac4\pi\,z-\sum_{k=1}^\infty a_kz^{2k+1},\qquad-1\leq z\leq1,
\]
with $a_k>0$ and $\sum_ka_k=4/\pi-1,$ the latter because $\rho(1)=1.$
When the expected product of the signs is a function of $\langle u,v\rangle$
of this form, Regev and Toner~\cite{RT09}, following Krivine~\cite{Kri79},
make it exactly $\langle u,v\rangle$ by applying a map built from tensor
powers to both vectors beforehand. In our proof, tensor powers serve only for the
comparison in~\eqref{eq:odd-moments}, and the rounding is applied to
$(u^t_x)$ and $(v^t_y)$ directly, which keeps its information cost small.
As in
the proof of Lemma~\ref{lem:rounding}, the sign $F(X,Y)$ is harmless, so
that $\Exp[F(X,Y)AB\mid X,Y,T=t]=\rho(F(X,Y)\langle u^t_X,v^t_Y\rangle)$
for $(X,Y)\sim\mu_t.$ By~\eqref{eq:odd-moments}, it follows that
\[
  \Exp[F(X,Y)AB\mid T=t]
  =\Exp_{\mu_t}\bigl[\rho(F(X,Y)\langle u^t_X,v^t_Y\rangle)\bigr]
  \geq\frac4\pi\,\Gamma_t-\Bigl(\frac4\pi-1\Bigr)\Gamma_t=\Gamma_t,
\]
and therefore $\err_\mu(\pi',F)\leq(1-\Exp\Gamma_T)/2\leq\epsilon.$

Finally, let $(X,Y)\sim\mu.$ The argument in the proof of
Lemma~\ref{lem:rounding}, applied conditionally on $T$ and with $v^T_Y$
replaced by $-v^T_Y$ if $\langle u^T_X,v^T_Y\rangle<0,$ shows that the
message of the rounding increases the information cost by at most
\[
  \frac2{\ln2}\,\Exp\bigl[1-\langle u^T_X,v^T_Y\rangle^2\bigr]
  \leq\frac2{\ln2}\bigl(1-(\Exp\Gamma_T)^2\bigr)
  \leq\frac8{\ln2}\,\epsilon(1-\epsilon).
\]
Here the first step uses the convexity of the square, since the signed
inner product $F(X,Y)\langle u^T_X,v^T_Y\rangle$ has mean $\Exp\Gamma_T,$ and
the second step uses~\eqref{eq:odd-moments} and $\epsilon\leq1/2.$ As
$\err_\mu(\pi',F)\leq\epsilon,$ this proves our claim.
\end{proof}

Our next result establishes~\eqref{eq:vector-comparison} from
the introduction, which compares vector, split, and announced outputs.

\begin{theorem}[Vector versus announced outputs]\label{thm:vector-comparison}
Let $F\colon\Xcal\times\Ycal\to\{-1,1,*\}$ be a communication problem, let
$\mu\in\mathcal P(\dom F),$ and let $0\leq\epsilon\leq1/2.$ Then
\begin{equation*}
  \VIC_\epsilon(F,\mu)\leq\SIC_\epsilon(F,\mu)\leq\IC_\epsilon(F,\mu)
  \leq\VIC_\epsilon(F,\mu)+5.
\end{equation*}
\end{theorem}

\begin{proof}
The first inequality holds because, by
Remark~\ref{rem:local-outputs}, the shares of a split-output protocol may
be assumed to be functions of the respective party's input and $T,$ and
the second because an announced answer can serve as Alice's share, with
Bob's share equal to $1.$ For the third, Lemma~\ref{lem:vector-rounding}
and the bound $\epsilon(1-\epsilon)\leq1/4$ give
$\SIC_\epsilon(F,\mu)\leq\VIC_\epsilon(F,\mu)+2/\ln2,$ and appending both
shares of a split-output protocol to the transcript, at a cost of at most
$2,$ gives $\IC_\epsilon(F,\mu)\leq\SIC_\epsilon(F,\mu)+2.$ Since
$2+2/\ln2<5,$ this proves the third inequality.
\end{proof}

\subsection{Pricing with vector outputs}\label{subsec:priced-vector}

We now adapt Section~\ref{subsec:xor-additive-loss} to vector outputs, with
deficits in place of errors. For a communication problem
$F\colon\Xcal\times\Ycal\to\{-1,1,*\},$ a distribution
$\mu\in\mathcal P(\dom F),$ and $\beta\geq0,$ define
\begin{equation}\label{eq:priced-vector}
  W^*_\beta(F,\mu)=\inf_\pi\bigl\{\IC(\pi,\mu)
     +\beta\Delta_\mu(\pi,F)\bigr\},
\end{equation}
where the infimum ranges over all protocols $\pi$ with vector outputs.
Constant orthogonal vectors
have deficit $1,$ so that $W^*_\beta(F,\mu)$ lies in $[0,\beta].$
Proposition~\ref{prop:cost-profile} and its proof remain valid for a
protocol $\pi$ with vector outputs, with the expected deficit
$\Exp[1-F(x,y)\langle u(x,T),v(y,T)\rangle]$ of $\pi$ on input $(x,y)$ in
place of $\Prob[\pi(x,y)\ne F(x,y)],$ and with $\Delta_{\mu'}(\pi,F)$ in
place of $\err_{\mu'}(\pi,F).$ We again call the resulting function the
cost profile of $\pi$ with respect to $\mu$ and $\beta.$
The counterpart of Lemma~\ref{lem:priced-product-xor} for vector outputs
reads as follows.

\begin{lemma}\label{lem:priced-product-vector}
Let $F_0\colon\Xcal_0\times\Ycal_0\to\{-1,1,*\}$ and
$F_1\colon\Xcal_1\times\Ycal_1\to\{-1,1,*\}$ be communication problems,
let $\mu_0\in\mathcal P(\dom F_0)$ and $\mu_1\in\mathcal P(\dom F_1),$ and
let $\beta>0.$ Then, with $W_j=W^*_\beta(F_j,\mu_j),$
\[
  W^*_\beta(F_0\oplus F_1,\mu_0\otimes\mu_1)
  \geq W_0+W_1-\frac{W_0W_1}\beta-2.
\]
\end{lemma}

The additive term here is $2,$ rather than $4$ as in
Lemma~\ref{lem:priced-product-xor}, because the
default outputs are now vectors, which need not be announced.
As with~\eqref{eq:priced-slack}, the normalized slack
$1-W^*_\beta/\beta$ is therefore submultiplicative up to an
additive term $2/\beta.$

\begin{proof}
We follow the proof of Lemma~\ref{lem:priced-product}, with protocols with
vector outputs, deficits, and the cost profiles just described in place of
protocols, errors, and the cost profiles of
Proposition~\ref{prop:cost-profile}. Only the default outputs change. Let
$\pi$ be a protocol for $F_0\oplus F_1$ with vector outputs
$u(X_0,X_1,T)$ and $v(Y_0,Y_1,T),$ where $T=(R,M),$ and put
$\Delta=\Delta_{\mu_0\otimes\mu_1}(\pi,F_0\oplus F_1).$ With
$V=(X_0,Y_1,R,M)$ as before and $B_j=F_j(X_j,Y_j),$ put
\[
  a=\Exp[B_1u(X_0,X_1,T)\mid V],\qquad
  b=\Exp[B_0v(Y_0,Y_1,T)\mid V],
\]
and let $\Delta_0=1-\|b\|$ and $\Delta_1=1-\|a\|.$ The default for block
$0$ consists of the vectors $b/\|b\|$ for Alice and $v(Y_0,Y_1,T)$ for Bob,
and the default for block $1$ consists of $u(X_0,X_1,T)$ for Alice and
$a/\|a\|$ for Bob, where a fixed unit vector replaces $a/\|a\|$ or
$b/\|b\|$ if $a=0$ or $b=0.$ These vectors are available locally: in
$\pi^{(0)},$ Alice knows $V,$ and Bob holds $Y_0$ and knows the public
$Y_1$ and $T;$ in $\pi^{(1)},$ Bob knows $V,$ and Alice holds $X_1$ and
knows the public $X_0$ and $T.$ Since $b$ is a function of $V,$ the default
for block $0$ has conditional deficit
$\Exp[1-B_0\langle b/\|b\|,v(Y_0,Y_1,T)\rangle\mid V]=1-\|b\|=\Delta_0$
given $V,$ and likewise the default for block $1$ has conditional deficit
$\Delta_1.$ Moreover, $B_0v(Y_0,Y_1,T)$ and $B_1u(X_0,X_1,T)$ are
functions of $(V,Y_0)$ and of $(V,X_1),$ respectively, so their conditional
independence given $V$ yields
$\Exp[B_0B_1\langle u(X_0,X_1,T),v(Y_0,Y_1,T)\rangle\mid V]
=\langle a,b\rangle\leq(1-\Delta_0)(1-\Delta_1).$ Averaging over $V,$ we
obtain, in place of~\eqref{eq:default-error},
\[
  \Exp\bigl[1-(1-\Delta_0)(1-\Delta_1)\bigr]\leq\Delta.
\]
The good event for block $j$ is now the event that $\beta\Delta_j\leq
C_j,$ and the output of $\pi_j$ is that of $\rho_j$ if $\rho_j$ is run, and
the default for block $j$ otherwise, with vectors padded by zeros to a
common dimension. With $\Delta_j$ and $\Delta$ in place of $E_j$ and
$\epsilon,$ the rest of the proof of Lemma~\ref{lem:priced-product}
applies verbatim.
\end{proof}

Iterating Lemma~\ref{lem:priced-product-vector} in the same way gives the
counterpart of Lemma~\ref{lem:block-choice-xor} for vector outputs.

\begin{lemma}\label{lem:block-choice-vector}
Let $F\colon\Xcal\times\Ycal\to\{-1,1,*\}$ be a communication problem, let
$\mu\in\mathcal P(\dom F)$ and $\beta>0,$ and let $n=k_1+k_2+\cdots+k_m$
for positive integers $k_1,k_2,\ldots,k_m.$ Then for some $i\in[m],$
\[
  \Bigl(1-\frac{W^*_\beta(F^{\oplus k_i},\mu^{k_i})}\beta\Bigr)^{n/k_i}
  \geq1-\frac{W^*_\beta(F^{\oplus n},\mu^n)+2(m-1)}\beta.
\]
\end{lemma}

\begin{proof}
The proof of Lemma~\ref{lem:block-choice} applies verbatim, with
$W^*_\beta,$ XORs, and Lemma~\ref{lem:priced-product-vector} in place of
$W_\beta,$ direct sums, and Lemma~\ref{lem:priced-product}.
\end{proof}

\subsection{Extraction with vector outputs}\label{subsec:vector-lift}

Let $F\colon\Xcal\times\Ycal\to\{-1,1,*\}$ be a communication problem, and let
$\mu\in\mathcal P(\dom F).$ For families $(w_x)_{x\in\Xcal}$ and
$(w'_y)_{y\in\Ycal}$ of unit vectors in $\mathbb R^d,$ their
\emph{deficit} under $\mu$ is defined to be 
$\Exp[1-F(X,Y)\langle w_X,w'_Y\rangle],$ where $(X,Y)\sim\mu.$ We put
\begin{equation}\label{eq:deficit}
  \Delta_\mu(F)=\min_{w,w'}\,\Exp\bigl[1-F(X,Y)\langle w_X,w'_Y\rangle\bigr],
\end{equation}
where the minimum ranges over all dimensions $d\geq1$ and all such
families. This quantity is the distributional analogue of margin:
by~\eqref{eq:margin-vector}, $1-\margin(F)$ is the same minimum with the
expectation replaced by a maximum over the legal inputs. In particular,
$\Delta_\mu(F)\leq1-\margin(F)\leq1.$ As with margin, we may take
$d=|\Xcal|+|\Ycal|,$ and the minimum is then attained by compactness.
Moreover, $1-\Delta_\mu(F)=\gamma_2^*(\mu\circ A_F)$ in the notation of
Appendix~\ref{app:norms}, by the description of the unit ball of $\gamma_2$
given there. Hence, the multiplicativity of $\gamma_2^*$ under tensor
products~\cite{CSUU08,LSS08}, stated in~\eqref{eq:gamma2-dual-tensor},
gives $1-\Delta_{\mu^n}(F^{\oplus n})=(1-\Delta_\mu(F))^n$ for every
integer $n\geq1,$ in parallel with Proposition~\ref{prop:margin-tensor}.

A protocol with vector outputs and no communication uses families of unit
vectors that depend only on its public randomness, so $\Delta_\mu(F)$ is the
least deficit $\Delta_\mu(\pi,F)$ of such a protocol $\pi.$ With
communication, $\Delta_\mu(\pi,F)$ can be much smaller than $\Delta_\mu(F).$
We now show that the gap is governed by the information that the transcript
reveals about the inputs: for a general protocol with vector outputs, the
dependence of the output vectors on the transcript can be removed at a cost
proportional to this information. To do so, we merge each party's vectors over all
transcripts, weighting them by the square roots of the conditional probabilities
of the public randomness and transcript given that party's input.

\begin{lemma}\label{lem:transcript-removal}
Let $F\colon\Xcal\times\Ycal\to\{-1,1,*\}$ be a communication problem, let
$\mu\in\mathcal P(\dom F),$ and let $\pi$ be a protocol with vector
outputs. Then
\[
  \Delta_\mu(F)\leq2\Delta_\mu(\pi,F)+2\IC(\pi,\mu).
\]
\end{lemma}

\begin{proof}\looseness=-1
By Proposition~\ref{prop:finite-fixed}, we may assume that $\pi$ has finite
public randomness. Let $(X,Y)\sim\mu,$ write $T=(R,M)$ for the public
randomness and transcript of $\pi,$ and let $u(X,T),v(Y,T)\in\mathbb R^d$ be
its vector outputs. Let $\mathcal T$ be the finite set of values of $T.$ For
$(x,y)\in\supp\mu,$ let $Q_{xy},$ $Q_x,$ and $Q_y$ be the conditional
distributions of $T$ given $(X,Y)=(x,y),$ given $X=x,$ and given $Y=y,$
respectively. Consider the Euclidean space $(\mathbb R^d)^{\mathcal T}$ of
families $w=(w_t)_{t\in\mathcal T}$ of vectors in $\mathbb R^d,$ with
$\|w\|^2=\sum_t\|w_t\|_2^2.$ For $x$ in the support of $X,$ let $\bar u_x$ be
the family $(\sqrt{Q_x(t)}\,u(x,t))_t;$ for $y$ in the support of $Y,$ let
$\bar v_y$ be the family $(\sqrt{Q_y(t)}\,v(y,t))_t;$ and let all other
$\bar u_x$ and $\bar v_y$ be arbitrary unit vectors in this space. All these
vectors have norm $1,$ because $\|\bar u_x\|^2=\sum_tQ_x(t)=1$ for $x$ in
the support of $X,$ and similarly for $\bar v_y.$

Fix $(x,y)\in\supp\mu,$ and abbreviate $s=F(x,y),$ $u_t=u(x,t),$ and
$v_t=v(y,t).$ Then for every $t\in\mathcal T,$ the component of
$\bar u_x-s\bar v_y$ indexed by $t$ is
\[
  \Bigl(\sqrt{Q_x(t)}-\sqrt{Q_{xy}(t)}\Bigr)\,u_t
  +\sqrt{Q_{xy}(t)}\,(u_t-sv_t)
  +s\Bigl(\sqrt{Q_{xy}(t)}-\sqrt{Q_y(t)}\Bigr)\,v_t.
\]
For any three vectors $w_1,w_2,w_3$ in a Euclidean space,
$\|w_1+w_2+w_3\|^2\leq2\|w_2\|^2
+2\|w_1+w_3\|^2\leq4\|w_1\|^2+2\|w_2\|^2+4\|w_3\|^2.$
As $u_t$ and $v_t$ are unit vectors, it follows that
\begin{align*}
  \|\bar u_x-s\bar v_y\|^2
  &\leq2\sum_tQ_{xy}(t)\,\|u_t-sv_t\|_2^2
     +4\sum_t\Bigl(\sqrt{Q_x(t)}-\sqrt{Q_{xy}(t)}\Bigr)^2\\
  &\qquad+4\sum_t\Bigl(\sqrt{Q_{xy}(t)}-\sqrt{Q_y(t)}\Bigr)^2.
\end{align*}
The first sum equals
$\Exp[\|u(X,T)-sv(Y,T)\|_2^2\mid(X,Y)=(x,y)]
=2\Exp[1-s\langle u(X,T),v(Y,T)\rangle\mid(X,Y)=(x,y)].$
The other two sums are bounded by relative entropies. Indeed, let $P$ and
$Q$ be distributions on $\mathcal T,$ and put $z=\sum_t\sqrt{P(t)Q(t)}.$
Then $\sum_t(\sqrt{P(t)}-\sqrt{Q(t)})^2=2-2z,$ and if $\kl PQ<\infty,$
Jensen's inequality gives
\[
  \klnat PQ=-2\sum_{t:\,P(t)>0}P(t)\ln\sqrt{\frac{Q(t)}{P(t)}}
  \geq-2\ln z\geq2-2z.
\]
Hence $\sum_t(\sqrt{P(t)}-\sqrt{Q(t)})^2\leq\klnat PQ\leq\kl PQ$ in all
cases. Applying this bound to the last two sums, we obtain
\begin{align*}
  \|\bar u_x-s\bar v_y\|^2
  &\leq4\Exp\bigl[1-s\langle u(X,T),v(Y,T)\rangle\bigm|(X,Y)=(x,y)\bigr]\\
  &\qquad+4\kl{Q_{xy}}{Q_x}+4\kl{Q_{xy}}{Q_y}.
\end{align*}
We now average over $(x,y)\sim\mu.$ By the first claim of
Proposition~\ref{prop:cost-profile} with $\beta=0,$ which does not
involve the outputs of $\pi,$ the relative entropies average to
$\IC(\pi,\mu).$ Since
$\|\bar u_x-s\bar v_y\|^2=2-2s\langle\bar u_x,\bar v_y\rangle,$ it follows
that the families $(\bar u_x)$ and $(\bar v_y)$ have deficit at most
$2\Delta_\mu(\pi,F)+2\IC(\pi,\mu)$ under $\mu.$ Finally, these finitely
many vectors span a subspace of dimension at most $|\Xcal|+|\Ycal|,$ and
identifying this subspace with a Euclidean space proves the lemma.
\end{proof}

We now come to the heart of the argument. Lemma~\ref{lem:vector-lift}
below converts a protocol with vector outputs for $F^{\oplus n}$ into one
for $F,$ dividing the information cost by $n/2.$ It plays the role of
Lemma~\ref{lem:two-checks} for direct sums. The deficit of the protocol
that it produces depends on the least deficit that the parties can achieve
without communication, which takes the place of the prediction errors
$p_A$ and $p_B.$

\begin{lemma}\label{lem:vector-lift}
Let $F\colon\Xcal\times\Ycal\to\{-1,1,*\}$ be a communication problem, let
$\mu\in\mathcal P(\dom F),$ let $n\geq1$ be an integer, and let $\sigma$ be
a protocol for $F^{\oplus n}$ with vector outputs and vector error
$\epsilon$ under $\mu^n.$ Then there is a protocol $\pi$ with vector outputs
such that
\begin{align}
  \IC(\pi,\mu)&\leq\frac{2\IC(\sigma,\mu^n)}n,\label{eq:lift-cost}\\
  \Delta_\mu(\pi,F)&\leq32\epsilon^2\Delta_\mu(F)+\frac{80\epsilon}n.
     \label{eq:lift-deficit}
\end{align}
\end{lemma}

To explain the key idea behind this result, consider the standard
embedding of $\sigma$ into a uniformly random coordinate $i.$ At the end of
it, Alice and Bob can compute conditional expectations $a$ and $b$ of Bob's
output vector, multiplied by the XORs of the first $i$ and the first $i-1$
coordinates of the completed input, respectively. If $\sigma$ has small
vector error, then $b$ is close to $F(x,y)\,a$ on average, where $(x,y)$ is
the actual input. To turn these conditional expectations into unit vectors,
we use the inverse stereographic projection, which generalizes the
rational parametrization of the unit circle: for a vector $a\in\mathbb R^d$
with $\|a\|\leq1$ and a unit vector $w\in\mathbb R^{d'},$ the vector
\[
  \Psi_a(w)=\frac1{1+\|a\|^2}\,\bigl(2a,(1-\|a\|^2)\,w\bigr)
  \in\mathbb R^{d+d'}
\]
is a unit vector, because $4\|a\|^2+(1-\|a\|^2)^2=(1+\|a\|^2)^2.$ It
equals $(0,w)$ for $a=0$ and $(a,0)$ for $\|a\|=1,$ so it retains $w$ when
$a$ is uninformative and commits to the direction of $a$ when $a$ is
decisive. For vectors $a,b\in\mathbb R^d$ of norm at most $1,$ unit vectors
$w,w'\in\mathbb R^{d'},$ and a sign $s\in\{-1,1\},$ a direct expansion gives
\begin{equation}\label{eq:lift-identity}
  1-s\langle\Psi_a(w),\Psi_b(w')\rangle
  =\frac{2\|sa-b\|^2}{(1+\|a\|^2)(1+\|b\|^2)}
   +\frac{(1-\|a\|^2)(1-\|b\|^2)}{(1+\|a\|^2)(1+\|b\|^2)}\,
     \bigl(1-s\langle w,w'\rangle\bigr).
\end{equation}
The first term is small when $b$ is close to $sa,$ and the coefficient of
the old deficit $1-s\langle w,w'\rangle$ is small when $b$ is close to
the unit sphere. As this coefficient lies in $[0,1-\|b\|^2],$ the
identity~\eqref{eq:lift-identity} implies
\begin{equation}\label{eq:lift-bound}
  1-s\langle\Psi_a(w),\Psi_b(w')\rangle
  \leq2\|sa-b\|^2+(1-\|b\|^2)\bigl(1-s\langle w,w'\rangle\bigr).
\end{equation}
In the standard embedding, a telescoping argument shows that
$\|F(x,y)\,a-b\|^2$ has expectation of order $\epsilon/n.$ Moreover, the
conditional expectation of $1-\|b\|^2$ given the actual input has average
of order $\epsilon.$ This conditional expectation may fluctuate, however,
and its product with the old deficit is not controlled by $\Delta_\mu(F)$
alone. We therefore run two independent embeddings and apply $\Psi$
twice, as in the proof of Lemma~\ref{lem:two-checks}. This squares the
conditional expectation, so that its fluctuations
enter only through its variance, which is of order $\epsilon/n$ by
Lemma~\ref{lem:variance}.

\begin{proof}
Let $(X,Y)\sim\mu^n,$ with coordinate pairs $(X_i,Y_i),$ let $T=(R,M)$
consist of the public randomness and transcript of $\sigma$ on input
$(X,Y),$ and let $u(X,T),v(Y,T)\in\mathbb R^d$ be its vector outputs. Put
\[
  L=\frac{1-F^{\oplus n}(X,Y)\,\langle u(X,T),v(Y,T)\rangle}2,
\]
so that $L$ takes values in $[0,1]$ and $\Exp L=\epsilon.$ For
$0\leq j\leq n,$ put
\[
  \zeta_j=\Exp\Bigl[\Bigl(\prod_{\ell\leq j}F(X_\ell,Y_\ell)\Bigr)\,v(Y,T)
     \Bigm|X,Y_{>j},T\Bigr],
  \qquad
  q_j=\Exp[L\mid X,Y_{>j},T].
\]
In particular, $\zeta_0=v(Y,T),$ and every $\zeta_j$ takes values in the
unit ball of $\mathbb R^d.$

We first record three properties of these quantities. For $0<j<n,$
applying~\eqref{eq:product-cross} to the blocks formed by the first $j$
and the last $n-j$ coordinates, with the distributions $\mu^j$ and
$\mu^{n-j},$ gives $X_{>j}\perp Y_{\leq j}\mid X_{\leq j},Y_{>j},T.$
Since $(\prod_{\ell\leq j}F(X_\ell,Y_\ell))\,v(Y,T)$ is a function of
$(X_{\leq j},Y,T),$ it follows that $\zeta_j$ is a function of
$(X_{\leq j},Y_{>j},T),$ which is trivially true also for $j\in\{0,n\}.$
Next, let $i\in[n].$ As $F(X_i,Y_i)$ is a function of $(X,Y_{\geq i}),$
the product $F(X_i,Y_i)\,\zeta_{i-1}$ is the conditional expectation of
$(\prod_{\ell\leq i}F(X_\ell,Y_\ell))\,v(Y,T)$ given $(X,Y_{\geq i},T),$
and the tower property gives
\begin{equation}\label{eq:bias-tower}
  \Exp\bigl[F(X_i,Y_i)\,\zeta_{i-1}\bigm|X,Y_{>i},T\bigr]=\zeta_i.
\end{equation}
Finally, $\prod_{\ell\leq j}F(X_\ell,Y_\ell)$ equals the product of
$F^{\oplus n}(X,Y)$ and $\prod_{\ell>j}F(X_\ell,Y_\ell).$ Since
$u(X,T)$ and the signs $F(X_\ell,Y_\ell)$ with $\ell>j$ are functions of
$(X,Y_{>j},T),$ it follows that
\[
  \langle u(X,T),\zeta_j\rangle
  =(1-2q_j)\prod_{\ell>j}F(X_\ell,Y_\ell).
\]
As $u(X,T)$ is a unit vector, the Cauchy--Schwarz inequality now gives
$\|\zeta_j\|\geq\abs{1-2q_j},$ and therefore
\begin{equation}\label{eq:bias-error-prob}
  1-\|\zeta_j\|^2\leq4q_j(1-q_j)\leq4q_j.
\end{equation}

For $i\in[n],$ consider now the standard embedding $\sigma^{(i)}$ into
coordinate $i$ of $\sigma$ with respect to $\mu$
(Definition~\ref{def:standard-embedding}). Its public randomness
$(R,X_{<i},Y_{>i})$ and the transcript $M$ are known to both parties.
Alice, who holds $X_i,$ can therefore compute $a_i=\zeta_i,$ which is a
function of $(X_{\leq i},Y_{>i},T).$ Likewise, Bob, who holds $Y_i,$ can
compute $b_i=\zeta_{i-1},$ which is a function of $(X_{<i},Y_{\geq i},T).$
By~\eqref{eq:bias-tower},
$\Exp[F(X_i,Y_i)\langle a_i,b_i\rangle]
=\Exp[\langle a_i,\Exp[F(X_i,Y_i)\,b_i\mid X,Y_{>i},T]\rangle]
=\Exp[\|a_i\|^2].$
Expanding the square, we obtain
$\Exp[\|F(X_i,Y_i)\,a_i-b_i\|^2]=\Exp[\|b_i\|^2]-\Exp[\|a_i\|^2].$ Summing
over $i,$ the right-hand side telescopes, and~\eqref{eq:bias-error-prob}
with $j=n$ gives
\begin{equation}\label{eq:bias-energy}
  \sum_{i=1}^n\Exp\bigl[\|F(X_i,Y_i)\,a_i-b_i\|^2\bigr]
  =\Exp\|\zeta_0\|^2-\Exp\|\zeta_n\|^2=\Exp\bigl[1-\|\zeta_n\|^2\bigr]
  \leq4\epsilon.
\end{equation}
Moreover, $(X_i,Y_i)$ is a function of $(X,Y_{\geq i}),$ so the tower
property and~\eqref{eq:bias-error-prob} with $j=i-1$ give
\begin{equation}\label{eq:bias-weight}
  \Exp\bigl[1-\|b_i\|^2\bigm|(X_i,Y_i)=(x,y)\bigr]
  \leq4\Exp\bigl[L\bigm|(X_i,Y_i)=(x,y)\bigr],
  \qquad(x,y)\in\supp\mu.
\end{equation}

Let $\rho$ be the protocol that publicly chooses a uniformly random
$i\in[n]$ and runs $\sigma^{(i)}$ on its input; at the end of $\rho,$
Alice knows $a=a_i$ and Bob knows $b=b_i.$ By~\eqref{eq:public-mixture-cost}
and Fact~\ref{fact:embedding-cost}, $\IC(\rho,\mu)=\IC(\sigma,\mu^n)/n.$
Fix $(x,y)\in\supp\mu,$ and run $\rho$ on input $(x,y).$ Conditional on
$i,$ the completed input, together with the public randomness and
transcript of the run of $\sigma,$ has the conditional distribution of
$(X,Y,R,M)$ given $(X_i,Y_i)=(x,y)$
(Definition~\ref{def:standard-embedding}). Hence~\eqref{eq:bias-weight}
shows that, on input $(x,y),$
\begin{equation}\label{eq:embedding-weight}
  \Exp\bigl[1-\|b\|^2\bigr]\leq4\bar\epsilon(x,y),\qquad
  \bar\epsilon(x,y)=\frac1n\sum_{i=1}^n\Exp\bigl[L\bigm|(X_i,Y_i)=(x,y)\bigr].
\end{equation}
Similarly, each $(X_i,Y_i)$ has distribution $\mu,$ so averaging
$\Exp[\|F(x,y)\,a-b\|^2]$ over $(x,y)\sim\mu$ gives
$\frac1n\sum_{i=1}^n\Exp[\|F(X_i,Y_i)\,a_i-b_i\|^2],$ which is at most
$4\epsilon/n$ by~\eqref{eq:bias-energy}. Lemma~\ref{lem:variance}, applied
to the coordinate pairs $(X_i,Y_i)$ and the random
variable $L,$ shows that
\begin{equation}\label{eq:local-error-moments}
  \Exp_\mu\bar\epsilon=\epsilon,\qquad\Var_\mu(\bar\epsilon)\leq\frac\epsilon n.
\end{equation}

Finally, let $(w_x)$ and $(w'_y)$ be families of unit vectors in
$\mathbb R^{d_0}$ that attain $\Delta_\mu(F),$ and put
$\delta(x,y)=1-F(x,y)\langle w_x,w'_y\rangle,$ so that $0\leq\delta\leq2$
and $\Exp_\mu\delta=\Delta_\mu(F).$ Let $\pi$ be the protocol that runs
two independent copies of $\rho$ on its input $(x,y),$ the first
producing vectors $a',b'$ and the second $a'',b'',$ and whose vector
outputs are
\[
  \Psi_{a''}(\Psi_{a'}(w_x)),\qquad\Psi_{b''}(\Psi_{b'}(w'_y)).
\]
These are unit vectors in $\mathbb R^{2d+d_0},$ and each of them is a
function of the respective party's input and of the public randomness and
transcripts of the two copies, as required. By
Lemma~\ref{lem:independent-executions},
$\IC(\pi,\mu)\leq2\IC(\rho,\mu)=2\IC(\sigma,\mu^n)/n,$ which
proves~\eqref{eq:lift-cost}.

To prove~\eqref{eq:lift-deficit}, fix $(x,y)\in\supp\mu$ and put
$s=F(x,y).$ Applying~\eqref{eq:lift-bound} first to $a'',$ $b'',$
$\Psi_{a'}(w_x),$ $\Psi_{b'}(w'_y)$ and then to $a',$ $b',$ $w_x,$ $w'_y,$ we
obtain
\begin{align*}
  &1-s\langle\Psi_{a''}(\Psi_{a'}(w_x)),\Psi_{b''}(\Psi_{b'}(w'_y))\rangle\\
  &\qquad\leq2\|sa''-b''\|^2+(1-\|b''\|^2)\bigl(2\|sa'-b'\|^2
     +(1-\|b'\|^2)\,\delta(x,y)\bigr)\\
  &\qquad\leq2\|sa'-b'\|^2+2\|sa''-b''\|^2
     +(1-\|b'\|^2)(1-\|b''\|^2)\,\delta(x,y).
\end{align*}
On input $(x,y),$ the two copies of $\rho$ are independent,
so~\eqref{eq:embedding-weight} bounds the expectation of the last term by
$16\bar\epsilon(x,y)^2\delta(x,y).$ Averaging over $(x,y)\sim\mu,$ and recalling
that $\Exp[\|F(x,y)\,a-b\|^2]$ averages to at most $4\epsilon/n$ for each
copy of $\rho,$ we obtain
\[
  \Delta_\mu(\pi,F)\leq\frac{16\epsilon}n+16\Exp_\mu[\bar\epsilon^2\delta].
\]
Since $\bar\epsilon^2\leq2\epsilon^2+2(\bar\epsilon-\epsilon)^2$ and $0\leq\delta\leq2,$
it follows from~\eqref{eq:local-error-moments} that
\[
  \Exp_\mu[\bar\epsilon^2\delta]\leq2\epsilon^2\Exp_\mu\delta+4\Var_\mu(\bar\epsilon)
  \leq2\epsilon^2\Delta_\mu(F)+\frac{4\epsilon}n.
\]
Combining the last two displays proves~\eqref{eq:lift-deficit}.
\end{proof}

Combining Lemma~\ref{lem:vector-lift} with
Lemma~\ref{lem:transcript-removal}, we obtain a distributional
counterpart of Theorem~\ref{thm:margin-ic}: if $F^{\oplus n}$ has small
vector-output information complexity under $\mu^n,$ then $F$ has families
of unit vectors with small deficit under $\mu.$ 

\begin{lemma}\label{lem:deficit}
For every communication problem $F\colon\Xcal\times\Ycal\to\{-1,1,*\},$
every $\mu\in\mathcal P(\dom F),$ and every integer $n\geq1,$
\[
  \Delta_\mu(F)\leq\frac{5\VIC_{1/100}(F^{\oplus n},\mu^n)+2}n.
\]
\end{lemma}

\begin{proof}
Let $\sigma$ be a protocol for $F^{\oplus n}$ with vector outputs and
vector error $\epsilon\leq1/100$ under $\mu^n,$ and put
$I=\IC(\sigma,\mu^n).$ Lemma~\ref{lem:vector-lift}
gives a protocol $\pi$ with vector outputs that
satisfies~\eqref{eq:lift-cost} and~\eqref{eq:lift-deficit}. By
Lemma~\ref{lem:transcript-removal},
\[
  \Delta_\mu(F)\leq2\Delta_\mu(\pi,F)+2\IC(\pi,\mu)
  \leq64\epsilon^2\Delta_\mu(F)+\frac{160\epsilon+4I}n
  \leq\frac{16}{2500}\,\Delta_\mu(F)+\frac{8+20I}{5n}.
\]
Rearranging, we obtain
\[
  \Delta_\mu(F)\leq\frac{2500}{2484}\cdot\frac{8+20I}{5n}\leq\frac{5I+2}n.
\]
Taking the infimum over $\sigma$ proves the lemma.
\end{proof}

We are now in a position to extract, from a protocol for $F^{\oplus n}$
with constant error, a protocol for a single copy with error of order
$1/n,$ with either vector outputs or split outputs. The proof combines
Lemmas~\ref{lem:block-choice-vector}, \ref{lem:vector-lift}, and~\ref{lem:deficit},
and, for split outputs, Lemma~\ref{lem:vector-rounding}. Protocols of
information cost at most $1/2$ are treated separately, as in the proof of
Theorem~\ref{main:dist-direct-sum}, so that no additive loss arises.

\begin{theorem}[Distributional extraction]\label{thm:dist-extraction}
Let $F\colon\Xcal\times\Ycal\to\{-1,1,*\}$ be a communication problem, and
let $\mu\in\mathcal P(\dom F).$ Then for every integer $n\geq1,$
\begin{align}
  \VIC_{3/(4n)}(F,\mu)
    &\leq\frac{34}n\VIC_{1/100}(F^{\oplus n},\mu^n),
    \label{eq:dist-extraction-vic}\\
  \SIC_{3/(4n)}(F,\mu)
    &\leq\frac{34}n\IC_{1/100}(F^{\oplus n},\mu^n).
    \label{eq:dist-extraction-sic}
\end{align}
\end{theorem}

\begin{proof}
Let $\pi$ be a protocol for $F^{\oplus n}$ with vector outputs and vector
error at most $1/100$ under $\mu^n,$ and put $I=\IC(\pi,\mu^n).$ A protocol
for $F^{\oplus n}$ that announces its output has vector outputs in
dimension $1$: Alice's vector is the announced sign, and Bob's is $1.$ Its
vector error is then its error. It therefore suffices to show that
$\VIC_{3/(4n)}(F,\mu)\leq34I/n$ for every such $\pi,$ and that
$\SIC_{3/(4n)}(F,\mu)\leq34I/n$ if $\pi$ announces its output; taking the
infimum over $\pi$ then proves the theorem.

Suppose first that $I\leq1/2.$ By Lemma~\ref{lem:deficit},
$\Delta_\mu(F)\leq(5I+2)/n<5/n.$ Lemma~\ref{lem:vector-lift}
therefore gives a protocol $\pi'$ with vector outputs such that
$\IC(\pi',\mu)\leq2I/n$ and, as the vector error of $\pi$ is at most
$1/100,$
\[
  \Delta_\mu(\pi',F)\leq\frac{32}{10^4}\cdot\frac5n+\frac{80}{100n}<\frac1n.
\]
Thus $\pi'$ has vector error less than $1/(2n)$ and information cost at
most $2I/n,$ which proves the claim for vector outputs. Now let $\pi$ announce its output. Let $R$ and $M$ be
the public randomness and transcript of $\pi$ on input $(X,Y)\sim\mu^n,$
and put $B=F^{\oplus n}(X,Y).$ For $x$ in the support of $X_1,$ put
\[
  \gamma(x)=\Exp[F(X_1,Y_1)\mid X_1=x],\qquad
  \Gamma=\Exp\abs{\gamma(X_1)}.
\]
We claim that $\Gamma^n>1/4.$ Conditional on $X,$ the inputs
$Y_1,\ldots,Y_n$ are independent, and therefore
$\Exp[B\mid X]=\prod_{i=1}^n\gamma(X_i).$ The function
$t\mapsto h_2((1-t)/2)$ is concave on $[-1,1],$ equals $1$ at $0,$ and
vanishes at $\pm1,$ so that $h_2((1-t)/2)\geq1-\abs t.$ Hence
\[
  H(B\mid X)\geq1-\Exp\prod_{i=1}^n\abs{\gamma(X_i)}=1-\Gamma^n.
\]
On the other hand, $B$ equals the output of $\pi,$ a function of $(R,M),$
unless $\pi$ errs, which happens with probability at most $1/100.$ Hence
$H(B\mid R,M)\leq h_2(1/100),$ and
\[
  H(B\mid X)=H(B\mid X,R)\leq I(B;M\mid X,R)+H(B\mid R,M)
  \leq I(M;Y\mid X,R)+h_2\Bigl(\frac1{100}\Bigr),
\]
where the equality holds because $R$ is independent of $(X,Y),$ and the
last step uses the data-processing inequality, as $B$ is a function of
$(X,Y).$ Since $I(M;Y\mid X,R)\leq I\leq1/2$ and $h_2(1/100)<1/4,$ it
follows that $1-\Gamma^n<3/4,$ which proves our claim. Now consider the
split-output protocol for $F$ without communication in which Alice, on
input $x,$ outputs the more likely value of $F(X_1,Y_1)$ given $X_1=x,$
and Bob outputs $1.$ By~\eqref{eq:prediction-error}, its error under $\mu$
is $(1-\Gamma)/2.$ Since
$\Gamma>4^{-1/n}=e^{-(\ln4)/n}\geq1-(\ln4)/n,$ this error is less than
$(\ln2)/n<3/(4n).$ As the protocol has information cost $0,$ we conclude that
$\SIC_{3/(4n)}(F,\mu)=0.$

In the complementary case, $I>1/2.$ Put $m=\min\{n,\lceil I\rceil\}.$ Then
$1\leq m\leq n,$ and $m\leq2I$ because $\lceil t\rceil\leq2t$ for
$t\geq1/2.$ Partition the $n$ coordinates into $m$ blocks of sizes
$\lfloor n/m\rfloor$ and $\lceil n/m\rceil;$ since $\lceil t\rceil\leq2t$
for $t\geq1,$ each block has size at most $2n/m.$ Put $\beta=100(I+2m),$
so that $\beta>250.$ The deficit of $\pi$ is at most $1/50,$ so that
$W^*_\beta(F^{\oplus n},\mu^n)\leq I+\beta/50,$ and
$(I+2(m-1))/\beta<1/100.$ Lemma~\ref{lem:block-choice-vector} therefore gives
the size $k$ of one of the blocks such that
\[
  \Bigl(1-\frac{W^*_\beta(F^{\oplus k},\mu^k)}\beta\Bigr)^{n/k}
  >1-\frac1{50}-\frac1{100}=\frac{97}{100}.
\]
Since $(1-t)^{n/k}\leq e^{-tn/k}$ for $0\leq t\leq1,$ and
$\ln(100/97)\leq3/97<1/30,$ it follows that
$W^*_\beta(F^{\oplus k},\mu^k)<\beta k/(30n).$ Hence there is a
protocol $\sigma$ for $F^{\oplus k}$ with vector outputs such that
\[
  \IC(\sigma,\mu^k)+2\beta\epsilon<\frac{\beta k}{30n},
\]
where $\epsilon$ is the vector error of $\sigma$ under $\mu^k.$ In
particular, $\epsilon<k/(60n)\leq1/(30m).$

We now bound $\Delta_\mu(F).$ By Lemma~\ref{lem:deficit} and the
inequality $5I+2<9I,$ we have $\Delta_\mu(F)<9I/n.$ If $m=\lceil I\rceil,$
this gives $\Delta_\mu(F)<9m/n.$ If instead $m=n,$ then
$\Delta_\mu(F)\leq1<9m/n$ trivially. In either case,
\[
  32\epsilon\Delta_\mu(F)<32\cdot\frac1{30m}\cdot\frac{9m}n
  =\frac{48}{5n}<\frac{10}k,
\]
where the last step uses $k\leq n.$ Lemma~\ref{lem:vector-lift}, applied to
$\sigma$ with $k$ in place of $n,$ now gives a protocol $\pi'$ with vector
outputs such that
$\IC(\pi',\mu)\leq2\IC(\sigma,\mu^k)/k$ and
\[
  \Delta_\mu(\pi',F)\leq32\epsilon^2\Delta_\mu(F)+\frac{80\epsilon}k
  \leq\frac{90\epsilon}k<\frac3{2n}.
\]
Thus $\pi'$ has vector error less than $3/(4n),$ and
$\IC(\pi',\mu)<\beta/(15n)\leq100I/(3n)<34I/n,$ where the
penultimate step uses $m\leq2I.$ This proves the claim for vector outputs.
For split outputs, let $\epsilon'$ be the vector error of $\pi',$ so that
$\epsilon'<3/(4n)$ and $\VIC_{\epsilon'}(F,\mu)\leq\IC(\pi',\mu).$
Lemma~\ref{lem:vector-rounding} and the inequality
$(8/\ln2)\,\epsilon'(1-\epsilon')\leq12\epsilon'=6\Delta_\mu(\pi',F)$ yield
\begin{align*}
  \SIC_{3/(4n)}(F,\mu)
  &\leq\IC(\pi',\mu)+6\Delta_\mu(\pi',F)\\
  &\leq\frac{2\IC(\sigma,\mu^k)+540\epsilon}k\\
  &\leq\frac{2(\IC(\sigma,\mu^k)+2\beta\epsilon)}k\\
  &<\frac\beta{15n}\\
  &<\frac{34I}n,
\end{align*}
where the third step uses $\beta\geq135,$ and the last uses $m\leq2I.$
This completes the proof.
\end{proof}

\subsection{Optimal distributional XOR lemma}\label{subsec:dist-xor}

We now prove the distributional XOR lemma for vector
outputs, the main result of this section. Its key ingredient
is Theorem~\ref{thm:dist-extraction}, which we extend
from constant error to any smaller error
$\epsilon.$ We then derive the consequences for the standard
model of announced outputs. We first restate the distributional XOR
lemma for convenience.

\begin{restated}{Theorem~\ref*{main:dist-xor}}{Distributional XOR lemma,
restated}
For every communication problem $F\colon\Xcal\times\Ycal\to\{-1,1,*\},$
every probability distribution $\mu$ on the domain of $F,$ every integer
$n\geq1,$ and every $\epsilon\in[0,1/400],$
\begin{equation*}
  \frac n{34}\VIC_{100\epsilon/n}(F,\mu)
    \leq\VIC_\epsilon(F^{\oplus n},\mu^n)
    \leq n\VIC_{\epsilon/n}(F,\mu).\tag{\ref*{eq:dist-xor}}
\end{equation*}
Moreover,
\begin{equation*}
  \VIC_0(F^{\oplus n},\mu^n)=n\VIC_0(F,\mu).\tag{\ref*{eq:dist-xor-zero}}
\end{equation*}
\end{restated}

\begin{proof}
As in the proofs of Theorems~\ref{thm:additive-conversion}
and~\ref{thm:extraction}, we run a given protocol independently on
disjoint blocks of coordinates. Let $\sigma$ be a protocol for
$F^{\oplus k}$ with vector outputs and vector error $\epsilon'$ under
$\mu^k,$ and let $r\geq1$ be an integer. Run $\sigma$ with independent
randomness on each of $r$ disjoint blocks of $k$ coordinate pairs, and let
each party output the tensor product of his or her $r$ output vectors.
Since
$\langle u_1\otimes\cdots\otimes u_r,v_1\otimes\cdots\otimes v_r\rangle
=\prod_j\langle u_j,v_j\rangle,$ the signed inner product of the new
output vectors with respect to $F^{\oplus rk}$ is the product of the
signed inner products of the $r$ runs. Under $\mu^{rk},$ the $r$ blocks
are independent with distribution $\mu^k,$ and so are the $r$ runs, so
that the expected signed inner product is
$(1-2\epsilon')^r\geq1-2r\epsilon'.$ The result is therefore a protocol
for $F^{\oplus rk}$ with vector outputs and vector error at most
$r\epsilon'$ under $\mu^{rk},$ and its information cost is at most
$r\IC(\sigma,\mu^k)$ by~\eqref{eq:independent-distinct-inputs}.

For the upper bound, fix $\eta>0$ and a protocol $\pi$ for $F$ with
vector outputs, vector error at most $\epsilon/n$ under $\mu,$ and
$\IC(\pi,\mu)\leq\VIC_{\epsilon/n}(F,\mu)+\eta.$ The construction just
described, applied to $\pi$ with $k=1$ and $r=n,$ gives a protocol for
$F^{\oplus n}$ with vector outputs, vector error at most $\epsilon$ under
$\mu^n,$ and information cost at most $n(\VIC_{\epsilon/n}(F,\mu)+\eta).$
Letting $\eta\to0$ proves the upper bound.

For the lower bound, suppose first that $\epsilon>0,$ and put
$r=\lfloor1/(100\epsilon)\rfloor.$ Then $r\epsilon\leq1/100,$ and
$r\geq3/(400\epsilon)$ because $\lfloor t\rfloor\geq3t/4$ for $t\geq4.$ Let
$\pi$ be a protocol for $F^{\oplus n}$ with vector outputs and vector error
at most $\epsilon$ under $\mu^n.$ The same construction, applied to $\pi$
with $k=n,$ gives a protocol for $F^{\oplus rn}$ with vector outputs, vector error
at most $1/100$ under $\mu^{rn},$ and information cost at most
$r\IC(\pi,\mu^n).$ Hence~\eqref{eq:dist-extraction-vic}, applied with $rn$
in place of $n,$ gives $\VIC_{3/(4rn)}(F,\mu)\leq34\IC(\pi,\mu^n)/n.$ Since
$3/(4rn)\leq100\epsilon/n,$ taking the infimum over $\pi$ proves the lower
bound.

Finally, let $\sigma$ be a protocol for $F^{\oplus n}$ with vector outputs
and vector error $0$ under $\mu^n,$ and let $\rho$ be the protocol in the
proof of Lemma~\ref{lem:vector-lift}. The random variable $L$
in that proof vanishes almost surely, so~\eqref{eq:bias-energy}
and~\eqref{eq:embedding-weight} show that, on every input
$(x,y)\in\supp\mu,$ the vectors $a$ and $b$ at the end of $\rho$ satisfy
$F(x,y)\,a=b$ and $\|b\|=1$ almost surely. Let Alice output $a$ and Bob
output $b,$ each replacing a vector whose norm is not $1$ by a fixed unit
vector. The result is a protocol for $F$ with vector outputs, vector error
$0$ under $\mu,$ and information cost $\IC(\sigma,\mu^n)/n.$ Hence
$\VIC_0(F,\mu)\leq\VIC_0(F^{\oplus n},\mu^n)/n.$ Together with the upper
bound, this proves~\eqref{eq:dist-xor-zero} and the lower bound
in~\eqref{eq:dist-xor} for $\epsilon=0.$
\end{proof}

For announced outputs, we obtain Theorem~\ref{main:xor-announced} from
the introduction, a distributional XOR lemma in which the single copy is
measured with vector outputs or with split outputs. We restate it for
convenience.

\begin{restated}{Theorem~\ref*{main:xor-announced}}{Distributional XOR
lemma for announced outputs, restated}
For every communication problem $F\colon\Xcal\times\Ycal\to\{-1,1,*\},$
every probability distribution $\mu$ on the domain of $F,$ every integer
$n\geq1,$ and every $\epsilon\in[0,1/400],$
\begin{align*}
  \frac n{34}\VIC_{100\epsilon/n}(F,\mu)
    &\leq\IC_\epsilon(F^{\oplus n},\mu^n)
    \leq n\VIC_{\epsilon/n}(F,\mu)+5,\tag{\ref*{eq:announced-xor-vic}}\\
  \frac n{34}\SIC_{100\epsilon/n}(F,\mu)
    &\leq\IC_\epsilon(F^{\oplus n},\mu^n)
    \leq n\SIC_{\epsilon/n}(F,\mu)+2.\tag{\ref*{eq:announced-xor-sic}}
\end{align*}
\end{restated}

\begin{proof}
The lower bound in~\eqref{eq:announced-xor-vic} follows from that
in~\eqref{eq:announced-xor-sic} and the first inequality in
Theorem~\ref{thm:vector-comparison}. For the upper bound
in~\eqref{eq:announced-xor-vic}, the last inequality in
Theorem~\ref{thm:vector-comparison} and the upper bound
in~\eqref{eq:dist-xor} give
$\IC_\epsilon(F^{\oplus n},\mu^n)\leq\VIC_\epsilon(F^{\oplus n},\mu^n)+5
\leq n\VIC_{\epsilon/n}(F,\mu)+5.$

It remains to prove~\eqref{eq:announced-xor-sic}. For the upper bound, fix
$\eta>0$ and a split-output protocol $\pi$ for
$F$ with $\err_\mu(\pi,F)\leq\epsilon/n$ and
$\IC(\pi,\mu)\leq\SIC_{\epsilon/n}(F,\mu)+\eta.$ As in the proof of
Theorem~\ref{main:characterization}, run independent copies of $\pi$ on
the $n$ coordinate pairs, let each party announce the XOR of his or her
$n$ shares, and let both parties output the XOR of the two announced
signs. Under $\mu^n,$ each coordinate pair has distribution $\mu,$ so the
union bound gives error at most $\epsilon.$ By
Lemma~\ref{lem:independent-executions}, the $n$ runs of $\pi$ have
information cost at most $n(\SIC_{\epsilon/n}(F,\mu)+\eta)$ under $\mu^n,$
and the two announced signs increase it by at most $2.$ Letting $\eta\to0$
proves the upper bound.

For the lower bound, suppose first that $\epsilon>0,$ let $r$ be as in the
proof of Theorem~\ref{main:dist-xor}, and let $\pi$ be a protocol for
$F^{\oplus n}$ with error at most $\epsilon$ under $\mu^n.$ Run $\pi$ with
independent randomness on each of $r$ disjoint blocks of $n$ coordinate
pairs, and output the XOR of the $r$ answers. By the union bound
and~\eqref{eq:independent-distinct-inputs}, the result is a protocol for
$F^{\oplus rn}$ with error at most $r\epsilon\leq1/100$ under $\mu^{rn}$
and information cost at most $r\IC(\pi,\mu^n).$ Hence~\eqref{eq:dist-extraction-sic},
applied with $rn$ in place of $n,$ gives
$\SIC_{3/(4rn)}(F,\mu)\leq34\IC(\pi,\mu^n)/n,$ and taking the infimum over
$\pi$ proves the lower bound. For $\epsilon=0,$
Lemma~\ref{lem:vector-rounding} and Theorem~\ref{thm:vector-comparison} give
$\SIC_0(F,\mu)=\VIC_0(F,\mu).$ By Theorems~\ref{main:dist-xor}
and~\ref{thm:vector-comparison}, it follows that
$n\SIC_0(F,\mu)=\VIC_0(F^{\oplus n},\mu^n)\leq\IC_0(F^{\oplus n},\mu^n),$
which implies the lower bound.
\end{proof}

\section{AI statement}

The author developed and refined the geometric approaches of
Sections~\ref{subsec:rounding} and~\ref{subsec:vector-lift}
with the assistance of OpenAI's GPT-6 Astra model, and also used the model
for literature exploration and editorial polishing. The author assumes full
responsibility for the content of this paper, including its mathematical
validity and bibliographic attributions.
\bigskip

\raggedbottom

\appendix
\section{Making public randomness finite}\label{app:finite}

This appendix shows that every protocol can be replaced by a protocol with
finite public randomness, with no change in its errors and with an
arbitrarily small change in its information cost. Recall from
Section~\ref{subsec:protocols} that a protocol $\pi$ with public randomness
$R$ runs the private-coin protocol $\pi_R,$ and that the probabilities of
error of $\pi$ and its information cost under every input distribution are
the averages over $R$ of the corresponding quantities for $\pi_R.$ It
therefore suffices to replace the distribution of $\pi_R$ by a distribution
on finitely many private-coin protocols with the same averages. The
following fact, a special case of a theorem of Richter~\cite{Ric57}, makes
this possible. We include its short proof, which uses only the separating
hyperplane theorem.

\begin{fact}[Finite support]\label{fact:finite-support}
Let $V$ be a bounded random vector in $\mathbb R^N,$ and let
$S\subseteq\mathbb R^N$ be a set with $\Prob[V\in S]=1.$ Then $\Exp V$ is a
convex combination of at most $N+1$ points of $S.$
\end{fact}

\begin{proof}
By Carath\'eodory's theorem, it suffices to show that $\Exp V$ lies in the
convex hull of $S.$ We proceed by induction on $N,$ the case $N=0$ being
trivial. Let $N\geq1,$ and suppose that $\Exp V$ lies outside the convex
hull of $S.$ By the separating hyperplane theorem, there is a nonzero
$\phi\in\mathbb R^N$ such that $\langle\phi,z\rangle\leq\langle\phi,\Exp V\rangle$
for every $z\in S.$ The random variable $\langle\phi,\Exp V-V\rangle$ is then
nonnegative with probability $1$ and has mean zero, so that it vanishes
with probability~$1.$ Thus, with probability $1,$ the vector $V$ lies in
the intersection of $S$ with the hyperplane
$\{z:\langle\phi,z\rangle=\langle\phi,\Exp V\rangle\},$ which contains
$\Exp V.$ Identifying this hyperplane with $\mathbb R^{N-1}$ and applying
the inductive hypothesis, we find that $\Exp V$ lies in the convex hull of
this intersection, a contradiction.
\end{proof}

To treat all output conventions at once, we define the \emph{loss} of a
protocol on a legal input $(x,y)$ of a communication problem $F$ as follows.
The loss of a protocol with vector outputs on $(x,y)$ is its expected deficit
$\Exp[1-F(x,y)\langle u(x,T),v(y,T)\rangle]$ on $(x,y),$ and the loss of any
other protocol is its probability of error on $(x,y)$ under the output
convention at hand. Losses lie in $[0,2],$ and two protocols with the same
loss on every legal input have the same worst-case error and the same
error or deficit under every input distribution. The loss of a protocol
$\pi$ on each legal input is the average over $R$ of the loss of $\pi_R$ on
that input.

\begin{proposition}\label{prop:finite-fixed}
Let $\pi$ be a protocol for a communication problem $F,$ with public
randomness $R,$ and let $\mu_1,\ldots,\mu_m\in\mathcal P(\Xcal\times\Ycal),$
where $m\geq0.$ Then there are values $r_1,\ldots,r_k$ of $R,$ where
$k\leq\abs{\dom F}+m+1,$ and probabilities $p_1,\ldots,p_k$ with sum $1,$ such
that the protocol $\widetilde\pi$ that publicly chooses $i$ with probability
$p_i$ and runs $\pi_{r_i}$ has the same loss as $\pi$ on every legal input
and satisfies $\IC(\widetilde\pi,\mu_j)=\IC(\pi,\mu_j)$ for $j=1,\ldots,m.$
\end{proposition}

\begin{proof}
For each value $r$ of $R,$ let $v(r)$ be the vector formed by the losses of
$\pi_r$ on the legal inputs and by
$\IC(\pi_r,\mu_1),\ldots,\IC(\pi_r,\mu_m).$ These $\abs{\dom F}+m$ coordinates
lie in $[0,2]$ and, by the proof of Fact~\ref{fact:concavity}, in
$[0,\log(|\Xcal||\Ycal|)],$ respectively. Thus $v(R)$ is a bounded random
vector, and its expectation is formed by the losses of $\pi$ and by
$\IC(\pi,\mu_1),\ldots,\IC(\pi,\mu_m),$ since these are averages over $R.$
Fact~\ref{fact:finite-support}, applied with $S$ the set of all vectors
$v(r),$ gives values $r_1,\ldots,r_k$ of $R,$ where $k\leq\abs{\dom F}+m+1,$
and probabilities $p_1,\ldots,p_k$ with sum $1,$ such that
$\Exp v(R)=\sum_ip_iv(r_i).$ Publicly choosing among protocols averages
their losses and information costs, as
in~\eqref{eq:public-mixture-cost} and~\eqref{eq:public-mixture-error}.
Hence the right-hand side is formed in the same way by the losses of
$\widetilde\pi$ and by $\IC(\widetilde\pi,\mu_1),\ldots,\IC(\widetilde\pi,\mu_m),$
which proves the proposition.
\end{proof}

For finitely many input distributions, Proposition~\ref{prop:finite-fixed} thus
preserves the information cost exactly. For all input distributions at
once, the continuity bound of Fact~\ref{fact:concavity}, which does
not depend on the protocol, reduces the problem to finitely many input
distributions.

\begin{proposition}\label{prop:finite-all}
Let $\pi$ be a protocol for a communication problem $F,$ and let $\eta>0.$
Then there is a protocol $\widetilde\pi$ that publicly chooses one of
finitely many values $r$ of the public randomness of $\pi,$ with suitable
probabilities, and runs $\pi_r,$ such that $\widetilde\pi$ has the same loss
as $\pi$ on every legal input and
\[
  \abs{\IC(\widetilde\pi,\mu)-\IC(\pi,\mu)}\leq\eta
\]
for every $\mu\in\mathcal P(\Xcal\times\Ycal).$
\end{proposition}

\begin{proof}
Choose $\Delta\in(0,1/2]$ with $2(\Delta\log(|\Xcal||\Ycal|)+2h_2(\Delta))\leq\eta.$ The
simplex $\mathcal P(\Xcal\times\Ycal)$ is compact, so it contains finitely
many distributions $\mu_1,\ldots,\mu_m$ such that every
$\mu\in\mathcal P(\Xcal\times\Ycal)$ satisfies $\TV(\mu,\mu_j)\leq \Delta$ for
some $j.$ Let $\widetilde\pi$ be the protocol of
Proposition~\ref{prop:finite-fixed} for these distributions. Then
$\widetilde\pi$ has the same loss as $\pi$ on every legal input, and
$\IC(\widetilde\pi,\mu_j)=\IC(\pi,\mu_j)$ for every $j.$ Now let $\mu$ be
arbitrary, and choose $j$ with $\TV(\mu,\mu_j)\leq \Delta.$ Since $h_2$ is
increasing on $[0,1/2],$ Fact~\ref{fact:concavity}, applied to
$\widetilde\pi$ and to $\pi,$ gives
\[
  \abs{\IC(\widetilde\pi,\mu)-\IC(\pi,\mu)}
  \leq\abs{\IC(\widetilde\pi,\mu)-\IC(\widetilde\pi,\mu_j)}
     +\abs{\IC(\pi,\mu_j)-\IC(\pi,\mu)}
  \leq\eta.\qedhere
\]
\end{proof}

Every information complexity measure in this paper is defined through
infima over protocols, in which a protocol enters only through its losses
on the legal inputs and its information cost under a single input
distribution or under every input distribution. By Propositions~\ref{prop:finite-fixed}
and~\ref{prop:finite-all}, none of these measures changes if the protocols
in its definition are required to have finite public
randomness. This conclusion holds for every
value of the error parameter, including zero. For the same reason, a proof
about a given protocol may assume that it has finite public randomness
whenever its conclusion depends on the protocol only through its losses
on the legal inputs and its information costs under finitely many input
distributions. The protocol $\widetilde\pi$ in either proposition runs
private-coin protocols of $\pi,$ so it inherits every property that all of
these share. For example, if the outputs of $\pi$ are functions of the
respective party's input, the public randomness, and the transcript, then
so are those of $\widetilde\pi;$ and $\widetilde\pi$ sends no messages if
$\pi$ sends none.

\section{Margins of partial matrices}\label{app:norms}

This appendix proves the facts about margins used in
Section~\ref{subsec:margin}: the multiplicativity of margin under tensor products
(Proposition~\ref{prop:margin-tensor}) and
the relationship between margin and discrepancy
(Proposition~\ref{prop:disc-margin}). The main tool is the factorization
norm $\gamma_2.$

For a real matrix $A\in\mathbb R^{\Xcal\times\Ycal},$ define 
\[
  \gamma_2(A)=\min_{u_x,v_y}\,\max_{x\in\Xcal,\,y\in\Ycal}\|u_x\|_2\,\|v_y\|_2,
\]
where the minimum ranges over all dimensions $d\geq1$ and all vectors
$u_x,v_y\in\mathbb R^d$ such that $A_{xy}=\langle u_x,v_y\rangle$ for all
$x$ and $y.$ It is well known that $\gamma_2$ is a norm on
$\mathbb R^{\Xcal\times\Ycal}.$ Its dual norm is
\[
  \gamma_2^*(A)=\max_{\gamma_2(B)\leq1}\langle A,B\rangle,
  \qquad
  \langle A,B\rangle=\sum_{x\in\Xcal}\sum_{y\in\Ycal}A_{xy}B_{xy}.
\]

We will use the following description of the unit ball of $\gamma_2.$
If $B_{xy}=\langle u_x,v_y\rangle$ for some unit vectors $u_x$ and $v_y,$
then $\gamma_2(B)\leq1$ by definition. Conversely, suppose that
$\gamma_2(B)\leq1,$ and let $B_{xy}=\langle u_x,v_y\rangle$ be a
factorization with $\|u_x\|_2\,\|v_y\|_2\leq1$ for all $x$ and $y.$
Multiplying all $u_x$ by a common positive factor and all $v_y$ by its
reciprocal, we may assume that $\|u_x\|_2\leq1$ and $\|v_y\|_2\leq1$ for
all $x$ and $y.$ Then the unit vectors
\[
  \Bigl(u_x,\sqrt{1-\|u_x\|_2^2},\,0\Bigr),
  \qquad
  \Bigl(v_y,\,0,\sqrt{1-\|v_y\|_2^2}\Bigr)
\]
in $\mathbb R^{d+2}$ have the same inner products as $u_x$ and $v_y.$
Thus $\gamma_2(B)\leq1$ if and only if $B_{xy}=\langle u_x,v_y\rangle$ for
some unit vectors $u_x$ and $v_y.$ In view
of~\eqref{eq:margin-definition}, the margin of a partial real matrix
$A\in(\mathbb R\cup\{*\})^{\Xcal\times\Ycal}$ can therefore be
alternatively written as
\begin{equation}\label{eq:margin-gamma2}
  \margin(A)=\max_{\gamma_2(B)\leq1}\,\min_{(x,y)\in\dom A}A_{xy}B_{xy},
\end{equation}
where $B$ ranges over the real matrices in $\mathbb R^{\Xcal\times\Ycal}.$

Both $\gamma_2$ and its dual norm are multiplicative under the tensor product of
Section~\ref{subsec:margin}: for all real matrices $A$ and $B,$
\begin{align}
  \gamma_2(A\otimes B)&=\gamma_2(A)\gamma_2(B),
       \label{eq:gamma2-tensor}\\
  \gamma_2^*(A\otimes B)&=\gamma_2^*(A)\gamma_2^*(B).
       \label{eq:gamma2-dual-tensor}
\end{align}
These identities are well known. For $\gamma_2,$ see the factorization
norm treatments of Linial, Mendelson, Schechtman, and
Shraibman~\cite{LMSS07} and Linial and Shraibman~\cite{LS09a}, in the
framework originating with Grothendieck~\cite{Gro53}. For $\gamma_2^*,$
see also the XOR-game product theorem of Cleve, Slofstra, Unger, and
Upadhyay~\cite{CSUU08}. Lee, Shraibman, and \v{S}palek~\cite{LSS08} give
simple proofs of both identities using semidefinite programming duality.

For total sign matrices, the following characterization of margin is due
to Linial, Mendelson, Schechtman, and Shraibman~\cite{LMSS07}, who state
it for the margin complexity $1/\margin(A);$ see also~\cite{LS09a}. We
generalize it to partial real matrices.

\begin{theorem}[Dual characterization of margin]\label{thm:margin-dual}
For every partial real matrix $A\in(\mathbb R\cup\{*\})^{\Xcal\times\Ycal},$
\begin{equation}\label{eq:margin-dual}
  \margin(A)=\min_{\mu\in\mathcal P(\dom A)}\gamma_2^*(\mu\circ A),
\end{equation}
where $\mu\circ A\in\mathbb R^{\Xcal\times\Ycal}$ is the real matrix with
entries $\mu(x,y)A_{xy}$ on $\dom A$ and zeros elsewhere.
\end{theorem}

\begin{proof}
Let $\Dcal=\dom A.$ For every $\mu\in\mathcal P(\Dcal)$ and every real
matrix $B$ with $\gamma_2(B)\leq1,$
\[
  \min_{(x,y)\in\Dcal}A_{xy}B_{xy}
  \leq\sum_{(x,y)\in\Dcal}\mu(x,y)A_{xy}B_{xy}
  =\langle\mu\circ A,B\rangle
  \leq\gamma_2^*(\mu\circ A).
\]
By~\eqref{eq:margin-gamma2}, it follows that
$\margin(A)\leq\gamma_2^*(\mu\circ A)$ for every $\mu.$

For the reverse inequality, write $A\circ B\in\mathbb R^{\Dcal}$ for the
vector with entries $A_{xy}B_{xy},$ where $(x,y)\in\Dcal,$ and consider
the sets
\begin{align*}
  K&=\{A\circ B:B\in\mathbb R^{\Xcal\times\Ycal},\ \gamma_2(B)\leq1\},\\
  O&=\{\omega\in\mathbb R^{\Dcal}:\omega_{xy}>\margin(A)
       \text{ for all }(x,y)\in\Dcal\}.
\end{align*}
The set $K$ is convex, being a linear image of the unit ball of
$\gamma_2,$ and the set $O$ is open and convex.
By~\eqref{eq:margin-gamma2}, the two sets are disjoint. By the separating
hyperplane theorem, there is a nonzero $\phi\in\mathbb R^{\Dcal}$ such
that
\[
  \langle\phi,A\circ B\rangle\leq\langle\phi,\omega\rangle
\]
whenever $\gamma_2(B)\leq1$ and $\omega\in O.$ Taking $B=0$ and letting a
single coordinate of $\omega$ tend to infinity shows that $\phi\geq0.$
Thus $\phi$ is a positive multiple of some $\mu\in\mathcal P(\Dcal),$ and
letting $\omega$ tend to the constant vector with entries $\margin(A)$
gives $\langle\mu,A\circ B\rangle\leq\margin(A)$ whenever
$\gamma_2(B)\leq1.$ Since
$\langle\mu,A\circ B\rangle=\langle\mu\circ A,B\rangle,$ it follows that
$\gamma_2^*(\mu\circ A)\leq\margin(A).$ Hence this $\mu$ attains the
minimum in~\eqref{eq:margin-dual}.
\end{proof}

\begin{proof}[Proof of Proposition~\ref{prop:margin-tensor}]
Let unit vectors $u_x,v_y$ attain $\margin(A),$ and let unit vectors
$u'_{x'},v'_{y'}$ attain $\margin(B).$ Then the unit vectors
$u_x\otimes u'_{x'}$ and $v_y\otimes v'_{y'}$ satisfy
\[
  A_{xy}B_{x'y'}\langle u_x\otimes u'_{x'},v_y\otimes v'_{y'}\rangle
  =A_{xy}\langle u_x,v_y\rangle\cdot B_{x'y'}\langle u'_{x'},v'_{y'}\rangle
  \geq\margin(A)\margin(B)
\]
for all $((x,x'),(y,y'))\in\dom(A\otimes B),$ because margins are
nonnegative (Section~\ref{subsec:margin}). Hence
$\margin(A\otimes B)\geq\margin(A)\margin(B).$ Conversely, let $\mu$ and
$\nu$ attain~\eqref{eq:margin-dual} for $A$ and $B.$ Then
$\mu\otimes\nu\in\mathcal P(\dom(A\otimes B))$ and
$(\mu\otimes\nu)\circ(A\otimes B)=(\mu\circ A)\otimes(\nu\circ B).$ By
Theorem~\ref{thm:margin-dual} and~\eqref{eq:gamma2-dual-tensor},
\[
  \margin(A\otimes B)
  \leq\gamma_2^*\bigl((\mu\circ A)\otimes(\nu\circ B)\bigr)
  =\gamma_2^*(\mu\circ A)\gamma_2^*(\nu\circ B)
  =\margin(A)\margin(B).
\]
The final assertion follows by induction on $n,$ since
$A_{F^{\oplus n}}=A_F^{\otimes n}.$
\end{proof}

\begin{proof}[Proof of Proposition~\ref{prop:disc-margin}]
For sets $S\subseteq\Xcal$ and $T\subseteq\Ycal,$ the matrix
$\1_S\1_T^{\mathsf T}$ has $\gamma_2$ norm at most $1,$ by its
one-dimensional factorization, and so does its negation. Hence, for
every $\mu\in\mathcal P(\dom F),$
\[
  \disc_\mu(F)
  =\max_{S,T}\,\abs{\langle\mu\circ A_F,\1_S\1_T^{\mathsf T}\rangle}
  \leq\gamma_2^*(\mu\circ A_F).
\]
Minimizing over $\mu$ and applying Theorem~\ref{thm:margin-dual} proves
$\disc(F)\leq\margin(F).$
\end{proof}

\section{A spectral lower bound for information complexity}\label{app:inner-product}

This appendix proves a lower bound on information complexity in terms of
the spectral norm, and deduces the linear lower bound for the inner
product function used in the proof of Lemma~\ref{lem:xor-linear}. For a
real matrix $A$ with rows indexed by $\Xcal$ and columns
indexed by $\Ycal,$ the \emph{spectral norm} of $A$ is
\[
  \|A\|=\max\bigl\{u^{\mathsf T}Av:u\in\mathbb R^{\Xcal},\
        v\in\mathbb R^{\Ycal},\ \|u\|_2=\|v\|_2=1\bigr\}.
\]
Thus $\abs{u^{\mathsf T}Av}\leq\|A\|\|u\|_2\|v\|_2$ for all
$u\in\mathbb R^{\Xcal}$ and $v\in\mathbb R^{\Ycal}.$ By the Cauchy--Schwarz
inequality, $\|A\|^2$ is at most the sum of the squares of the entries of
$A.$ 

\begin{theorem}[Spectral lower bound]\label{thm:spectral-lower}
Let $F\colon\Xcal\times\Ycal\to\{-1,1\}$ be given, and let $\mu$ be the uniform distribution on
$\Xcal\times\Ycal.$ Then for every $0\leq\epsilon<1/2,$
\begin{equation}\label{eq:spectral-lower}
  \IC_\epsilon(F)\geq\IC_\epsilon(F,\mu)
  \geq\frac{1-2\epsilon}4\log\frac{|\Xcal||\Ycal|}{\|A_F\|^2}-3,
\end{equation}
where $A_F=[F(x,y)]_{x\in\Xcal,y\in\Ycal}$ is the
characteristic matrix of $F.$ 
\end{theorem}

For error $1/3,$ a bound of the form~\eqref{eq:spectral-lower}, with
unspecified constants, also follows from the discrepancy bound of
Braverman and Weinstein~\cite{BW} (Fact~\ref{fact:discrepancy}):
applying the definition of $\|A_F\|$ to the indicator vectors of a
rectangle shows that the discrepancy of $F$ under the uniform distribution
is at most $\|A_F\|/\sqrt{|\Xcal||\Ycal|}.$ The direct
proof below works for every error $\epsilon<1/2$ and gives explicit constants.
Through Lemma~\ref{lem:xor-linear}, they enter the explicit constant in
Theorem~\ref{main:xor}.

\begin{proof}[Proof of Theorem~\ref{thm:spectral-lower}]
The first inequality is immediate from the definitions of
$\IC_\epsilon(F)$ and $\IC_\epsilon(F,\mu).$
For the second, fix a protocol $\pi$ with $\err_\mu(\pi,F)\leq\epsilon;$ by
Proposition~\ref{prop:finite-fixed}, we may assume that $\pi$ has finite
public randomness. Let $(X,Y)\sim\mu,$ and let $R$ and $M$ be the public randomness and
transcript of $\pi$ on input $(X,Y).$ The marginals $\mu_{\Xcal}$ and
$\mu_{\Ycal}$ of $\mu$ are uniform, and $\mu=\mu_{\Xcal}\times\mu_{\Ycal}.$
For values $r$ of $R$ and $m$ of $M,$ the rectangle
property~\eqref{eq:rectangle} shows that the conditional probability of
$(X,Y)=(x,y)$ given $(R,M)=(r,m)$ is proportional to
$a_{r,m}(x)b_{r,m}(y),$ because $\mu$ is uniform and $R$ is independent of
the input. Hence this conditional law is a product
$p_{r,m}\times q_{r,m}$ of distributions $p_{r,m}\in\mathcal P(\Xcal)$ and
$q_{r,m}\in\mathcal P(\Ycal).$ In particular, $X$ and $Y$ are independent
conditional on $(R,M),$ as well as unconditionally.

We first express the information cost of $\pi$ through these
distributions. 
Using~\eqref{eq:information-cost-R}, we have 
\[\IC(\pi,\mu)=I(R,M;X\mid Y)+I(R,M;Y\mid X).\]
Moreover,
\begin{align*}
  I(R,M;X\mid Y)&=H(X)-H(X\mid R,M)\\
  &=I(R,M;X)\\
&=\Exp\kl{p_{R,M}}{\mu_{\Xcal}},
\end{align*}
where the first step uses the two independence
statements, and the third step is justified by 
the remark following~\eqref{eq:kl-mixture} since
the distributions $p_{R,M}$ average
to $\mu_{\Xcal}.$ An analogous
argument applies to $I(R,M;Y\mid X).$ As a result,
\begin{equation}\label{eq:spectral-cost}
  \IC(\pi,\mu)=\Exp d_{R,M},
  \qquad
  d_{r,m}=\kl{p_{r,m}}{\mu_{\Xcal}}+\kl{q_{r,m}}{\mu_{\Ycal}}.
\end{equation}

Next, we relate the error of $\pi$ to the conditional bias of $F(X,Y).$ The
output of $\pi$ is a function of $(R,M).$ By~\eqref{eq:prediction-error},
conditional on $(R,M),$ this output differs from $F(X,Y)$ with probability
at least $\bigl(1-\abs{\Exp[F(X,Y)\mid R,M]}\bigr)/2.$ Averaging over
$(R,M)$ and using $\err_\mu(\pi,F)\leq\epsilon,$ we obtain
\begin{equation}\label{eq:spectral-bias}
  \Exp\abs[\big]{\Exp[F(X,Y)\mid R,M]}\geq1-2\epsilon.
\end{equation}

We now bound the conditional bias given $(R,M)=(r,m)$ in terms of
$d_{r,m}.$ Abbreviate $p=p_{r,m}$ and $q=q_{r,m},$ and regard them as
vectors in $\mathbb R^{\Xcal}$ and $\mathbb R^{\Ycal}.$ Then
\[
  \Exp[F(X,Y)\mid(R,M)=(r,m)]=p^{\mathsf T}A_Fq.
\]
Let $\theta\geq1$ be a parameter to be set later, and consider the sets
\[
  S=\Bigl\{x\in\Xcal:p(x)>\frac{\theta}{|\Xcal|}\Bigr\},
  \qquad
  T=\Bigl\{y\in\Ycal:q(y)>\frac{\theta}{|\Ycal|}\Bigr\},
\]
on which $p$ and $q$ exceed $\theta$ times the uniform probabilities. These
sets have small probability unless $d_{r,m}$ is large:
\begin{equation}\label{eq:spectral-heavy}
  \bigl(p(S)+q(T)\bigr)\log\theta\leq d_{r,m}+2.
\end{equation}
Indeed, if $S$ is nonempty, then $|S|\,\theta/|\Xcal|<p(S)\leq1,$ so that
$\mu_{\Xcal}(S)<1/\theta.$ The data-processing inequality, applied to the
indicator of $S,$ then gives
\begin{align*}
  \kl{p}{\mu_{\Xcal}}
    &\geq p(S)\log\frac{p(S)}{\mu_{\Xcal}(S)}
         +\bigl(1-p(S)\bigr)\log\frac{1-p(S)}{1-\mu_{\Xcal}(S)}\\
    &=p(S)\log\frac1{\mu_{\Xcal}(S)}
         +\bigl(1-p(S)\bigr)\log\frac1{1-\mu_{\Xcal}(S)}
         -h_2\bigl(p(S)\bigr)\\
    &\geq p(S)\log\theta-1,
\end{align*}
where the last inequality uses $\mu_{\Xcal}(S)<1/\theta$ and $h_2\leq1.$
The same bound $\kl{p}{\mu_{\Xcal}}\geq p(S)\log\theta-1$ holds
trivially if $S$ is empty. An analogous argument gives
$\kl{q}{\mu_{\Ycal}}\geq q(T)\log\theta-1,$ and adding the two bounds
proves~\eqref{eq:spectral-heavy}.

Next, let $p'\in\mathbb R^{\Xcal}$ agree with $p$ outside $S$ and vanish on
$S,$ and define $q'\in\mathbb R^{\Ycal}$ from $q$ and $T$ in the same way.
The entries of $p'$ lie in $[0,\theta/|\Xcal|]$ and sum to at most one, so
that $\|p'\|_2^2\leq\theta/|\Xcal|;$ similarly,
$\|q'\|_2^2\leq\theta/|\Ycal|.$ It follows that
\[
  \abs{p'^{\mathsf T}A_Fq'}\leq\|A_F\|\|p'\|_2\|q'\|_2
  \leq\frac{\theta\|A_F\|}{\sqrt{|\Xcal||\Ycal|}}.
\]
Since the entries of $A_F$ have absolute value one, we also have
$\abs{(p-p')^{\mathsf T}A_Fq}\leq p(S)$ and
$\abs{p'^{\mathsf T}A_F(q-q')}\leq q(T).$ Writing
$p^{\mathsf T}A_Fq=p'^{\mathsf T}A_Fq'+(p-p')^{\mathsf T}A_Fq
+p'^{\mathsf T}A_F(q-q'),$ we conclude that
\[
  \abs[\big]{\Exp[F(X,Y)\mid(R,M)=(r,m)]}\leq p(S)+q(T)+
      \frac{\theta\|A_F\|}{\sqrt{|\Xcal||\Ycal|}}.
\]
We multiply this inequality by $\log\theta\geq0$ and apply
\eqref{eq:spectral-heavy}. Since $(r,m)$ was
arbitrary, this yields
\[
  \abs[\big]{\Exp[F(X,Y)\mid R,M]}\log\theta\leq d_{R,M}+
      2+\frac{\|A_F\|\,\theta\log\theta}{\sqrt{|\Xcal||\Ycal|}}.
\]
Taking expectations and applying~\eqref{eq:spectral-cost}
and~\eqref{eq:spectral-bias}, we obtain
\[
  (1-2\epsilon)\log\theta\leq\IC(\pi,\mu)+
      2+\frac{\|A_F\|\,\theta\log\theta}{\sqrt{|\Xcal||\Ycal|}}.
\]
Finally, set $\theta=(|\Xcal||\Ycal|/\|A_F\|^2)^{1/4};$ the remark
preceding the theorem gives $\|A_F\|^2\leq|\Xcal||\Ycal|,$ so that
$\theta\geq1$ as required. Then
$\|A_F\|\,\theta/\sqrt{|\Xcal||\Ycal|}=1/\theta,$ and the previous
display becomes
$(1-2\epsilon)\log\theta\leq\IC(\pi,\mu)+2+(\log\theta)/\theta
\leq\IC(\pi,\mu)+3.$ Taking the infimum over
$\pi$ proves~\eqref{eq:spectral-lower}.
\end{proof}

\begin{corollary}\label{cor:ip-lower}
Let $n\geq1$ be an integer, and let
$\IP_n\colon\{0,1\}^n\times\{0,1\}^n\to\{-1,1\}$ be the inner product
function, given by $\IP_n(x,y)=(-1)^{\sum_{i=1}^nx_iy_i}.$ Then for every
$0\leq\epsilon<1/2,$
\[
  \IC_\epsilon(\IP_n)
  \geq\frac{(1-2\epsilon)\bigl(1-h_2(\epsilon)\bigr)}{16}\,n.
\]
\end{corollary}

\begin{proof}
The characteristic matrix $A_{\IP_n}$ of $\IP_n$ satisfies
$A_{\IP_n}^{\mathsf T}A_{\IP_n}=2^n I,$ whence $\|A_{\IP_n}\|=2^{n/2}.$ Now
Theorem~\ref{thm:spectral-lower} gives
\[ \IC_\epsilon(\IP_n)\geq\frac{1-2\epsilon}4\,n-3.\]
Since $\IP_n$ is not constant,
Proposition~\ref{prop:local-variation} implies the
alternative lower bound 
\[\IC_\epsilon(\IP_n)\geq1-h_2(\epsilon).\]
The claimed result follows from the former bound if
$n\geq16/(1-2\epsilon),$ and from the latter bound
otherwise.
\end{proof}

\end{document}